\documentstyle[11pt,axodraw]{article}

\newcommand{\be}{\begin{equation}}
\newcommand{\ee}{\end{equation}}
\newcommand{\bea}{\begin{eqnarray}}
\newcommand{\eea}{\end{eqnarray}}

\newcommand{\nn}{\nonumber}
\newcommand{\eq}[1]{Eq.~(\ref{#1})}

\newcommand{\NPB}[3]{Nucl.\ Phys.\ {\bf B{#1}} (19{#2}) {#3}}
\newcommand{\PRD}[3]{Phys.\ Rev.\ {\bf D{#1}} (19{#2}) {#3}}
\newcommand{\PLB}[3]{Phys.\ Lett.\ {\bf B{#1}} (19{#2}) {#3}}
\newcommand{\PRL}[3]{Phys.\ Rev.\ Lett.\ {\bf {#1}} (19{#2}) {#3}}
\newcommand{\RMP}[3]{Rev.\ Mod.\ Phys.\ {\bf {#1}} (19{#2}) {#3}}
\newcommand{\AP}[3]{Ann.\ Phys.\ {\bf {#1}} (19{#2}) {#3}}
\newcommand{\ZPC}[3]{Z.\ Phys.\ {\bf C{#1}} (19{#2}) {#3}}

\newcommand{\Lcl}{{\cal L}_{\mbox{\scriptsize cl}}}
\newcommand{\Lgf}{{\cal L}_{\mbox{\scriptsize gf}}}
\newcommand{\Lgh}{{\cal L}_{\mbox{\scriptsize gh}}}
\newcommand{\Lclint}{{\cal L}_{\mbox{\scriptsize cl}}^{
\mbox{\scriptsize int}}}
\newcommand{\eeff}{e_{\mbox{\scriptsize eff}}}
\newcommand{\geff}{g_{\mbox{\scriptsize eff}}}
\newcommand{\aeff}{\alpha_{\mbox{\scriptsize eff}}}
\newcommand{\aseff}{\alpha_{\mbox{\scriptsize s\,eff}}}
\newcommand{\CUV}{C_{\mbox{\scriptsize UV}}}
\newcommand{\MSbar}{\overline{\mbox{MS}}}
\newcommand{\Tr}{\mbox{Tr}}
\newcommand{\bfq}{\mbox{\bf q}} 
\newcommand{\ks}{{\mbox k \!\!\! /}}
\newcommand{\ps}{{\mbox p \!\!\! /}}
\newcommand{\qs}{{\mbox q \!\!\! /}}
\newcommand{\sdot}{\!\cdot\!}

\begin{document}

\begin{titlepage}

\begin{center}

{\Large Centre de Physique Th\'eorique
- CNRS - Luminy, Case 907}

{\Large F-13288 Marseille Cedex 9 - France}

\vspace{1 cm}

{\Large {\bf THE GAUGE-INDEPENDENT QCD EFFECTIVE CHARGE}}

\vspace{1 cm}

{\large\bf N.J. Watson}\footnote{Present address:
Institut de Physique Nucl\'eaire,
Universit\'e Paris-Sud,
F-91406 Orsay Cedex,
France.
Email: watson@ipno.in2p3.fr }

\vspace{1 cm}

{\bf Abstract}
\end{center}

It is shown how the QED concept of a gauge-, scale- and scheme-independent
one-loop effective charge can be extended directly 
at the diagrammatic level to QCD,
thus justifying explicitly the ``naive non-abelianization''
prescription used in renormalon calculus.
It is first argued that, for on-shell external fields and
at the strictly one-loop level,
the required gluon self-energy-like function is precisely that
obtained from S-matrix elements
via the pinch technique. The generalization of the pinch
technique to explicitly off-shell processes is then introduced.
It is shown how, as a result of a fundamental cancellation
among conventional perturbation theory diagrams, encoded
in the QCD Ward identities,
the pinch technique one-loop gluon self-energy 
$i\hat{\Pi}_{\mu\nu}^{ab}(q)$
remains gauge-independent and universal
regardless of the fact that the ``external'' fields
in the given process are off-shell.
This demonstration involves a simple technique enabling
the isolation, in an arbitrary gauge, of 
$i\hat{\Pi}_{\mu\nu}^{ab}(q)$
from subclasses of up to several hundred diagrams at once.
Furthermore, it is shown how this
one-loop cancellation mechanism iterates for the
subclasses of $n$-loop diagrams containing implicitly the
Dyson chains of $n$ one-loop self-energies
$i\hat{\Pi}_{\mu\nu}^{ab}(q)$.
The gauge cancellation required for the Dyson
summation of  $i\hat{\Pi}_{\mu\nu}^{ab}(q)$
is thus demonstrated explicitly in a general class of ghost-free
gauges for all orders $n$.

\bigskip


\bigskip

\noindent Number of figures: 13

\bigskip

\noindent June 1996 (revised Jan.\ 1997)

\noindent CPT-96/P.3347

\bigskip

\noindent anonymous ftp or gopher: cpt.univ-mrs.fr

\end{titlepage}


\setcounter{equation}{0}
\def\theequation{1.\arabic{equation}}

{\Large\bf 1. Introduction}

\vspace{5pt}

\noindent
In quantum electrodynamics (QED), the renormalized
photon propagator $i\Delta_{R\mu\nu}(q)$ depends on a function
$d_{R}(q^{2})$ which is gauge-independent at all $q^{2}$:
\be\label{qedprop}
i\Delta_{R\mu\nu}(q)
=
\frac{i}{q^{2} + i\epsilon}\Biggl\{
\biggl(-g_{\mu\nu} +\frac{q_{\mu}q_{\nu}}{q^{2}}\biggr)d_{R}(q^{2}) 
- \xi \frac{q_{\mu}q_{\nu}}{q^{2}}\Biggr\}
\ee
where $\xi$ is the gauge parameter in the class of covariant 
gauges ($\xi = 0$ is the Landau gauge)
and $R$ denotes renormalized quantities.
The function $d_{R}$ is given by the infinite Dyson series
in the one-particle-irreducible photon self-energy $\Pi_{R}$,
illustrated in Fig.\ 1. 

\begin{center}
\begin{picture}(400,100)(11,-60)

\Photon(10, 0)(30, 0){5}{2}
\GCirc(40,0){10}{0}
\Photon(50, 0)(70, 0){5}{2}

\put( 90,0){\makebox(0,0)[c]{\large =}}

\Photon(110, 0)(150, 0){5}{4}

\put(165,0){\makebox(0,0)[c]{\large +}}

\Photon(180, 0)(200, 0){5}{2}
\GCirc(210,0){10}{0.7}
\Photon(220, 0)(240, 0){5}{2}

\put(255,0){\makebox(0,0)[c]{\large +}}

\Photon(270, 0)(290, 0){5}{2}
\GCirc(300,0){10}{0.7}
\Photon(310, 0)(330, 0){5}{2}
\GCirc(340,0){10}{0.7}
\Photon(350, 0)(370, 0){5}{2}

\put(383,0){\makebox(0,0)[l]{\large +\,\,.\,\,.\,\,.}}

\put(200,-55){\makebox(0,0)[c]{\footnotesize
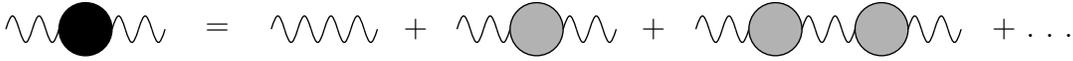
Fig.\ 1. The Dyson series in the 1PI photon self-energy
$\Pi_{R}(q^{2})$.}}

\end{picture}
\end{center}

\noindent
Summing this series, the propagator Eq.\ (\ref{qedprop}) then
naturally defines an {\em effective charge} for the theory 
\cite{gellmannlow}:
\be\label{qedeeff}
e_{R}^{2}d_{R}(q^{2})
=
\frac{e_{R}^{2}}{1 - \Pi_{R}(q^{2})}
=
\eeff^{2}(q^{2})
=       
4\pi\aeff(q^{2}).
\ee
This effective charge has the following properties:
\begin{itemize}
\item It is gauge-independent, since the photon self-energy is
gauge-independent to all orders.
\item It is both renormalization scale- ($\mu$-) and scheme-independent.
This is a direct result of the QED Ward identity giving the relation
$Z_{1} = Z_{2}$: 
\be
\eeff^{2}(q^{2}) 
= 
e_{R}^{2}d_{R}(q^{2})
=
\biggl(\frac{Z_{2}^{2}Z_{3}}{Z_{1}^{2}}e^{2}\biggr)
\biggl(\frac{1}{Z_{3}}d(q^{2})\biggr)
= 
e^{2}d(q^{2})
\ee
so that $\eeff^{2}(q^{2})$
can be expressed entirely in terms of bare quantities. 
\item At $-q^{2}/m^{2} \to \infty$, where $m$ is the fermion mass,
it matches on to the {\em running coupling}
$\bar{e}(q^{2})$ defined\footnote{
It is important to distinguish between the effective charge,
defined from the radiative corrections to the propagator
$i\Delta_{R\mu\nu}(q)$ by Eq.\ (\ref{qedeeff}), and the running coupling,
defined from the renormalization group $\beta$-function by
$d\bar{e}(t,e_{R})/dt = \beta(\bar{e})$,
with $t = \frac{1}{2}\log(-q^{2}/\mu^{2})$ and boundary condition
$\bar{e}(t=0,e_{R}) = e_{R}$: only for asymptotic $q^{2}$
do they coincide.}
from the renormalization group: at the one-loop level, 
\be\label{qederun}
\lim_{-q^{2}/m^{2} \to \infty}\eeff^{2}(q^{2})  
= 
\bar{e}^{2}(q^{2})
=
\frac{e_{R}^{2}}
{1- (e_{R}^{2}/16\pi^{2})\beta_{1}\log (-q^{2}/\mu^{2})}
\ee
where 
$\beta_{1} = +4/3$ is the coefficient of the first term
$e^{3}/16\pi^{2}$
in the perturbative expansion of the QED $\beta$-function.
\item At $q^{2} = 0$ (the Thomson limit), it matches on to the fine
structure constant: 
$\aeff(0) = \alpha = 1/137.036\ldots$. 
\item For $-q^{2}/m^{2} \ll 1$, it gives the correction
to the Coulomb Law interaction between two static heavy charges.
\end{itemize}
(For an account of renormalization schemes in QED, see \cite{coq}.)
Using this effective charge, it is then possible
to account for a well-defined, infinite, 
gauge-independent subset of radiative corrections 
to a photon mediating the tree level interaction 
between two fermion currents
essentially just by making the replacement
$e \rightarrow \eeff(q^{2})$
in the tree level photon-fermion-fermion vertices.

In quantum chromodynmanics (QCD), in addition to couplings to fermions
similar to that in QED, the gauge bosons also couple directly to one
another in triple and quadruple gauge vertices.
As a result of these self-couplings, the gauge boson self-energy,
while it remains transverse as required by a Slavnov-Taylor identity,
becomes gauge-dependent: for SU($N$) QCD 
with $n_{\!f}$ flavours of massless fermion,
at the one-loop level,
\be
\Pi(\xi,q^{2})
=
\frac{g^{2}}{16\pi^{2}}\Biggl\{
\biggl[\biggl(-\frac{13}{6} + \frac{\xi}{2}\biggr)N 
+ \frac{2}{3}n_{\!f}\biggr]
\biggl[ -\CUV + \ln\biggl(\frac{-q^{2}}{\mu^{2}}\biggr)\biggr]
+ \biggl(\frac{97}{36} +\frac{\xi}{2} + \frac{\xi^{2}}{4}\biggr)N
- \frac{10}{9}n_{\!f} \Biggr\}
\label{covPi}
\ee
where $\CUV = 1/\epsilon + \ln (4\pi) -\gamma_{E}$ with
$\gamma_{E}$ Euler's constant (we work always in $d = 4-2\epsilon$
dimensions and with 't Hooft mass scale $\mu$).
Furthermore, in QCD the Slavnov-Taylor identities do not require 
the relation $Z_{1} = Z_{2}$. As a result of these differences,
although it is possible to sum the renormalized gluon self-energy
in a Dyson series to give a radiatively-corrected gluon
propagator, the quantity defined by analogy with the QED
effective charge Eq.\ (\ref{qedeeff})
is in general gauge-, scale- and scheme-dependent,
and at asymptotic $q^{2}$ 
does not\footnote{An exception \cite{baucoq} is for $\xi =  -3$.} 
match on to the QCD running coupling
$\bar{g}(q^{2})$ defined from the renormalization group.
The simple QED correspondence between the gauge boson self-energy
and an effective charge for the theory is therefore lost. 

The existence of an effective charge for QCD 
analogous to that of QED is nevertheless explicitly assumed in
renormalon calculus \cite{renormalon}. 
The usual framework for renormalon analyses is
the $1/n_{\!f}$ expansion. In this framework, the leading order
corrections to the tree level gluon propagator are given by
chains of single fermion loops, i.e.\ precisely the Dyson series
illustrated in Fig.\ 1 with the
blobs here each representing a single fermion loop. 
The unrenormalized gluon self-energy due to such a loop is given by
\be
\Pi^{(f)}(q^{2})
=
\frac{g^{2}n_{\!f}}{16\pi^{2}}\Biggl\{ 
b_{1}\biggl[
-\CUV + \ln\biggl(\frac{-q^{2}}{\mu^{2}}\biggr)\biggr]
-\frac{10}{9}\Biggr\} 
\label{nfPi}
\ee
where $b_{1} = +2/3$. After renormalization,
this self-energy may be summed, exactly as in QED, to give a
renormalized gluon propagator, the effects of which may be accounted for
by an effective charge at the vertices at each end of the gluon line.
However, at this order in the $1/n_{\!f}$
expansion, gluons do not contribute to the $\beta$-function for the
rescaled coupling $gn_{\!f}^{1/2}$, 
so that $b_{1}$ is positive and the fundamental
QCD property of asymptotic freedom is absent. In order to introduce
this property, the usual procedure is simply to replace
$b_{1} = 2/3$ with the full
one-loop value $b_{1} = 2/3 - 11N/(3n_{\!f})$. Although
asymptotic freedom is thereby recovered in the $1/n_{\!f}$ 
framework, this ``naive non-abelianization'' prescription 
for the QCD effective charge 
leaves unanswered the following two basic questions:
\begin{enumerate}
\item What gauge-independent combination of gluon and ghost
loop corrections to the tree level gluon propagator is one summing,
together with the fermion loop corrections, to obtain this effective
charge with the full non-abelian one-loop $\beta$-function 
coefficient for the logarithmic term?
Although it follows from operator product expansion arguments that, 
at least at asymptotic $q^{2}$, such a gauge-independent 
combination must exist, a direct diagrammatic interpretation
is lacking. The absence of such a direct interpretation obstructs
a proper understanding of the approximations made in
renormalon calculus.
\item What is the contribution of
this gauge-independent combination of gluon and ghost fields
to the constant term (in a given renormalization scheme)
in the corresponding 
one-loop self-energy-like function? The $\MSbar$
fermion loop contribution $-10/9$ is well-defined, 
but in the absence of an unambiguous extension of the QED concept
of an effective charge to QCD, the contribution from gluon and ghost
loops remains undefined. This ambiguity in the value of the constant
term leads to different estimates of renormalon
contributions to physical observables (see e.g.\ \cite{gamsir}). 
\end{enumerate}

One possible approach to these questions is provided by the background
field method (BFM) \cite{abbott}.
In this approach, the gauge fields are split into
background and quantum components, and the gauge-fixing for the quantum
fields then chosen such that the effective action remains explicitly
invariant under gauge transformations of the background 
fields.\footnote{It is important to distinguish between
background gauge invariance, i.e.\ invariance with respect to
gauge transformations of the background fields, 
and quantum gauge independence, i.e.\ independence with respect to
changes in the value of the quantum field gauge fixing
parameter $\xi_{Q}$: the former does not imply the latter.}
As a result of this exact background gauge invariance, 
the 1PI background field $n$-point functions obey to all
orders in perturbation theory the same set of Ward identities as
the corresponding tree level functions. In particular, this
results in the QED-like identity $Z_{1} = Z_{2}$. 
Thus, the effective
charge constructed in the BFM via the Dyson summation of 
the background gluon self-energy is 
renormalization scale- and scheme-independent, just as
in QED, and at asymptotic $q^{2}$ matches on to the
QCD running coupling $\bar{g}(q^{2})$ i.e.\ 
the coefficient of
the logarithm is given by the full one-loop coefficient
$\beta_{1} = -\frac{11}{3}N + \frac{2}{3}n_{\!f}$ of the 
SU($N$) QCD $\beta$-function. 
However, while the ultra-violet divergent parts of
the background field $n$-point functions,
and hence the renormalization counterterms, are independent
of the quantum gauge fixing parameter $\xi_{Q}$
(Kallosh's theorem \cite{kallosh}), the finite
parts are quantum gauge-dependent:
for the background gluon self-energy at the one-loop level,
\be
\tilde{\Pi}(\xi_{Q},q^{2})
=
\frac{g^{2}}{16\pi^{2}}\Biggl\{ 
\beta_{1}
\biggl[-\CUV + \ln\biggl(\frac{-q^{2}}{\mu^{2}}\biggr)\biggr] 
+ \biggl(\frac{67}{9} - \frac{(1-\xi_{Q})(7+\xi_{Q})}{4}\biggr)N
- \frac{10}{9}n_{\!f}
\Biggl\}.
\label{bfmPi}
\ee
The BFM effective charge, though renormalization
scale- and scheme-independent, therefore
remains gauge-dependent.
Thus, while the BFM specifies the one-loop corrections
to the background gluon propagator which result,
independent of $\xi_{Q}$, in the
coefficient $\beta_{1}$ for the logarthmic term
in the self-energy Eq.\ (\ref{bfmPi}),
it does not specify a unique gluon and ghost
contribution to the constant term.
Furthermore, and more fundamentally still, in the BFM formalism
the background fields by construction do not 
propagate inside loops. The BFM effective charge therefore
cannot be used to account for the radiative corrections to a
quantum gluon propagating across, for example, 
a fermion loop.\footnote{Recall that in the BFM, the Feynman rules
for the interactions purely among quantum fields are identical to 
those in a conventional covariant gauge.}

The most promising approach to the above questions
is provided by the pinch technique (PT).
Originally introduced by Cornwall 
\cite{cornwall1}-\cite{cornwall4}, 
and since much developed by
Papavassiliou and collaborators \cite{papa1}-\cite{papa5},
the PT is based on the observation that one-loop diagrams which
appear to give only vertex or box corrections to tree level
processes in fact implicitly contain self-energy-like
components. Exploiting this observation, the PT provides
a well-defined algorithm for the rearrangement of the
conventional gauge-dependent one-loop contributions
to S-matrix elements into individually
gauge-independent one-loop self-energy-like, vertex-like and
box-like contributions. 

In order to illustrate this, consider the S-matrix element for the
four-fermion scattering process
$\psi_{i}^{\scriptscriptstyle (f)}(p_{1})
\psi_{i'}^{\scriptscriptstyle (f')}(p_{1}') 
\rightarrow \psi_{j}^{\scriptscriptstyle (f)}(p_{2})
\psi_{j'}^{\scriptscriptstyle (f')}(p_{2}')$ 
in SU($N$) QCD. With flavours $f\neq f'$,
the one-loop diagrams for this process are as shown in 
Fig.\ 2. The contribution of the diagram 

\pagebreak

\begin{center}
\begin{picture}(425,595)(73,-100)



\put(235,510){\makebox(0,0)[r]
{\small $\psi_{j'}^{\scriptscriptstyle{(f')}}(p_{2}')$}}
\put(235,450){\makebox(0,0)[r]
{\small $\psi_{i'}^{\scriptscriptstyle{(f')}}(p_{1}')$}}
\ArrowLine(245,450)(245,480)
\ArrowLine(245,480)(245,510)
\Photon(245,480)(265,480){2.5}{2}
\GCirc(285,480){20}{0.7}
\Photon(305,480)(325,480){2.5}{2}
\ArrowLine(325,450)(325,480)
\ArrowLine(325,480)(325,510)
\put(285,440){\makebox(0,0)[c]{\small (a)}}
\put(335,510){\makebox(0,0)[l]
{\small $\psi_{j}^{\scriptscriptstyle{(f)}}(p_{2})$}}
\put(335,450){\makebox(0,0)[l]
{\small $\psi_{i}^{\scriptscriptstyle{(f)}}(p_{1})$}}


\Line(100,360)(100,420)
\Photon(100,390)(140,390){2.5}{4}
\Photon(140,390)(180,410){2.5}{4.5}
\Photon(140,390)(180,370){-2.5}{4.5}
\Line(180,360)(180,420)
\put(140,350){\makebox(0,0)[c]{\small (b)}}

\put(285,405){\makebox(0,0)[c]{pinch}}
\put(270,390){\vector(1, 0){30}}


\Line(100,280)(100,340)
\Photon(100,287.5)(180,287.5){2.5}{8}
\Line(180,280)(180,340)
\PhotonArc(180,315)(17.5,-90,90){2.5}{6.5}
\put(140,270){\makebox(0,0)[c]{\small (c)}}


\Line(100,200)(100,260)
\Photon(100,230)(180,230){2.5}{8}
\Line(180,200)(180,260)
\PhotonArc(180,230)(17.5,-90,90){2.5}{6.5}
\put(140,190){\makebox(0,0)[c]{\small (d)}}

\put(235,230){\makebox(0,0)[c]
{$ \left. \begin{array}{c}  \\ \\ \\ \\ \\ \\ \\ \\ \\ \\ \\ \\ \\ \\ 
\end{array} \right\}$}}
\put(285,245){\makebox(0,0)[c]{\small pinch}}
\put(270,230){\vector(1, 0){30}}


\Line(100,120)(100,180)
\Photon(100,172.5)(180,172.5){2.5}{8}
\Line(180,120)(180,180)
\PhotonArc(180,145)(17.5,-90,90){2.5}{6.5}
\put(140,110){\makebox(0,0)[c]{\small (e)}}


\Line(100, 40)(100,100)
\Photon(100, 87.5)(180, 87.5){2.5}{8}
\Photon(100, 52.5)(180, 52.5){2.5}{8}
\Line(180, 40)(180,100)
\put(140, 30){\makebox(0,0)[c]{\small (f)}}

\put(285, 85){\makebox(0,0)[c]{\small pinch}}
\put(270, 70){\vector(1, 0){30}}


\Line(390,360)(390,420)
\Photon(390,390)(430,390){2.5}{4}
\PhotonArc(450,390)(17.5,22.5,382.5){2.5}{12}
\Line(470,360)(470,420)
\put(430,350){\makebox(0,0)[c]{\small (g)}}


\Line(390,200)(390,260)
\Photon(390,230)(470,230){2.5}{8}
\Line(470,200)(470,260)
\PhotonArc(490,230)(17.5,22.5,382.5){2.5}{12}
\put(430,190){\makebox(0,0)[c]{\small (h)}}


\Line(390, 40)(390,100)
\PhotonArc(410,70)(17.5,90,-90){2.5}{6.5}
\Photon(410, 87.5)(450, 87.5){-2.5}{4.5}
\Photon(410, 52.5)(450, 52.5){ 2.5}{4.5}
\PhotonArc(450,70)(17.5,-90,90){2.5}{6.5}
\Line(470, 40)(470,100)
\put(430, 30){\makebox(0,0)[c]{\small (i)}}


\ArrowLine(245,-50)(245,-20)
\ArrowLine(245,-20)(245, 10)
\Photon(245,-20)(265,-20){2.5}{2}
\GCirc(285,-20){20}{0.7}
\put(285,-20){\makebox(0,0)[c]{\large PT}} 
\Photon(305,-20)(325,-20){2.5}{2}
\ArrowLine(325,-50)(325,-20)
\ArrowLine(325,-20)(325, 10)
\put(285,-60){\makebox(0,0)[c]{\small (j)}}

\put(285,-85){\makebox(0,0)[c]{\footnotesize 
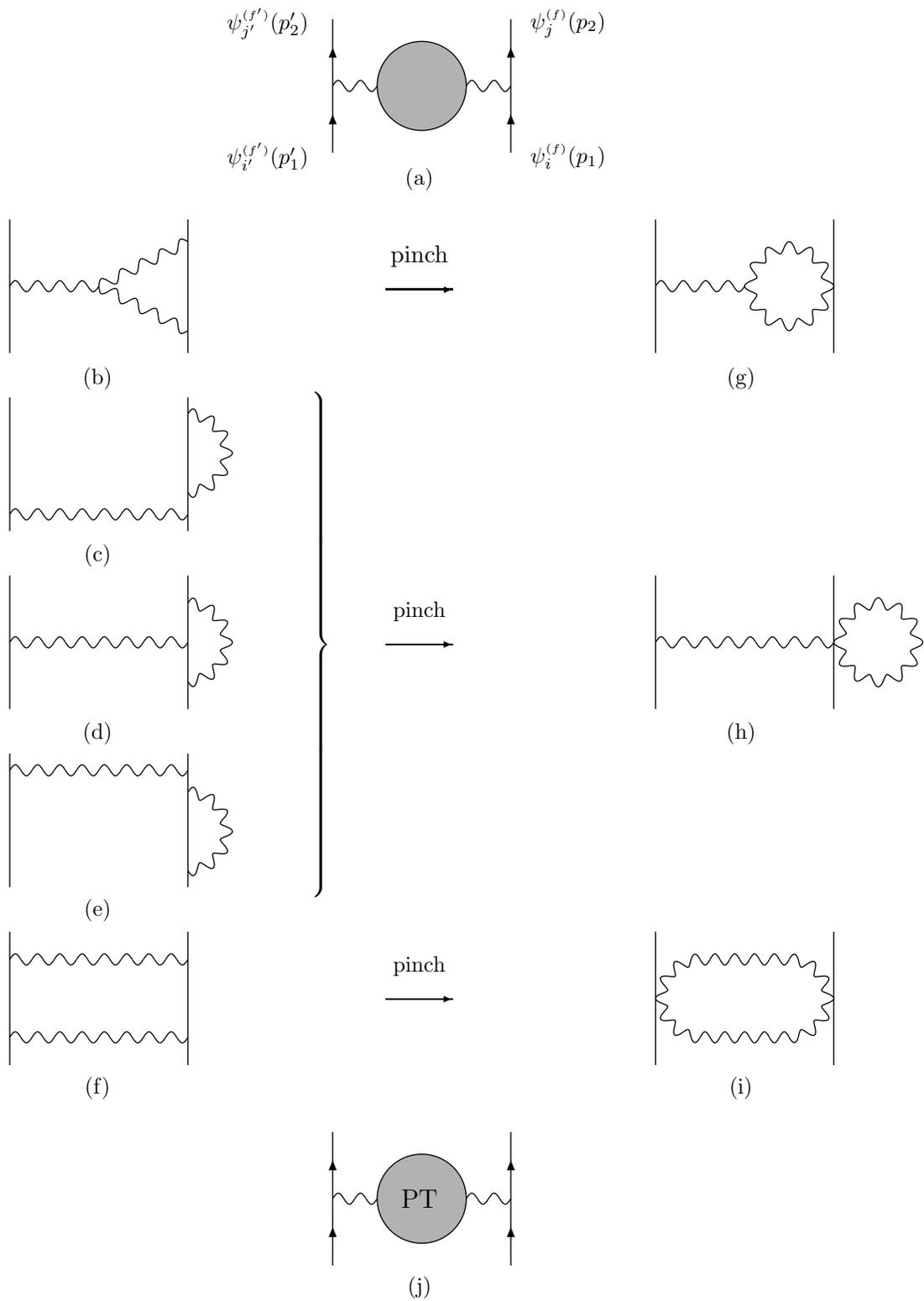
Fig.\ 2. The complete set of one-loop radiative corrections to the
four-fermion }}
\put(285,-95){\makebox(0,0)[c]{\footnotesize 
process 
$\psi_{i}^{\scriptscriptstyle{(f)}}(p_{1})
 \psi_{i'}^{\scriptscriptstyle{(f')}}(p_{1}')
\rightarrow
\psi_{j}^{\scriptscriptstyle{(f)}}(p_{2})
\psi_{j'}^{\scriptscriptstyle{(f')}}(p_{2}')$, 
together with their pinch parts.}}

\end{picture}
\end{center}

\noindent
in Fig.\ 2(a) involving the
conventional gauge boson self-energy Eq.\ (\ref{covPi}) is given by
\begin{equation}\label{fig2a}
\mbox{Fig.\,\,2(a)}
=
\Bigl(\overline{u}_{j'}ig\gamma_{\mu}T_{j'i'}^{m}u_{i'}\Bigr)
\,\frac{-i}{q^{2}}\Pi(\xi,q^{2})\,
\Bigl(\overline{u}_{j}ig\gamma_{\mu}T_{ji}^{m}u_{i}\Bigr)
\end{equation}
where $q = p_{2}-p_{1} = p_{1}'-p_{2}'$
(the colour indices $i,i',j,j'$ are not summed).
The effect of the PT algorithm is to isolate the self-energy-like
components of the remaining diagrams in Fig.\ 2,
{\it defined as those parts of the Feynman integrands for the
diagrams which have the form of 
a (gauge-dependent) function 
of $q$ and the loop integration variable only,
between two tree level gluon-fermion-fermion vertices}. 
These are the pinch parts of the diagrams, shown in Figs.\ 2(g)-(i).
These pinch parts arise \cite{cornwall1}-\cite{papa5}
when, in the Feynman integrands, 
factors of longitudinal gluon four-momentum $k_{\mu}$ occur
contracted into the Dirac matrices $\gamma_{\mu}$
associated with the gluon-fermion-fermion vertices,
triggering the elementary Ward identity
\be
\ks
=
S^{-1}(p+k,m) - S^{-1}(p,m)
\ee
where $S^{-1}(p,m) = \ps -m$ is the inverse fermion propagator.
The effect of these inverse propagators is to 
cancel the fermion propagators in the integrand.
Adding these pinch parts of the vertex and box diagrams 
to the diagram Fig.\ 2(a) involving the
conventional gauge boson two-point function 
and carrying out the loop integrations gives the full one-loop
self-energy-like contribution to the four fermion process,
illustrated in Fig.\ 2(j), and defines the PT
gauge boson self-energy $\hat{\Pi}(q^{2})$:
\begin{equation}\label{fig2j}
\mbox{Fig.\,\,2(j)}
=
\Bigl(\overline{u}_{j'}ig\gamma_{\mu}T_{j'i'}^{m}u_{i'}\Bigr)
\,\frac{-i}{q^{2}}\hat{\Pi}(q^{2})\,
\Bigl(\overline{u}_{j}ig\gamma_{\mu}T_{ji}^{m}u_{i}\Bigr).
\end{equation}

Up to a trivial dependence on the external spinors, the component
Eq.\ (\ref{fig2j}) of the S-matrix element for the four-fermion
process depends only on the $t$-channel momentum transfer $q^{2}$, and
not on the $s$-channel momentum transfer $(p_{1}+p_{1}')^{2}$ 
or the external
fermions' masses. It must therefore be individually gauge-independent,
as can be verified by explicit calculation. 
One obtains \cite{cornwall1}\cite{cornwall2}
\be
\hat{\Pi}(q^{2})
=
\frac{g^{2}}{16\pi^{2}}
\Biggl\{ \beta_{1}
\biggl[-\CUV + \ln\biggl(\frac{-q^{2}}{\mu^{2}}\biggr)\biggr] 
+ \frac{67}{9}N - \frac{10}{9}n_{\!f}
\Biggl\}.
\label{ptPi}
\ee
Thus the PT definition Eq.\ (\ref{fig2j}) 
of the QCD$\,$\footnote{
In QED, the pinch parts of the corresponding 
vertex and box diagrams vanish
identically as a result of the abelian structure of the theory
and the PT self-energy
reduces precisely to the conventional photon self-energy.}
gauge boson self-energy
results in the full one-loop $\beta$-function coefficient $\beta_{1}$
as the coefficient of the logarithmic
term, together with an unambiguous value for the constant term.
It is emphasized that {\em all} terms in 
Eq.\ (\ref{ptPi}) are fully gauge-independent.

The PT gauge-independent one-loop
improper three- and four-point functions 
\cite{cornwall4}\cite{papa2}
are defined in a similar way
to the two-point function Eq.\ (\ref{fig2j}). 
The resulting functions, in addition to being gauge-independent, 
display many theoretically desirable properties. In particular, they
satisfy the same Ward identities as the corresponding tree level 
quantities. Thus, $Z_{1} = Z_{2}$ in the PT framework.
Furthermore, it has been shown by Degrassi and Sirlin 
\cite{degsir} that the PT
algorithm in fact corresponds to a systematic use of current algebra,
thus demonstrating explicitly the PT's basis in the underlying
gauge symmetry of the theory. 
Also, the PT gluon self-energy has been shown to be universal,
i.e.\ independent of the species of external field in the
S-matrix \cite{njw1}.
Lastly, it has been observed that the PT gauge-independent
one-loop $n$-point functions {\em coincide} with the
background field $n$-point functions computed in the
Feynman quantum gauge $\xi_{Q} = 1$, both in QCD 
\cite{hashimoto}
and the standard electroweak model 
\cite{denner}\cite{edrnjw}.

In general, the PT avoids the gauge-dependence of conventional
$n$-point functions by working directly with S-matrix elements
for the interaction of on-shell fields. 
Using the fact that S-matrix elements are known from general
proofs to be gauge-independent,
it then follows from simple kinematic arguments that the
various one-loop functions obtained in the PT must themselves be
individually gauge-independent.
However, in QED the photon self-energy is gauge-independent
to all orders in perturbation theory regardless of whether or not
it occurs as a component of an S-matrix element.
Thus, the renormalized QED propagator Eq.\ (\ref{qedprop}), and hence
the QED effective charge, may be used in processes in which the
photon couples to fermions which are explicitly {\em off-shell}.

Furthermore, the construction of the effective charge in QED
involves, through the Dyson summation, diagrams occurring at
all orders in perturbation theory. This summation is essential
if the effective charge is to satisfy the known constraints from
the renormalization group.
An heuristic outline of the Dyson summation 
of the PT self-energy has been given by Papavassiliou and Pilaftsis 
\cite{papapil1} involving the re-allocation ``by hand'' 
of the pinch contributions required 
from multi-loop diagrams 
in the Feynman gauge to form the Dyson chains.
However, a direct 
demonstration of the required generalization of
the PT gauge cancellation mechanism 
from one loop to all orders is lacking.

Lastly, the fundamental criticism \cite{denner}
of the PT is that it is merely a prescription for the division of S-matrix
elements into individually gauge-independent components: because
the PT $n$-point functions are extracted from S-matrix elements,
rather than directly from the path integral 
according to some basic field-theoretic principle,
they apparently represent only a particular choice for this decomposition.
In the particular case of the gluon two-point function,
this issue of uniqueness leads to the question of whether, 
away from the asymptotic region governed by the renormalization group,
the QED concept of an effective charge can be
extended unambiguously to QCD at all.

In this paper, it is shown how the QED 
concept of a gauge-, scale- and scheme-independent
effective charge may be extended directly and unambiguously 
at the diagrammatic level to QCD. 
The starting point (Sec.~2) is a simple
re-analysis of the basic idea of an effective charge.
It is argued that for on-shell external fields and at the
strictly one-loop level,
the required self-energy-like function is precisely 
that given by the S-matrix PT. 
After listing the tree level SU($N$) $n$-point functions and
their Ward identities needed subsequently (Sec.~3), 
the generalization of the PT to 
arbitrary off-shell processes
is introduced\footnote{The original 
construction by Cornwall of the PT gluon self-energy 
in fact involved the couplings of gluons to off-shell 
scalar fields $\Phi^{a}$ contributing to the manifestly
gauge-invariant Green's function $G(x,y) = \langle 0|T[
\Tr\,\Phi^{\dagger}(x)\Phi(x)
\Tr\,\Phi^{\dagger}(y)\Phi(y)]|0\rangle$.
However, this particular approach was not pursued.}
(Sec.~4).
The PT one-loop gauge boson self-energy,
or ``effective'' two-point function, is defined (Sec.~4.1)
for the general case of explicitly off-shell processes,
entirely independent of any embedding in S-matrix
elements (or any other a priori gauge-independent quantity).
It is then shown explicitly how
the PT one-loop gluon ``effective'' two-point function
remains gauge-independent (Sec.\ 4.2) and universal (Sec.\ 4.3),
despite the gauge-dependence of the various off-shell processes
of which it is a component.
This involves the consideration 
in both the class of linear covariant
gauges and the class of non-covariant gauges $n\sdot A^{a} = 0$
of subclasses of up to several hundred one-loop diagrams.
By writing these diagrams as ``products''
of tree level four- and five-point
functions and exploiting the Ward identities satisfied by these
$n$-point functions, the demonstration of the off-shell
gauge independence and universality is made simple
and the underlying cancellation mechanism responsible
is directly identified.
Lastly, it is shown (Sec.\ 4.4) how this one-loop cancellation
mechanism extends to the $n$-loop diagrams
involved in the Dyson summation of the PT self-energy.
This involves the consideration in the class of non-covariant
gauges $n\sdot A^{a} = 0$ (to avoid ghosts) of
subclasses of diagrams occurring at all orders in perturbation
theory. It is shown explicitly how, using an iterative procedure,
the chains of $n$ PT gauge-independent one-loop
self-energies can be isolated in the Feynman integrands for
the relevant subclasses of diagrams for all $n$. 
The paper finishes with a summary and conclusions (Sec.\ 5).
Technical details, together with
a discussion of the relation between the effective charge defined here 
and that obtained from the static heavy quark potential,
are given in three appendices.


\setcounter{equation}{0}
\def\theequation{2.\arabic{equation}}

\vspace{10pt}

{\Large\bf 2. Effective Charges}

\vspace{5pt}

\noindent
In QED, the interaction part of the classical lagrangian is given by
\be\label{qedli}
\Lclint
=
eJ_{\mu}A_{\mu}
\ee
where $J_{\mu}$ is the electromagnetic current.

At tree level, the interaction
between electromagnetic
currents at points $x_{1}$ and $x_{2}$ 
is mediated by a single photon and has $x$-space amplitude given by
\be\label{qedjdj}
ieJ_{\mu}(x_{1})iD_{\mu\nu}(x_{1} - x_{2})ieJ_{\nu}(x_{2})
\ee
where $iD_{\mu\nu}(x_{1} - x_{2})$ is the Fourier transform of the tree
level photon propagator $iD_{\mu\nu}(q)$:
\be\label{Dmunu}
iD_{\mu\nu}(q)
=
\frac{i}{q^{2} + i\epsilon}
\biggl(-g_{\mu\nu} + (1 - \xi)\frac{q_{\mu}q_{\nu}}{q^{2}}\biggr).
\ee

Beyond tree level, the
renormalized interaction between the two currents at $x_{1}$ and
$x_{2}$ is given by
\be\label{qedjdeltaj}
ie_{R}J_{\mu}(x_{1})i\Delta_{R\mu\nu}(x_{1} - x_{2})ie_{R}J_{\nu}(x_{2})
\ee
where $i\Delta_{R\mu\nu}(x_{1} - x_{2})$ is the Fourier transform of
the renormalized photon propagator Eq.\ (\ref{qedprop}), 
involving the Dyson summation of the 1PI photon self-energy 
$\Pi_{R}(q^{2})$.
There are also of course vertex and box 
corrections to the tree level four-fermion process 
Eq.\ (\ref{qedjdj}), but 
in QED these make no contribution to the two-point
current-current component of the interaction in (\ref{qedjdeltaj}).
Precisely because in (\ref{qedjdeltaj})
the interaction vertices of the 
photon with the currents $J_{\mu}(x_{1})$, $J_{\nu}(x_{2})$ remain
as specified by the interaction part Eq.\ (\ref{qedli})
of the QED classical lagrangian,
the radiative corrections to the tree level propagator $iD_{\mu\nu}$
included in the renormalized propagator $i\Delta_{R\mu\nu}$ can be fully
accounted for just 
by making for the transverse part of the interaction the
replacement $e \rightarrow \eeff(q^{2})$ 
in the tree level photon-fermion-fermion vertices.

In QCD, the interaction part of the classical lagrangian can be
written
\be\label{qcdli}
\Lclint
=
g\Bigl(J_{\mu}^{m} + T_{\mu}^{m} + gQ_{\mu}^{m}\Bigr)A_{\mu}^{m}
\ee
where $J_{\mu}^{m}$ is the chromoelectric current,
$T_{\mu}^{m}$ denotes the pair of gauge bosons 
coming from the triple gauge vertex, 
and $Q_{\mu}^{m}$ denotes the 
triplet of gauge bosons coming from the quadruple gauge vertex:
\bea
J_{\mu}^{m}
&=&
{\textstyle \sum_{f = 1}^{n_{\!f}}}
\overline{\psi}_{j}^{\scriptscriptstyle (f)}\gamma_{\mu}T_{ji}^{m}
\psi_{i}^{\scriptscriptstyle (f)} \\
T_{\mu}^{m}
&=&
{\textstyle -\frac{1}{3}}f^{mnr}\Bigl(
A_{\nu}^{n}(\partial_{\mu}A_{\nu}^{r})
+
(\partial_{\nu} A_{\mu}^{n})A_{\nu}^{r} 
-
\partial_{\nu} (A_{\nu}^{n}A_{\mu}^{r}) \Bigr)
\\
Q_{\mu}^{m}
&=&
{\textstyle -\frac{1}{4}}f^{mnr}f^{rst}A_{\nu}^{n}A_{\mu}^{s}A_{\nu}^{t}
\eea
(the derivative has been symmetrized in $T_{\mu}^{m}A_{\mu}^{m}$).

At tree level, the interaction between any pair of terms in
the interaction part Eq.\ (\ref{qcdli}) 
of the classical lagrangian at points 
$x_{1}$ and $x_{2}$ mediated by a single gluon has amplitude
given by the appropriate term from
\be\label{qcdjdj}
\left.
\begin{array}{r}
igJ_{\mu}^{m}(x_{1}) \\
\\
igT_{\mu}^{m}(x_{1}) \\
\\
ig^{2}Q_{\mu}^{m}(x_{1}) \\
\end{array}
\right\}
\times iD_{\mu\nu}^{mn}(x_{1} - x_{2}) \times
\left\{
\begin{array}{l}
igJ_{\nu}^{n}(x_{2}) \\
\\
igT_{\nu}^{n}(x_{2}) \\
\\
ig^{2}Q_{\nu}^{n}(x_{2}) \\     
\end{array}
\right.
\ee
where $iD_{\mu\nu}^{mn}(x_{1} - x_{2})$ is the Fourier transform of
the tree level gluon propagator $i\delta^{mn}D_{\mu\nu}(q)$.

If the concept of an effective charge is to be extended 
directly at the diagrammatic level from QED to QCD, 
then it must be shown \cite{njw2}
that, beyond tree level, the renormalized 
interaction between any pair of terms from $\Lclint$ 
at $x_{1}$, $x_{2}$ can be written 
\be\label{qcdjdeltaj}
\left.
\begin{array}{r}
ig_{R}J_{\mu}^{m}(x_{1}) \\
\\
ig_{R}T_{\mu}^{m}(x_{1}) \\
\\
ig_{R}^{2}Q_{\mu}^{m}(x_{1}) \\
\end{array}
\right\}
\times i\hat{\Delta}_{R\mu\nu}^{mn}(x_{1} - x_{2})\times
\left\{
\begin{array}{l}
ig_{R}J_{\nu}^{n}(x_{2}) \\
\\
ig_{R}T_{\nu}^{n}(x_{2}) \\
\\
ig_{R}^{2}Q_{\nu}^{n}(x_{2}). \\
\end{array}
\right.
\ee
By {\em definition}, $i\hat{\Delta}_{R\mu\nu}^{mn}(x_{1} - x_{2})$
is the gauge boson propagator-like function
the effects of which can be fully accounted for
just by appropriately changing the coupling appearing in
the tree level vertices specified by
the interaction part Eq.\ (\ref{qcdli}) of the QCD classical lagrangian.
It is important to note that $i\hat{\Delta}_{R\mu\nu}^{mn}(x_{1} - x_{2})$
in (\ref{qcdjdeltaj})
is defined in terms of the two-point interaction 
between vertices only from the classical 
lagrangian---combinations of ``external'' fields, 
e.g.\ ghost-ghost, from tree level vertices which originate
from the gauge fixing procedure are not included.

What then is this function $i\hat{\Delta}_{R\mu\nu}^{mn}(x_{1} - x_{2})$?
At the strictly one-loop level (i.e.\ without any Dyson summation)
and for on-shell external fields, it is precisely 
the Fourier transform of the gauge boson self-energy-like function 
obtained in the PT: by construction, the
PT gauge boson self-energy captures from the integrands for all of 
the relevant Feynman diagrams the full one-loop interaction 
between any pair of combinations of on-shell fields 
from the interaction part $\Lclint$ of the classical lagrangian
at two points $x_{1}$, $x_{2}$.
{\em It is this feature which distinguishes the PT from all
other prescriptions for the rearrangement of S-matrix elements
into contributions from individually gauge-independent components.}
For the case of the interaction between
fermion fields, this is made particularly transparent in the
Degrassi-Sirlin current algebra formulation of the PT \cite{degsir}.
The fact that for on-shell external fields the PT self-energy is 
indeed universal in this way was shown explicitly in \cite{njw1}.

However, if the effective charge is to coincide 
at high energies with the running coupling $\bar{g}(q^{2})$
defined from the renormalization group,
then the function $i\hat{\Delta}_{R\mu\nu}^{mn}$ in
(\ref{qcdjdeltaj}) must involve not just one
PT self-energy correction to the tree level
gluon propagator, but rather 
the infinite Dyson series in $\hat{\Pi}_{R}$. Thus, in $q$-space,
\be\label{qcdprop}
i\hat{\Delta}_{R\mu\nu}^{mn}(q)
=
\frac{i\delta^{mn}}{q^{2} + i\epsilon}\Biggl\{
\biggl(-g_{\mu\nu} +\frac{q_{\mu}q_{\nu}}{q^{2}}\biggr)
\hat{d}_{R}(q^{2}) 
- \xi \frac{q_{\mu}q_{\nu}}{q^{2}}\Biggr\}
\ee
where 
\be\label{ptdR}
\hat{d}_{R}(q^{2})
= 
\sum_{n=0}^{\infty}\Bigl(\hat{\Pi}_{R}(q^{2})\Bigr)^{n}
= \frac{1}{1 - \hat{\Pi}_{R}(q^{2})}.
\ee
Furthermore, for the corresponding effective charge
to be generally applicable as in QED, 
the interaction (\ref{qcdjdeltaj}) 
between {\em any} of the pairs of terms from $\Lclint$ must 
remain {\em uniquely described} at all (perturbative) 
$q^{2}$ by the gauge-independent 
function $\hat{d}_{R}(q^{2})$ when the external fields
in (\ref{qcdjdeltaj}) are {\em off-shell}. 
This infinite gauge-independent subset of QCD radiative
corrections may then be fully accounted for
by the gauge-, scale- and scheme-independent QCD effective charge
\be\label{qcdgeff}
\geff^{2}(q^{2})
=
g_{R}^{2}\hat{d}_{R}(q^{2})
=
\frac{g_{R}^{2}}{1 - \hat{\Pi}_{R}(q^{2})}
=
4\pi\aseff(q^{2}).
\ee

Thus, we see that the QED concept of an effective charge has nothing
to do with the conventionally-defined gauge boson propagator per se.
Rather, we argue that it depends on the existence of a                   
unique, gauge-independent subset of radiative corrections 
to the tree level interaction between sets of fields
from $\Lclint$ at two points. 
The corresponding propagator $i\hat{\Delta}_{R\mu\nu}^{mn}$
may be referred to as the Dyson-summed
{\em ``effective''} gauge boson two-point function.
In the case of QED,
this in fact corresponds to the conventional renormalized
gauge boson propagator Eq.\ (\ref{qedprop})
only because the theory is abelian. But in a non-abelian theory such as
QCD, this function also receives
contributions (pinch parts) from the conventionally-defined 
vertex and box diagrams. 

The remainder of this paper is devoted to
showing explicitly how the function 
$\hat{d}_{R}(q^{2})$ occurs in QCD. 


\setcounter{equation}{0}
\def\theequation{3.\arabic{equation}}

\vspace{10pt}

{\Large\bf 3. QCD Tree Level Ward Identities}

\vspace{5pt}

\noindent
The PT rearrangement of perturbation theory diagrams
is based on the systematic use of the tree level
Ward identites \cite{wt} of the theory.
Before introducing the off-shell PT, it is convenient
to collect together the SU($N$) 
QCD tree level $n$-point functions and
their associated Ward identities~\cite{stbrs}. 
We consider $n_{\!f}$ flavours of fermion with mass $m_{\!f}$
in the fundamantal representation of SU($N$), 
with hermitian generator matrices $T^{a}$ satisfying
$\Tr \, T^{a}T^{b} = \frac{1}{2}\delta^{ab}$ and
$[T^{a},T^{b}] = if^{abc}T^{c}$. 

The tree level gluon propagator may be written in the general form
\be
iD_{\mu\nu}(q)
=
\frac{i}{q^{2}+i\epsilon}\biggl( -g_{\mu\nu} 
+ a_{\mu}(q)q_{\nu} + q_{\mu}a_{\nu}(q) + b(q)q_{\mu}q_{\nu}\biggr)
\label{prop}
\ee
(trivial colour indices are omitted).
We shall consider the following two classes of gauge:

\vspace{5pt}

{\em a) Linear covariant gauges.} 
The class of linear covariant gauges is obtained 
from the path integral by adding the
gauge-fixing term $\Lgf = -(\partial\sdot A^{a})^{2}/2\xi$ to $\Lcl$.
The gluon propagator is thus specified by \eq{prop} with
\be\label{abcov}
a_{\mu}(q) = 0\,, \qquad b(q) = \frac{1-\xi}{q^{2}}\,.
\ee
Then $\xi=1$ is the Feynman gauge and the limit $\xi \rightarrow 0$ 
is the Landau gauge.

The associated Fadeev-Popov ghost term is
$\Lgh = -\bar{\eta}^{a}\partial\sdot D^{ab}\eta^{b}$
where $D_{\mu}$ is the covariant derivative.
The ghost propagator is thus given by
\be
iG(q) = \frac{i}{q^{2}+i\epsilon}
\ee
(trivial colour indices are again omitted).
The gluon-ghost-ghost vertex is 
\be
g\Gamma_{\alpha}^{abc}(q_{1},q_{2},q_{3})
=
-gf^{abc}q_{3\alpha}
\ee
with $q_{1} + q_{2} = q_{3}$, where the four-momentum
$q_{1}$ of the gluon $A_{\alpha}^{a}(q_{1})$ is incoming.

\vspace{5pt}

{\em b) Non-covariant gauges $n\sdot A^{a} = 0$.}
The class of non-covariant gauges $n\sdot A^{a} = 0$, 
where $n_{\mu}$ is an arbitrary constant four-vector, is obtained from the
path integral by adding the gauge-fixing term
$\Lgf = -(n\sdot A^{a})^{2}/2\lambda$ to $\Lcl$
and then taking the limit $\lambda \rightarrow 0$.
The gluon propagator is thus specified by \eq{prop} with
\be\label{abaxial}
a_{\mu}(q) = \frac{n_{\mu}}{n\sdot q}\,, \qquad 
b(q) = -\frac{n^{2}}{(n\sdot q)^{2}}\,.
\ee
Then $n^{2} < 0$ is the class of pure axial gauges,
$n^{2} = 0$ is the light-cone gauge and
$n^{2} > 0$ [$n_{\mu} = (1,0,0,0)$] is the temporal gauge.
(For a review of non-covariant gauges, see \cite{liebbrandt}.) 

The associated Fadeev-Popov ghost term is
$\Lgh = -\bar{\eta}^{a}n\sdot D^{ab}\eta^{b}$.
The ghost propagator is thus given by
\be
iG(q) = \frac{i}{n\sdot q}\,.
\ee
The gluon-ghost-ghost vertex is 
\be
g\Gamma_{\alpha}^{abc}(q_{1},q_{2},q_{3})
=
-gf^{abc}n_{\alpha}\,.
\ee
A fundamental property of non-covariant gauges is that the ghost fields
decouple from S-matrix elements. Ghosts are however required in the
discussion of the Ward identities.

\vspace{5pt}

In these two classes of linear gauge, the remaining
tree level vertices originate only from $\Lclint$ and
are as follows (all gluon four-momenta $q_{i}$ are incoming; 
for the purely gluonic $n$-point functions, $\sum_{i=1}^{n}q_{i} = 0$):

i) The gluon-fermion-fermion vertex:
\be
ig\gamma_{\mu}T_{ji}^{m}.
\label{gff}
\ee

ii) The triple gluon vertex:
\be
g\Gamma_{\alpha\beta\gamma}^{abc}(q_{1},q_{2},q_{3})
= 
gf^{abc}\Bigl(
 (q_{2} - q_{3})_{\alpha}g_{\beta\gamma}
+(q_{3} - q_{1})_{\beta}g_{\gamma\alpha}
+(q_{1} - q_{2})_{\gamma}g_{\alpha\beta}\Bigr).
\label{tgv}
\ee

iii) The quadruple gluon vertex:
\bea
-ig^{2}\Gamma_{\alpha\beta\gamma\delta}^{abcd}(q_{1},q_{2},q_{3},q_{4})
&=&
-ig^{2}\Bigl(
f^{rab}f^{rcd}(g_{\alpha\gamma}g_{\beta\delta} 
             - g_{\alpha\delta}g_{\beta\gamma})
\nonumber\\
& &
\hspace{20pt}+
f^{rac}f^{rdb}(g_{\alpha\delta}g_{\gamma\beta} 
             - g_{\alpha\beta}g_{\gamma\delta})
\nonumber\\
\label{qgv}
& &
\hspace{20pt}+
f^{rad}f^{rbc}(g_{\alpha\beta}g_{\delta\gamma} 
             - g_{\alpha\gamma}g_{\delta\beta})\Bigr).
\eea

In addition to the vertices originating directly from
$\Lclint + \Lgh$,
it will be very convenient to define three further 
$n$-point functions, constructed from the vertices in 
$\Lclint$ together with the gluon propagator
$iD_{\mu\nu}(q)$ and the fermion propagator 
$iS(q,m) = i(\qs -m + i\epsilon)^{-1}$:

\begin{center}
\begin{picture}(400,120)(-200,-70)

\put(-150, 30){\makebox(0,0)[r]{\footnotesize $A_{\alpha}^{a}(q_{1})$}}
\put(-150,-30){\makebox(0,0)[r]{\footnotesize $A_{\beta}^{b}(q_{2})$}}
\put(-90,-30){\makebox(0,0)[l]{\footnotesize 
$\psi_{i}^{\scriptscriptstyle{(f)}}(q_{3})$}}
\put(-90, 30){\makebox(0,0)[l]{\footnotesize 
$\psi_{j}^{\scriptscriptstyle{(f)}}(q_{4})$}}
\Photon(-140,-20)(-140,20){2.5}{4}
\Photon(-140,0)(-100,0){2.5}{4}
\ArrowLine(-100,-20)(-100,0)
\ArrowLine(-100,0)(-100,20)

\put(-60,0){\makebox(0,0)[c]{\large +}}

\Photon(-20,-20)(0,-20){2.5}{2}
\Photon(-20, 20)(0, 20){-2.5}{2}
\ArrowLine(0,-20)(0,20)
\ArrowLine(20,-20)(0,-20)
\ArrowLine(0,20)(20,20)

\put( 60,0){\makebox(0,0)[c]{\large +}}

\Photon(100,-20)(107,-6){2.5}{1.5}
\Photon(113,  6)(120,20){2.5}{1.5}
\Photon(100, 20)(120,-20){2.5}{5}
\ArrowLine(120,-20)(120,20)
\ArrowLine(140,-20)(120,-20)
\ArrowLine(120,20)(140,20)

\put(0,-70){\makebox(0,0)[c]{\footnotesize{
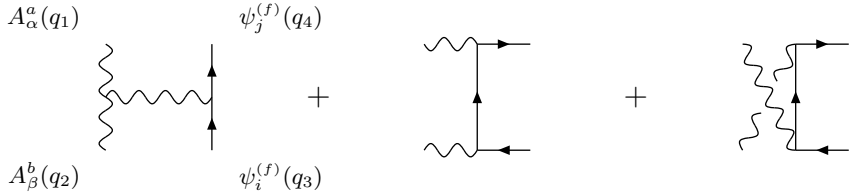
Fig.\ 3. The three diagrams contributing to the four-point function 
$G_{\alpha\beta ij}^{ab(f)}(q_{1},q_{2},q_{3},q_{4})$.}}}

\end{picture}
\end{center}

iv) The connected four-point function
$G_{\alpha\beta ij}^{ab(f)}(q_{1},q_{2},q_{3},q_{4})$
specifying the tree level coupling of a pair of gluons
$A_{\alpha}^{a}(q_{1})$, $A_{\beta}^{b}(q_{2})$ to a pair
of fermions $\psi_{i}^{\scriptscriptstyle{(f)}}(q_{3})$, 
$\psi_{j}^{\scriptscriptstyle{(f)}}(q_{4})$ 
shown in Fig.\ 3:
\bea
\lefteqn{
{\rm Fig.\, 3} =
ig^{2}G_{\alpha\beta ij}^{ab(f)}
(q_{1},q_{2},q_{3},q_{4}) 
= } \hspace{80pt}\nonumber \\
& &
+g\Gamma_{\alpha\beta\rho'}^{abr}(q_{1},q_{2},-q_{1}\!-\!q_{2})
\,iD_{\rho'\rho}(q_{1}\!+\!q_{2})\,
ig\gamma_{\rho}T_{ji}^{r} \nonumber \\
& &
+ig\gamma_{\beta}T_{jk}^{b}\,iS(q_{1}\!+\!q_{3},m_{\!f})\,
ig\gamma_{\alpha}T_{ki}^{a}
\nonumber \\
& &
+ig\gamma_{\alpha}T_{jk}^{a}\,iS(q_{4}\!-\!q_{1},m_{\!f})\,
ig\gamma_{\beta}T_{ki}^{b}
\label{ggff}
\eea
with $\sum_{i=1}^{3}q_{i} = q_{4}$.

\begin{center}
\begin{picture}(400,120)(-200,-70)

\put(-90, 30){\makebox(0,0)[r]{\footnotesize $A_{\alpha}^{a}(q_{1})$}}
\put(-90,-30){\makebox(0,0)[r]{\footnotesize $A_{\beta}^{b}(q_{2})$}}
\put(-30,-30){\makebox(0,0)[l]{\footnotesize $A_{\gamma}^{c}(q_{3})$}}
\put(-30, 30){\makebox(0,0)[l]{\footnotesize $A_{\delta}^{d}(q_{4})$}}
\put(-60,-35){\makebox(0,0)[c]{\footnotesize 3 perms}}
\Photon(-80,-20)(-80,20){2.5}{4}
\Photon(-80,0)(-40,0){2.5}{4}
\Photon(-40,-20)(-40,20){2.5}{4}

\put(0,0){\makebox(0,0)[c]{\large +}}

\Photon(40,-20)(80, 20){2.5}{6}
\Photon(40, 20)(80,-20){2.5}{6}

\put(0,-70){\makebox(0,0)[c]{\footnotesize
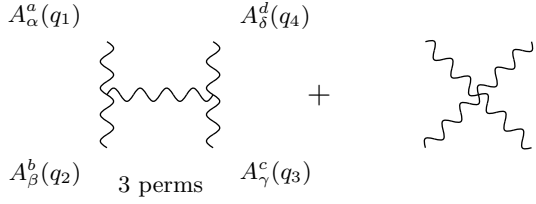
Fig.\ 4. The four diagrams contributing to the four-point function 
$G_{\alpha\beta\gamma\delta}^{abcd}(q_{1},q_{2},q_{3},q_{4})$.}}

\end{picture}
\end{center}

v) The connected four-point function
$G_{\alpha\beta\gamma\delta}^{abcd}(q_{1},q_{2},q_{3},q_{4})$
specifying the tree level coupling of four gluons
$A_{\alpha}^{a}(q_{1})$, $A_{\beta}^{b}(q_{2})$,
$A_{\gamma}^{c}(q_{3})$, $A_{\delta}^{d}(q_{4})$
shown in Fig.\ 4:
\bea
\lefteqn{
{\rm Fig.\, 4} =
ig^{2}G_{\alpha\beta\gamma\delta}^{abcd}(q_{1},q_{2},q_{3},q_{4}) 
= } \hspace{80pt}\nonumber \\
& &
+g\Gamma_{\alpha\beta\rho'}^{abr}(q_{1},q_{2},-q_{1}\!-\!q_{2})
\,iD_{\rho'\rho}(q_{1}\!+\!q_{2})\,
g\Gamma_{\rho\gamma\delta}^{rcd}(q_{1}\!+\!q_{2},q_{3},q_{4})
\nonumber \\
& &
+g\Gamma_{\alpha\gamma\rho'}^{acr}(q_{1},q_{3},-q_{1}\!-\!q_{3})
\,iD_{\rho'\rho}(q_{1}\!+\!q_{3})\,
g\Gamma_{\rho\delta\beta}^{rdb}(q_{1}\!+\!q_{3},q_{4},q_{2})
\nonumber \\
& &
+g\Gamma_{\alpha\delta\rho'}^{adr}(q_{1},q_{4},-q_{1}\!-\!q_{4})
\,iD_{\rho'\rho}(q_{1}\!+\!q_{4})\,
g\Gamma_{\rho\beta\gamma}^{rbc}(q_{1}\!+\!q_{4},q_{2},q_{3})
\nonumber \\
& &
-ig^{2}\Gamma_{\alpha\beta\gamma\delta}^{abcd}(q_{1},q_{2},q_{3},q_{4}).
\label{gggg}
\eea

\begin{center}
\begin{picture}(400,120)(-200,-70)

\put(-90, 30){\makebox(0,0)[r]{\footnotesize $A_{\alpha}^{a}(q_{1})$}}
\put(-112, 0){\makebox(0,0)[r]{\footnotesize $A_{\beta}^{b}(q_{2})$}}
\put(-90,-30){\makebox(0,0)[r]{\footnotesize $A_{\gamma}^{c}(q_{3})$}}
\put(-30,-30){\makebox(0,0)[l]{\footnotesize $A_{\delta}^{d}(q_{4})$}}
\put(-30, 30){\makebox(0,0)[l]{\footnotesize $A_{\epsilon}^{e}(q_{5})$}}
\put(-60,-35){\makebox(0,0)[c]{\footnotesize 10 perms}}
\Photon(-80,-20)(-80,20){2.5}{4}
\Photon(-100,0)(-40,0){2.5}{6}
\Photon(-40,-20)(-40,20){2.5}{4}

\put(5,0){\makebox(0,0)[c]{\large +}}

\Photon( 50,-20)( 50,20){2.5}{4}
\Photon( 80,  0)( 80,20){2.5}{2}
\Photon(110,-20)(110,20){2.5}{4}
\Photon(50,0)(110,0){2.5}{6}
\put(80,-35){\makebox(0,0)[c]{\footnotesize 15 perms}}

\put(0,-70){\makebox(0,0)[c]{\footnotesize
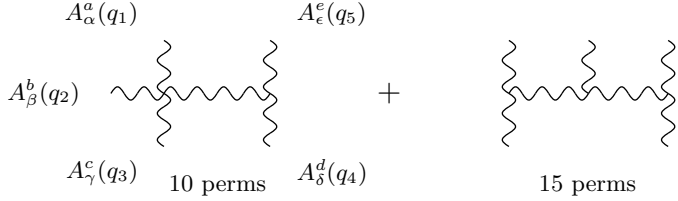
Fig.\ 5. The twenty-five diagrams contributing to the five-point function 
$G_{\alpha\beta\gamma\delta\epsilon}^{abcde}
(q_{1},q_{2},q_{3},q_{4},q_{5})$.}}

\end{picture}
\end{center}

vi) Lastly, the connected five-point function
$G_{\alpha\beta\gamma\delta\epsilon}^{abcde}
(q_{1},q_{2},q_{3},q_{4},q_{5})$
specifying the tree level coupling of five gluons
$A_{\alpha}^{a}(q_{1})$, $A_{\beta}^{b}(q_{2})$,
$A_{\gamma}^{c}(q_{3})$, $A_{\delta}^{d}(q_{4})$,
$A_{\epsilon}^{e}(q_{5})$
shown in Fig.\ 5:
\be
{\rm Fig.\, 5}
=
-g^{3}G_{\alpha\beta\gamma\delta\epsilon}^{abcde}
(q_{1},q_{2},q_{3},q_{4},q_{5})
\label{ggggg}
\ee
where the explicit form of 
$G_{\alpha\beta\gamma\delta\epsilon}^{abcde}$,
involving twenty-five terms, is not recorded here.

In order to express the Ward identities obeyed by these $n$-point
functions, we first define the transverse projection operator
\be
t_{\mu\nu}(q)
=
g_{\mu\nu} - \frac{q_{\mu}q_{\nu}}{q^{2}}.
\ee
Also, it is convenient to define a quantity
$\tilde{G}_{\mu}(q)$ from the product of the gluon-ghost-ghost
vertex and the propagator for the outgoing ghost via
$f^{abc}\tilde{G}_{\alpha}(q_{3})
= \Gamma_{\alpha}^{abc}(q_{1},q_{2},q_{3})\,G(q_{3})$.
Thus, 
\be
\tilde{G}_{\mu}(q) = 
\left\{
\begin{array}{ll}
{\displaystyle -q_{\mu}/q^{2}}    & \qquad \mbox{linear covariant gauges} \\
{\displaystyle -n_{\mu}/n\sdot q} & \qquad n\sdot A^{a}=0 \mbox{{} gauges}.
\end{array}
\right.
\ee
Note that $q^{2}t_{\mu\rho}(q)\,D_{\rho\nu}(q) =
-g_{\mu\nu} - \tilde{G}_{\mu}(q)\,q_{\nu}$.
The Ward identities are then as follows~\cite{stbrs}:

i) For the gluon-fermion-fermion vertex
(with $q_{1} + q_{2} = q_{3}$):
\be\label{fermionwid}
q_{1\mu}i\gamma_{\mu}T_{ji}^{m}
=
-iT_{ji}^{m}\Bigl(S^{-1}(q_{2},m) - S^{-1}(q_{3},m) \Bigr).
\ee

ii) For the triple gluon vertex:
\be\label{tgvwid}
q_{1\alpha}\Gamma_{\alpha\beta\gamma}^{abc}(q_{1},q_{2},q_{3})
=
-f^{abc}\Bigl(q_{2}^{2}t_{\beta\gamma}(q_{2})
             -q_{3}^{2}t_{\beta\gamma}(q_{3}) \Bigr).
\ee

iii) For the quadruple gluon vertex:
\bea
q_{1\alpha}\Gamma_{\alpha\beta\gamma\delta}^{abcd}
(q_{1},q_{2},q_{3},q_{4})
&=&
-f^{rab}\Gamma_{\beta\gamma\delta}^{rcd}(q_{1}\!+\!q_{2},q_{3},q_{4})
\nonumber\\
& & 
-f^{rac}\Gamma_{\gamma\delta\beta}^{rdb}(q_{1}\!+\!q_{3},q_{4},q_{2})
\nonumber\\
& &
-f^{rad}\Gamma_{\delta\beta\gamma}^{rbc}(q_{1}\!+\!q_{4},q_{2},q_{3}).
\label{qgvwid}
\eea

iv) For the gluon pair--fermion pair four-point function:
\bea
\lefteqn{
q_{1\alpha}G_{\alpha\beta ij}^{ab(f)}(q_{1},q_{2},q_{3},q_{4})
=} \nonumber \\
& &
-f^{rab}\Bigl(q_{2}^{2}t_{\beta\rho'}(q_{2})
D_{\rho'\rho}(q_{1}\!+\!q_{2}) \,+\, 
\tilde{G}_{\beta}(q_{1}\!+\!q_{2})\,(q_{1}\!+\!q_{2})_{\rho} \Bigr)
\,i\gamma_{\rho}T_{ji}^{r} \hspace{10pt}\nonumber \\
& &
- i\gamma_{\beta}T_{jk}^{b}\,S(q_{1}\!+\!q_{3},m_{\!f})\,
S^{-1}(q_{3},m_{\!f})
\,iT_{ki}^{a}
\nonumber \\
& &
+ iT_{jk}^{a}\,S^{-1}(q_{4},m_{\!f})\,
S(q_{4}\!-\!q_{1},m_{\!f})
\,i\gamma_{\beta}T_{ki}^{b}.
\label{ggffwid}
\eea

v) For the gluon four-point function:
\bea
\lefteqn{
q_{1\alpha}G_{\alpha\beta\gamma\delta}^{abcd}(q_{1},q_{2},q_{3},q_{4})
=} \nonumber \\
& &
-f^{rab}\Bigl(q_{2}^{2}t_{\beta\rho'}(q_{2})D_{\rho'\rho}(q_{1}\!+\!q_{2}) 
\,+\, \tilde{G}_{\beta}(q_{1}\!+\!q_{2})\,(q_{1}\!+\!q_{2})_{\rho} \Bigr)
\,\Gamma_{\rho\gamma\delta}^{rcd}(q_{1}\!+\!q_{2},q_{3},q_{4})\nonumber \\
& &
-f^{rac}\Bigl(q_{3}^{2}t_{\gamma\rho'}(q_{3})D_{\rho'\rho}(q_{1}\!+\!q_{3}) 
\,+\, \tilde{G}_{\gamma}(q_{1}\!+\!q_{3})\,(q_{1}\!+\!q_{3})_{\rho} \Bigr)
\,\Gamma_{\rho\delta\beta}^{rdb}(q_{1}\!+\!q_{3},q_{4},q_{2}) \nonumber \\
& &
-f^{rad}\Bigl(q_{4}^{2}t_{\delta\rho'}(q_{4})D_{\rho'\rho}(q_{1}\!+\!q_{4}) 
\,+\, \tilde{G}_{\delta}(q_{1}\!+\!q_{4})\,(q_{1}\!+\!q_{4})_{\rho} \Bigr)
\,\Gamma_{\rho\beta\gamma}^{rbc}(q_{1}\!+\!q_{4},q_{2},q_{3}).
\label{ggggwid}
\eea

vi) For the gluon five-point function:
\bea
\lefteqn{
q_{1\alpha}G_{\alpha\beta\gamma\delta\epsilon}^{abcde}
(q_{1},q_{2},q_{3},q_{4},q_{5}) = }\nonumber \\
& &
-f^{rab}\Bigl(q_{2}^{2}t_{\beta\rho'}(q_{2})D_{\rho'\rho}(q_{1}\!+\!q_{2}) 
\,+\, \tilde{G}_{\beta}(q_{1}\!+\!q_{2})\,(q_{1}\!+\!q_{2})_{\rho} \Bigr)
\,G_{\rho\gamma\delta\epsilon}^{rcde}(q_{1}\!+\!q_{2},q_{3},q_{4},q_{5}) 
\nonumber \\
& &
-f^{rac}\Bigl(q_{3}^{2}t_{\gamma\rho'}(q_{3})D_{\rho'\rho}(q_{1}\!+\!q_{3}) 
\,+\, \tilde{G}_{\gamma}(q_{1}\!+\!q_{3})\,(q_{1}\!+\!q_{3})_{\rho} \Bigr)
\,G_{\rho\delta\epsilon\beta}^{rdeb}(q_{1}\!+\!q_{3},q_{4},q_{5},q_{2}) 
\nonumber \\
& &
-f^{rad}\Bigl(q_{4}^{2}t_{\delta\rho'}(q_{4})D_{\rho'\rho}(q_{1}\!+\!q_{4}) 
\,+\, \tilde{G}_{\delta}(q_{1}\!+\!q_{4})\,(q_{1}\!+\!q_{4})_{\rho} \Bigr)
\,G_{\rho\epsilon\beta\gamma}^{rebc}(q_{1}\!+\!q_{4},q_{5},q_{2},q_{3}) 
\nonumber \\
& &
-f^{rae}\Bigl(q_{5}^{2}t_{\epsilon\rho'}(q_{5})D_{\rho'\rho}(q_{1}\!+\!q_{5}) 
\,+\, \tilde{G}_{\epsilon}(q_{1}\!+\!q_{5})\,(q_{1}\!+\!q_{5})_{\rho} \Bigr)
\,G_{\rho\beta\gamma\delta}^{rbcd}(q_{1}\!+\!q_{5},q_{2},q_{3},q_{4}).
\qquad \label{gggggwid}
\eea
It is important to note the similarities among these identities.

\pagebreak


\setcounter{equation}{0}
\def\theequation{4.\arabic{equation}}

{\Large\bf 4. The Off-Shell Pinch Technique}

\vspace{5pt}

\noindent
In this section, the generalization of the PT to explicitly off-shell
processes is introduced. The 
one-loop gluon ``effective'' two-point 
function is first defined. It is then shown that
this function is gauge-independent, universal and that it
may be summed in a Dyson series.

\vspace{10pt}

{\bf 4.1. Definition of $i\hat{\Pi}_{\mu\nu}^{mn}(q)$}

\noindent
We consider the complete set of one-loop corrections to the 
tree level gluonic interaction (\ref{qcdjdj})
between any pair of the combinations of fields
in $\Lclint$. All fields are taken to be off-shell. 
Then, for example, in the particular
case of a pair of fermion currents, the set of diagrams
are again those shown in Fig.\ 2, but now with the external fermions
all off-shell.

In the off-shell PT, the PT one-loop gauge boson ``effective''
two-point function $i\hat{\Pi}_{\mu\nu}^{mn}(q)$ is defined from
{\em the coefficient $\hat{\Pi}'(k,q)$ of the
component of the Feynman integrands for the one-loop
interaction which has the Lorentz and colour structure of 
the transverse projection operator contracted
with two tree level gluon propagators
between the appropriate pair of tree level vertices:}
\be\label{ptPidef1}
\mu^{2\epsilon}\int\frac{d^{d}k}{(2\pi)^{d}}
\left\{
\begin{array}{c}
ig\gamma_{\mu}T_{j'i'}^{m} \\
\\
g\Gamma_{\mu\alpha'\beta'}^{ma'b'} \\
\\
-ig^{2}\Gamma_{\mu\alpha'\beta'\gamma'}^{ma'b'c'} \\
\end{array}
\right\}
iD_{\mu\mu'}(q)
\Bigl\{iq^{2}t_{\mu'\nu'}(q)\hat{\Pi}'(k,q)\Bigr\}
iD_{\nu'\nu}(q)
\left\{
\begin{array}{c}
ig\gamma_{\nu}T_{ji}^{m} \\
\\
g\Gamma_{\nu\alpha\beta}^{mab} \\
\\
-ig^{2}\Gamma_{\nu\alpha\beta\gamma}^{mabc} \\
\end{array}
\right\}
\ee
where the tree level gluon propagator $iD_{\mu\nu}(q)$ is given
in \eq{prop}. Thus,
\be
iD_{\mu\mu'}(q)iq^{2}t_{\mu'\nu'}(q)iD_{\nu'\nu}(q) 
=
\frac{i}{q^{2}\!+\!i\epsilon} \!
\left\{ \!\!
\begin{array}{ll}
{\displaystyle -g_{\mu\nu} + \frac{q_{\mu}q_{\nu}}{q^{2}}} & 
\mbox{covariant gauges}
\\
{\displaystyle -g_{\mu\nu} + \frac{n_{\mu}q_{\nu} + q_{\mu}n_{\nu}}{n\sdot q} 
- \frac{n^{2}q_{\mu}q_{\nu}}{(n\sdot q)^{2}}} &  
n\sdot A^{a} = 0 \mbox{{} gauges}.
\end{array}
\right.
\label{ptPidef2}
\ee
Then
\be
i\hat{\Pi}_{\mu\nu}^{mn}(q)
=
i\delta^{mn}q^{2}t_{\mu\nu}(q)\,\hat{\Pi}(q^{2}) 
=
i\delta^{mn}q^{2}t_{\mu\nu}(q)
\,\mu^{2\epsilon}\int\frac{d^{d}k}{(2\pi)^{d}}
\,\hat{\Pi}'(k,q)
\label{ptPidef3}
\ee
i.e.\ $i\hat{\Pi}_{\mu\nu}^{mn}(q)$ is transverse by definition.

For QED and for fermion (and scalar) loop contributions
in QCD, this definition reproduces
the conventional gauge boson self-energy defined
from the Fourier transform of the conventional
two-point Green's function 
$\langle 0| T(A_{\mu}^{m}(x_{1})A_{\nu}^{n}(x_{2})) |0\rangle$.
For gauge boson loop contributions however, 
this definition includes not only the contributions to the
conventional self-energy but also pinch contributions.
It is emphasized that the definition 
of the PT ``effective'' two-point function
is in terms of the Feynman {\em integrands} corresponding
to the diagrams for the interactions: all 
rearrangements in the PT are carried out under the 
loop momentum integral sign(s).

The definition (\ref{ptPidef1})-(\ref{ptPidef3}) of
$i\hat{\Pi}_{\mu\nu}^{mn}(q)$
is as in the S-matrix PT, except that
i) the ``external'' fields in (\ref{ptPidef1})
are explicitly off-shell and not contracted into the
corresponding spinors and polarization vectors, and
ii) the tensor structures in (\ref{ptPidef2}) are non-trivial and not
just proportional to $g_{\mu\nu}$.
In each case,  with the external fields all off-shell,
the corresponding set of one-loop diagrams no
longer represents the one-loop contribution to an S-matrix element
as in the S-matrix PT, and
the overall one-loop amplitude for the given (sub)process 
is gauge-dependent. However, we will show that, 
using the definition (\ref{ptPidef1})-(\ref{ptPidef3}), it
remains possible to  
identify unambiguously at the level of the Feynman integrals
a gauge-independent and universal
one-loop gluon ``effective'' two-point component
$i\hat{\Pi}_{\mu\nu}^{mn}(q)$ of
these (sub)processes, exactly analogous to that in QED,
and that this self-energy is identical to that obtained
in the S-matrix PT.

\vspace{10pt}

{\bf 4.2. Gauge-Independence of $i\hat{\Pi}_{\mu\nu}^{mn}(q)$}

\noindent
In order to prove the gauge-independence of the ``effective''
two-point function
$i\hat{\Pi}_{\mu\nu}^{mn}(q)$ defined in the off-shell PT,
we consider the interaction 
$\psi_{i}^{\scriptscriptstyle{(f)}}(p_{1})
\psi_{i'}^{\scriptscriptstyle{(f')}}(p_{1}') 
\rightarrow 
\psi_{j}^{\scriptscriptstyle{(f)}}(p_{2})
\psi_{j'}^{\scriptscriptstyle{(f')}}(p_{2}')$
between two off-shell fermion pairs with flavours $f \neq f'$.
Of the possible combinations of fields in (\ref{ptPidef1}), 
this is the simplest case. 
The complete set of one-loop 
radiative corrections to this tree level process is 
then as shown in Fig.\ 2, where now the fermions are all off-shell.

We consider first the diagrams shown in Figs.\ 6(a), (b) and (c)
(symmetry factors have been included explicitly). In order to exploit
the Ward identites of the previous section, it is very convenient to
express the sum of these five diagrams as the sum of all possible
``products'' of diagrams on the r.h.s.\ of Fig.\ 6, where the wavy
lines represent the two gauge field propagators 
$iD_{\rho'\rho}(k_{1})$ and $iD_{\sigma'\sigma}(k_{2})$ associated with
the gluons propagating in the loops. The sums of diagrams 
Figs.\ 6(d)+(e)+(f) and (g)+(h)+(i) 
are each precisely the connected four-point
function defined in Eq.\ (\ref{ggff}),
consisting of all possible ways of coupling a pair of gluons to a pair
of fermions at tree level. Thus, in any gauge, the sum of the five 
diagrams on the l.h.s.\ of Fig.\ 6 can be written in the compact form
\be
\mbox{Figs.\ 6(a)+(b)+(c)}
=
{\textstyle \frac{1}{2}}
\mu^{2\epsilon}\int\frac{d^{d}k}{(2\pi)^{d}}
\,ig^{2}G_{\rho'\sigma' i'j'}^{rs(f')}
\,iD_{\rho'\rho}(k_{1})\,iD_{\sigma'\sigma}(k_{2})
\,ig^{2}G_{\rho\sigma ij}^{rs(f)}
\label{fig6abc}
\ee
where the overall factor $\frac{1}{2}$ accounts for the
symmetry of $G_{\rho'\sigma' i'j'}^{rs(f')}$ and
$G_{\rho\sigma ij}^{rs(f)}$ under interchange of the gluons
propagating in the loops
(arguments for the four-point functions are omitted for brevity;
$k_{2}-k_{1} = q$ with, e.g., $k_{1} =k$). 
In order to identify the PT ``effective''
two-point component of these diagrams,
it is then necessary to isolate {\em all} of the 
factors of longitudinal
four-momentum $k_{1\rho}$, $k_{2\sigma}$
associated with the gluons 
$A_{\rho}^{r}(k_{1})$, $A_{\sigma}^{s}(k_{2})$
propagating in the loops.

\begin{center}
\begin{picture}(400,240)(-200,-130)

\put(-175,70){\makebox(0,0)[c]{\Large $\frac{1}{2}$}}
\Line(-160, 50)(-160,90)
\Photon(-160,70)(-142,70){2}{2}\PhotonArc(-130,70)(10,34,394){2}{8}
\Photon(-118,70)(-100,70){2}{2}
\Line(-100, 50)(-100,90)
\put(-130,35){\makebox(0,0)[c]{\footnotesize (a)}}

\Line(-160,-20)(-160,20)
\Photon(-160,0)(-130,0){2}{3}
\Photon(-130,0)(-100, 10){2}{3.5}
\Photon(-130,0)(-100,-10){-2}{3.5}
\Line(-100,-20)(-100,20)
\put(-130,-23){\makebox(0,0)[c]{\footnotesize + reversed}}
\put(-130,-35){\makebox(0,0)[c]{\footnotesize (b)}}

\Line(-160,-90)(-160,-50)
\Photon(-160,-60)(-100,-60){2}{6}
\Photon(-160,-80)(-100,-80){2}{6}
\Line(-100,-90)(-100,-50)
\put(-130,-93){\makebox(0,0)[c]{\footnotesize + crossed}}
\put(-130,-105){\makebox(0,0)[c]{\footnotesize (c)}}

\put(-80,0){\makebox(0,0)[c]{$ \left. \begin{array}{c} 
\\ \\ \\ \\ \\ \\ \\ \\ \\ \\ \\ \\   \end{array} \right\} $}}
\put(-55,0){\makebox(0,0)[c]{\large =}}
\put(-40,0){\makebox(0,0)[c]{\Large $\frac{1}{2}$}}
\put(-20,0){\makebox(0,0)[c]{$ \left\{\begin{array}{c} 
\\ \\ \\ \\ \\ \\ \\ \\ \\ \\ \\ \\   \end{array} \right. $}}

\Line(0, 50)(0,90)
\Photon(0,70)(8,70){2}{1}
\PhotonArc(20,70)(10,90,-90){2}{4.5}
\put(10,35){\makebox(0,0)[c]{\footnotesize (d)}}

\Line(0,-20)(0,20)
\Photon(0,-10)(20,-10){2}{2}
\Photon(0, 10)(20, 10){2}{2}
\put(10,-35){\makebox(0,0)[c]{\footnotesize (e)}}

\Line(0,-90)(0,-50)
\Photon(0,-80)(20,-60){-2}{3}
\Photon(0,-60)(7, -67){2}{1}
\Photon(14,-73)(20,-80){2}{1}
\put(10,-105){\makebox(0,0)[c]{\footnotesize (f)}}

\put( 40,0){\makebox(0,0)[c]{$ \left. \begin{array}{c} 
\\ \\ \\ \\ \\ \\ \\ \\ \\ \\ \\ \\   \end{array} \right\} $}}
\Photon(70,-10)(90,-10){2}{2}
\Photon(70, 10)(90, 10){2}{2}
\put(120,0){\makebox(0,0)[c]{$ \left\{ \begin{array}{c} 
\\ \\ \\ \\ \\ \\ \\ \\ \\ \\ \\ \\    \end{array} \right. $}}

\PhotonArc(140,70)(10,-90,90){2}{4.5}
\Photon(152,70)(160,70){2}{1}
\Line(160,50)(160, 90)
\put(150,35){\makebox(0,0)[c]{\footnotesize (g)}}

\Photon(140,-10)(160,-10){2}{2}
\Photon(140, 10)(160, 10){2}{2}
\Line(160,-20)(160, 20)
\put(150,-35){\makebox(0,0)[c]{\footnotesize (h)}}

\Photon(140,-80)(160,-60){-2}{3}
\Photon(140,-60)(147, -67){2}{1}
\Photon(154,-73)(160,-80){2}{1}
\Line(160,-90)(160,-50)
\put(150,-105){\makebox(0,0)[c]{\footnotesize (i)}}

\put(0,-130){\makebox(0,0)[c]{\footnotesize
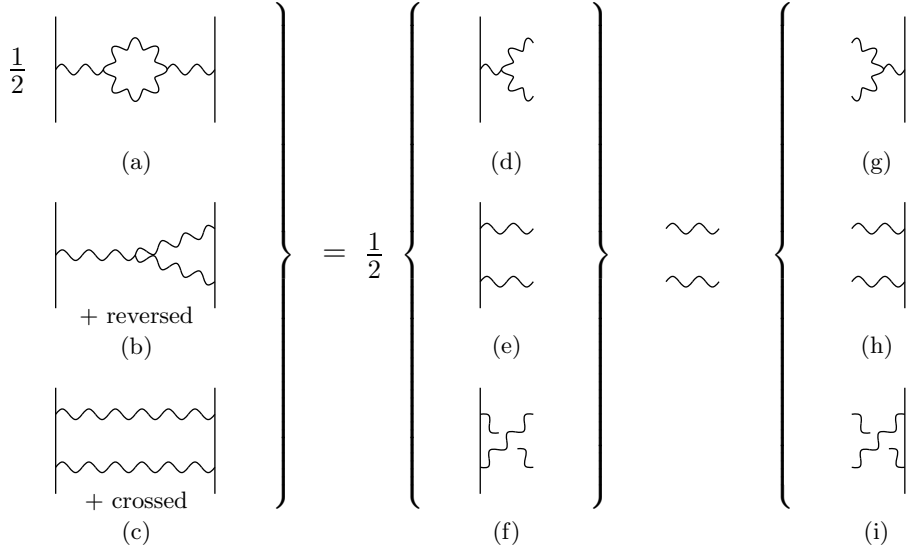
Fig.\ 6. The five one-loop diagrams formed from the ``product'' of
four-point functions 
$G_{\rho'\sigma' i'j'}^{rs(f')}$, $G_{\rho\sigma ij}^{rs(f)}$.}}

\end{picture}
\end{center}

i) {\em Feynman gauge}

\noindent
To begin with, we consider the Feynman gauge. The propagators
$iD_{\rho'\rho}(k_{1})$, $iD_{\sigma'\sigma}(k_{2})$
are then proportional to $g_{\rho'\rho}$, $g_{\sigma'\sigma}$
respectively. Thus, in the Feynman gauge, the
only sources of longitudinal factors $k_{1\rho}$, $k_{2\sigma}$
are the triple gluon vertices in Figs.\ 6(d) and (g). 

The triple gluon vertex in Fig.\ 6(g) may be decomposed as
\be\label{gammathooft}
\Gamma_{\nu\rho\sigma}^{nrs}
=
f^{nrs}\Bigl(\Gamma_{\nu\rho\sigma}^{F} + \Gamma_{\nu\rho\sigma}^{P}\Bigr)
\ee
where
\bea
\Gamma_{\nu\rho\sigma}^{F}(-q;-k_{1},k_{2})
&=&
-(k_{1}+k_{2})_{\nu}g_{\rho\sigma}
+2q_{\rho}g_{\sigma\nu} 
-2q_{\sigma}g_{\rho\nu} \\
\Gamma_{\nu\rho\sigma}^{P}(-q;-k_{1},k_{2})
&=&
k_{1\rho}g_{\sigma\nu} + k_{2\sigma}g_{\rho\nu}.
\label{GammaP}
\eea
The part $\Gamma_{\nu\rho\sigma}^{F}$ 
contributes no factors of
longitudinal loop four-momentum and 
obeys a simple Ward identity 
$q_{\nu}\Gamma_{\nu\rho\sigma}^{F}(-q;-k_{1},k_{2})=
(k_{1}^{2} - k_{2}^{2})g_{\rho\sigma}$
involving the difference of
two inverse gauge field propagators in the Feynman gauge.
The part $\Gamma_{\nu\rho\sigma}^{P}$ by contrast 
depends only on the longitudinal loop four-momenta.

In order to disentangle the effects of the two triple gluon vertices,
it is then convenient to make a similar decomposition for the two
four-point functions in Eq.\ (\ref{fig6abc}). Thus,
the four-point function $G_{\rho\sigma ij}^{rs(f)}$ represented
by the sum of diagrams Figs.\ 6(g)+(h)+(i) may be decomposed as
\be
G_{\rho\sigma ij}^{rs(f)}
=
G_{\rho\sigma ij}^{Frs(f)} + G_{\rho\sigma ij}^{Prs(f)}
\ee
where $G_{\rho\sigma ij}^{Prs(f)}$ is the part of Fig.\ 6(g)
proportional to $\Gamma^{P}$,
\be
G_{\rho\sigma ij}^{Prs(f)}(-k_{1},k_{2};p_{1},p_{2})
=
f^{nrs}\Gamma_{\nu'\rho\sigma}^{P}(-q;-k_{1},k_{2})
\,D_{\nu'\nu}(q)\,i\gamma_{\nu}T_{ji}^{n}
\label{ggffP}
\ee
and $G_{\rho\sigma ij}^{Frs(f)}$ is all of the remainder
of $G_{\rho\sigma ij}^{rs(f)}$, i.e.\ the 
part of Fig.\ 6(g) proportional to $\Gamma^{F}$ plus 
Figs.\ 6(h)+(i). The part 
$G_{\rho\sigma ij}^{Frs(f)}$ then satisfies the identity
\bea
\lefteqn{k_{1\rho}
G_{\rho\sigma ij}^{Frs(f)}(-k_{1},k_{2};p_{1},p_{2})
=} \nn \\
& &
f^{nrs}\Bigl(
[(k_{2}^{2} - k_{1}^{2})g_{\sigma\nu'} 
- k_{2\sigma}(k_{1} + k_{2})_{\nu'}]D_{\nu'\nu}(q) 
\,+\, \tilde{G}_{\mu}(q)\,q_{\nu}
\Bigr)\,i\gamma_{\nu}T_{ji}^{n} \nonumber \\
& &
+i\gamma_{\sigma}T_{jk}^{s}\,S(p_{1}\!-\!k_{1},m_{\!f})\,
S^{-1}(p_{1},m_{\!f})\,iT_{ki}^{r}
\nonumber\\ 
& &
-iT_{jk}^{r}\,S^{-1}(p_{2},m_{\!f})\,
S(p_{2}\!+\!k_{1},m_{\!f})\,i\gamma_{\sigma}T_{ki}^{s},
\label{ggffFwid}
\eea
easily obtained by writing $G^{F} = G - G^{P}$ 
and using the Ward identity Eq.\ (\ref{ggffwid}) for $G$
together with the above definition of $G^{P}$.
An exactly similar decomposition can be made for 
$G_{\rho'\sigma' i'j'}^{rs(f')}$.

Thus, in the Feynman gauge, we can write
\be
\left.\mbox{Figs.\ 6(a)+(b)+(c)}\right|_{\rm Feyn}
= 
{\textstyle \frac{1}{2}}g^{4}
\mu^{2\epsilon}\int\frac{d^{d}k}{(2\pi)^{d}}\frac{1}{k_{1}^{2}k_{2}^{2}}
\Bigl(G_{\rho\sigma i'j'}^{Frs(f')}
+G_{\rho\sigma i'j'}^{Prs(f')}\Bigr)
\Bigl( G_{\rho\sigma ij}^{Frs(f)}
      +G_{\rho\sigma ij}^{Prs(f)}\Bigr).
\label{fig6abcxi=1}
\ee
The isolation of the PT ``effective'' two-point component of the
diagrams Figs.\ 6(a), (b) and (c) 
then involves contracting together the terms
in the above equation and identifying the resulting 
two-point components according to the definition 
(\ref{ptPidef1})-(\ref{ptPidef3}) of $i\hat{\Pi}_{\mu\nu}^{mn}(q)$.

The term $G^{P}G^{P}$ is obtained immediately from the definitions
Eqs.\ (\ref{GammaP}) and (\ref{ggffP}):
\be
G_{\rho\sigma i'j'}^{Prs(f')}\,
G_{\rho\sigma ij}^{Prs(f)}
=
N\Bigl(i\gamma_{\mu}T_{j'i'}^{m}\Bigr)\frac{-1}{q^{2}}
\Bigl(
-(k_{1}^{2} + k_{2}^{2})g_{\mu\nu}
-k_{1\mu}k_{2\nu}
-k_{2\mu}k_{1\nu} \Bigr)
\frac{-1}{q^{2}}\Bigl(i\gamma_{\nu}T_{ji}^{m}\Bigr).
\label{GPGP}
\ee
{}From the definition (\ref{ptPidef1}), the transverse projection of
all of these terms contributes to the PT self-energy.

The term $G^{P}G^{F}$ is obtained using the identity 
Eq.\ (\ref{ggffFwid}):
\bea
G_{\rho\sigma i'j'}^{Prs(f')}\,
G_{\rho\sigma ij}^{Frs(f)}
&=&
\Bigl(i\gamma_{\mu}T_{j'i'}^{m}\Bigr)\frac{-1}{q^{2}}
f^{mrs}\Gamma_{\mu\rho\sigma}^{P}(q;k_{1},-k_{2})
G_{\rho\sigma ij}^{Frs(f)}(-k_{1},k_{2};p_{1},p_{2}) 
\label{GPGF0} \\
&=&
\Bigl(i\gamma_{\mu}T_{j'i'}^{m}\Bigr)\frac{-1}{q^{2}}\biggl\{
N\Bigl( (k_{1}+k_{2})_{\mu}(k_{1}+k_{2})_{\nu}\frac{-1}{q^{2}}
\,-\, 2\tilde{G}_{\mu}(q)\,q_{\nu}\Bigr)\Bigl(i\gamma_{\nu}T_{ji}^{m}\Bigr)
\nonumber \\
& &
\hspace{10pt}
-{\textstyle \frac{1}{2}}Ni\gamma_{\mu}T_{ji}^{m}
\Bigl(S(p_{1}\!-\!k_{1},m_{\!f}) 
\,+\, S(p_{1}\!+\!k_{2},m_{\!f})\Bigr)S^{-1}(p_{1})
\nonumber \\
& &
\hspace{10pt}
-{\textstyle \frac{1}{2}}NS^{-1}(p_{2},m_{\!f})
\Bigl(S(p_{1}\!-\!k_{1},m_{\!f}) 
\,+\, S(p_{1}\!+\!k_{2},m_{\!f})\Bigr)
i\gamma_{\mu}T_{ji}^{m}\biggr\}.
\label{GPGF}
\eea
{}From the definitions (\ref{ptPidef1}), (\ref{ptPidef2}), only the
term above proportional
to $(k_{1}+k_{2})_{\mu}(k_{1}+k_{2})_{\nu}$ contributes to the PT
self-energy: the term in Eq.\ (\ref{GPGF})
proportional to $\tilde{G}_{\mu}(q) q_{\nu}$ 
is orthogonal to the transverse tensor in (\ref{ptPidef2}), 
while the terms proportional to $S^{-1}(p_{1},m_{\!f})$ 
and $S^{-1}(p_{2},m_{\!f})$
involve the fermion propagators appearing in Figs.\ 6(h) and (i). 
It should be noted that
when the fermions are on-shell as in the S-matrix PT,
these last three terms each vanish identically
as a result of the Ward identity 
Eq.\ (\ref{fermionwid}) and the equations of motion
$\overline{u}(p_{2},m_{\!f})S^{-1}(p_{2},m_{\!f}) 
= S^{-1}(p_{1},m_{\!f})u(p_{1},m_{\!f}) = 0$
for the spinors
$\overline{u}(p_{2},m_{\!f})$  and 
$u(p_{1},m_{\!f})$.
When the fermions are off-shell, these terms do not vanish, but
they make no contribution to 
the ``effective'' two-point function defined in 
(\ref{ptPidef1})-(\ref{ptPidef3}) in the off-shell PT. 

An exactly similar expression to Eq.\ (\ref{GPGF}) is obtained
from the contraction $G^{F}G^{P}$, resulting in an identical 
contribution to the PT self-energy. 

Lastly, there is the term $G^{F}G^{F}$. By definition, 
the two $G^{F}$'s do not involve any factors of longitudinal
loop four-momentum. 
When contracted together, they therefore do 
not trigger any Ward identities, and so do
not lead to the
cancellation (pinching) of any propagators. 
This term can therefore be written
\be
G_{\rho\sigma i'j'}^{Frs(f')}
G_{\rho\sigma ij}^{Frs(f)}
=
N\Bigl(i\gamma_{\mu}T_{j'i'}^{m}\Bigr)\frac{-1}{q^{2}}
\Bigl(\Gamma_{\mu\rho\sigma}^{F}(q;k_{1},-k_{2})
\Gamma_{\nu\rho\sigma}^{F}(-q;-k_{1},k_{2})\Bigr)
\frac{-1}{q^{2}}\Bigl(i\gamma_{\nu}T_{ji}^{m}\Bigr) \,+\, \ldots
\label{GFGF}
\ee
where the ellipsis represents the terms involving the
fermion propagators appearing in Figs.\ 6(e), (f), (h) and (i),
and which constitute vertex and box corrections to the four-fermion
processs.

Adding up the above expressions, we obtain
\bea
\lefteqn{\mbox{Figs.\ 6(a)+(b)+(c)}|_{\rm Feyn} = }\nonumber \\
& &
{\textstyle \frac{1}{2}}
Ng^{2}\mu^{2\epsilon}\int\frac{d^{d}k}{(2\pi)^{d}}
\frac{1}{k_{1}^{2}k_{2}^{2}} 
\Bigl(ig\gamma_{\mu}T_{j'i'}^{m}\Bigr)\frac{-1}{q^{2}}
\biggl\{
\Gamma_{\mu\rho\sigma}^{F}(q;k_{1},-k_{2})
\Gamma_{\nu\rho\sigma}^{F}(-q;-k_{1},k_{2}) 
\nonumber \\
& & 
+\,2(k_{1}\!+\!k_{2})_{\mu}(k_{1}\!+\!k_{2})_{\nu} 
-(k_{1}^{2}\!+\!k_{2}^{2})g_{\mu\nu} 
-k_{1\mu}k_{2\nu}
-k_{2\mu}k_{1\nu}
\biggr\}
\frac{-1}{q^{2}}\Bigl(ig\gamma_{\nu}T_{ji}^{m}\Bigr)
\,+\, \ldots \qquad
\label{fig6abc2ptF}
\eea
where the ellipsis represents terms which do not contribute
to the PT self-energy.

In the Feynman gauge, the diagrams Figs.\ 2(c), (d) and (e)
make no contribution to the PT self-energy. This is because
there are no factors of longitudinal
four-momentum associated with the gluon propagating in the loop
to trigger the Ward identity Eq.\ (\ref{fermionwid})
required to pinch the two fermion propagators.
Thus, in this gauge, the entire gluonic contribution to
$i\hat{\Pi}_{\mu\nu}^{mn}(q)$ comes from the diagrams
Figs.\ 6(a), (b) and (c).
It therefore just remains to include the standard
linear covariant gauge ghost loop contribution
\be\label{Pighlincov}
\left.i\hat{\Pi}_{\mu\nu}^{{\rm (gh)}mn}(q)\right|_{\rm lin\,cov}
=
-N\delta^{mn}g^{2}
\mu^{2\epsilon}\int\frac{d^{d}k}{(2\pi)^{d}}
\frac{k_{1\mu}k_{2\nu}}{k_{1}^{2}k_{2}^{2}},
\ee
together with the fermion contribution
$i\Pi_{\mu\nu}^{{\rm (f)}mn}(q)$. Using the dimensional regularization 
rule $\int d^{d}k\,k^{-2} = 0$, we finally obtain
for the off-shell PT one-loop gluon ``effective'' two-point function 

\pagebreak

\begin{center}
\begin{picture}(400,260)(-30,-30)


\ArrowLine(100,160)(100,190)
\ArrowLine(100,190)(100,220)
\Photon(60,210)(80,190){2.5}{2.5}
\Photon(60,170)(80,190){-2.5}{2.5}
\Photon(80,190)(100,190){2.5}{2}
\put(35,210){\makebox(0,0)[r]{$k_{2\sigma}$}}
\put(40,210){\vector(1, 0){10}}
\put(35,170){\makebox(0,0)[r]{$-k_{1\rho}$}}
\put(40,170){\vector(1, 0){10}}

\put(170,205){\makebox(0,0)[c]{\small pinch}}
\put(155,190){\vector(1, 0){30}}

\put(220,190){\makebox(0,0)[c]{\Large $+$}}
\Photon(240,210)(280,190){2.5}{4.5}
\Photon(240,170)(280,190){-2.5}{4.5}
\ArrowLine(280,160)(280,190)
\ArrowLine(280,190)(280,220)
\put(300,190){\makebox(0,0)[l]{\Large $+ \cdots$}}

\ArrowLine(100, 80)(100,140)
\Photon(60,128)(100,128){2.5}{4}
\Photon(60, 92)(100, 92){2.5}{4}
\put(35,128){\makebox(0,0)[r]{$k_{2\sigma}$}}
\put(40,128){\vector(1, 0){10}}
\put(35,92){\makebox(0,0)[r]{$-k_{1\rho}$}}
\put(40,92){\vector(1, 0){10}}

\put(120,70){\makebox(0,0)[c]
{$ \left. \begin{array}{c}  \\ \\ \\ \\ \\ \\ \\ \\ \\
\end{array} \right\}$}}
\put(170,85){\makebox(0,0)[c]{\small pinch}}
\put(155,70){\vector(1, 0){30}}

\ArrowLine(100,  0)(100, 60)
\Photon(60, 50)(100, 10){2.5}{4}
\Photon(60, 10)( 73, 23){-2.5}{1.5}
\Photon(87, 37)(100, 50){2.5}{1.5}
\put(35,50){\makebox(0,0)[r]{$k_{2\sigma}$}}
\put(40,50){\vector(1, 0){10}}
\put(35,10){\makebox(0,0)[r]{$-k_{1\rho}$}}
\put(40,10){\vector(1, 0){10}}

\put(220,70){\makebox(0,0)[c]{\Large $-$}}
\Photon(240,90)(280,70){2.5}{4.5}
\Photon(240,50)(280,70){-2.5}{4.5}
\ArrowLine(280,40)(280, 70)
\ArrowLine(280,70)(280,100)
\put(300,70){\makebox(0,0)[l]{\Large $+ \cdots$}}

\put(170,-30){\makebox(0,0)[c]{\footnotesize 
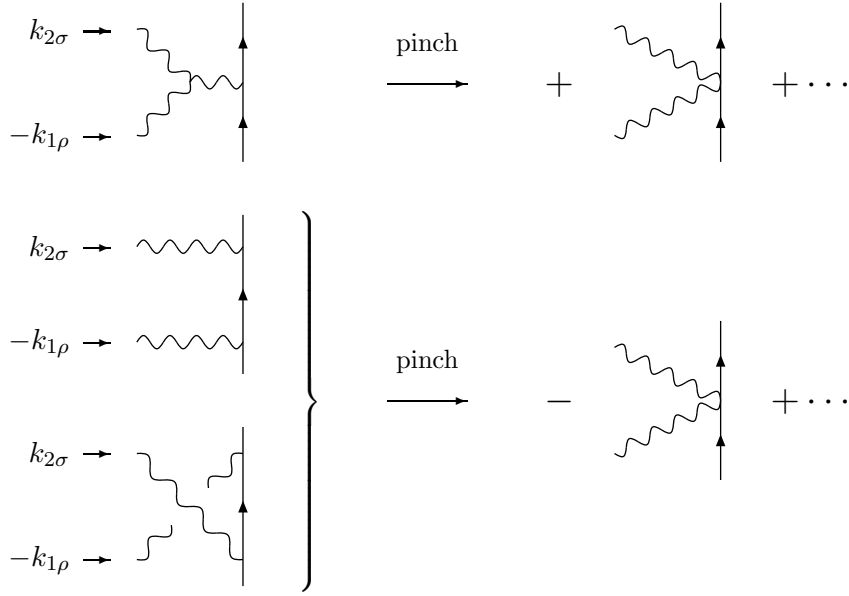
Fig.\ 7. The fundamental PT cancellation, expressed in the
Ward identity Eq.\ (\ref{ggffwid}).}}

\end{picture}
\end{center}
\bea
i\hat{\Pi}_{\mu\nu}^{mn}(q)
&=&
i\Pi_{\mu\nu}^{{\rm (f)}mn}(q)
\,\,-\,\,N\delta^{mn}g^{2}
\mu^{2\epsilon}\int\frac{d^{d}k}{(2\pi)^{d}}\frac{1}{k_{1}^{2}k_{2}^{2}}
\times 
\nonumber \\
& &
\Bigl\{{\textstyle \frac{1}{2}}
\Gamma_{\mu\rho\sigma}^{F}(q;k_{1},-k_{2})
\Gamma_{\nu\rho\sigma}^{F}(-q;-k_{1},k_{2}) 
\,+\,(k_{1}\!+\!k_{2})_{\mu}(k_{1}\!+\!k_{2})_{\nu} \Bigr\}
\label{ptPi3} \\
&=&
i\delta^{mn}q^{2}t_{\mu\nu}(q)\hat{\Pi}(q^{2})
\eea
where the function $\hat{\Pi}(q^{2})$
is identical to that obtained in the S-matrix PT,
given in the introduction in Eq.\ (\ref{ptPi}).\footnote{The
function Eq.\ (\ref{ptPi}) is actually for massless fermions
for the sake of simplicity.}

We make several remarks:
\begin{itemize}
\item In obtaining the PT self-energy Eq.\ (\ref{ptPi3}),
an exact cancellation has occurred
in the contributions from $G^{P}G^{F}$ and $G^{F}G^{P}$
Eq.\ (\ref{GPGF0}) between
i) components of the conventional self-energy
diagram Fig.\ 6(a), generated by factors of longitudinal
gluon loop four-momentum $k_{1\rho}$, $k_{2\sigma}$ from
the part $\Gamma^{P}$ of the triple gluon vertices,
in which
the gluon propagators $q^{-2}$ have been pinched, and
ii) the components of the conventional vertex diagrams,
Fig.\ 6(b) generated by the {\em same} factors
$k_{1\rho}$, $k_{2\sigma}$, in which the fermion propagtors
have been pinched.
This cancellation, illustrated schematically in Fig.\ 7,
is expressed succinctly in the Ward identity Eq.\ (\ref{ggffwid})
for the gluon pair--fermion pair four-point function
(more precisely, in the above Feynman gauge calculation, the
cancellation was expressed in the identity \eq{ggffFwid}
for the part $G^{F}$ of the four-point function).
By dealing with the {\em set} of diagrams
Figs.\ 6(a)+(b)+(c), as opposed to
individual diagrams, we have been able to
make this cancellation simply and immediately. 
We thus see that the PT algorithm amounts to the
identification of a {\em fundamental cancellation} among
contributions to one-loop processes generated by factors
of longitudinal four-momentum associated with the gauge fields
propagating in loops. This cancellation, expressed in the 
Ward identities of the theory, is independent of whether the
``external'' fields are on-shell, as in an S-matrix element,
or off-shell, as here. 
\item An exact cancellation has also occurred between the
contribution to $i\hat{\Pi}_{\mu\nu}^{mn}(q)$ of i) the
ghost component $i\Pi_{\mu\nu}^{{\rm (gh)}mn}(q)|_{\rm lin\,cov}$
of the conventional linear covariant gauge self-energy,
and ii) the component $G^{P}G^{P}$ of Fig.\ 6(a),
involving the contraction of the longitudinal parts $\Gamma^{P}$ of the 
triple gluon vertices (cf.\ Eqs.\ (\ref{GPGP}) and (\ref{Pighlincov})).
Furthermore, the components $G^{P}G^{F} + G^{F}G^{P}$
of Figs.\ 6(a)+(b)+(c)
contribute a term to $i\hat{\Pi}_{\mu\nu}^{mn}(q)$ 
identical to that of a set of scalar fields in the
adjoint representation, but with an overall minus sign.
\item The expression Eq.\ (\ref{ptPi3}) exactly
{\em coincides} with the background gluon
self-energy computed in the Feynman quantum gauge.
This is a particular 
example of the general correspondence 
\cite{hashimoto}-\cite{edrnjw} between the
PT gauge-independent one-loop $n$-point functions and the
background field Feynman gauge one-loop $n$-point functions
mentioned in the introduction. Thus, there exists a set
of Feynman rules which {\em reproduce} the PT one-loop $n$-point
functions.
\item As a result of the cancellation
described in the first remark,
the PT self-energy Eq.\ (\ref{ptPi3}) couples
to the external fermion lines via two single gluon propagators
in exactly the same way as any scalar or fermion contributions
to the self-energy. Thus, despite initial appearances to 
the contrary in Figs.\ 2(g), (h) and (i),
the diagrammatic concept of one-particle irreducibility 
for the self-energy in the PT is in fact retained. 
\end{itemize}

ii) {\em Arbitrary gauge}

\noindent
The next step is to show how the fundamental PT cancellation
described above operates in an arbitrary gauge,
i.e.\ when the terms $a$ and $b$ in the expression \eq{prop}
for the tree level gluon propagator are non-zero.
The task is to show that, regardless of the fact that the external fermions
are off-shell, the additional contributions to $i\hat{\Pi}_{\mu\nu}^{mn}(q)$
due to the longitudinal $a$ and $b$ terms in the gluon propagators exactly
cancel among themselves, leaving always the result \eq{ptPi3}.
This task is greatly facilitated by the decomposition of
diagrams shown in Fig.~6.
We consider both the class of linear covariant gauges and the class of
non-covariant gauges $n\sdot A^{a} = 0$, as described in Sec.~3.
For the sake of overall continuity, the detailed 
account of the cancellation
mechanism is relegated to App.~A.

The analysis in App.~A can be summarized in two remarks:
\begin{itemize}
\item The contribution 
$i\hat{\Pi}_{\mu\nu}^{{\rm (a)}mn}(q)|_{n\cdot A = 0}$ 
to the PT self-energy due to the $a$ terms in the propagators
$iD_{\rho'\rho}(k_{1})$, $iD_{\sigma'\sigma}(k_{2})$
in the amplitude (\ref{fig6abc}) for the diagrams
Figs.~6(a), (b) and (c)
in the class of non-covariant gauges $n\sdot A^{a} = 0$
is {\em identical} to the ghost contribution
$i\Pi_{\mu\nu}^{{\rm (gh)}mn}(q)|_{\rm lin\,cov}$ 
to the self-energy in the class of linear covariant gauges
(cf.\ Eqs.~(\ref{acontrib}) and (\ref{Pighlincov})). Given that
$i\hat{\Pi}_{\mu\nu}^{{\rm (gh)}mn}(q)|_{n\cdot A = 0} =
 i\Pi_{\mu\nu}^{{\rm (a)}mn}(q)|_{\rm lin\,cov} = 0$,
this equality can be expressed as the gauge-independent statement
\be\label{agh}
i\Pi_{\mu\nu}^{{\rm (gh)}mn}(q) + i\hat{\Pi}_{\mu\nu}^{{\rm (a)}mn}(q) 
=
-N\delta^{mn}g^{2}
\mu^{2\epsilon}\int\frac{d^{d}k}{(2\pi)^{d}}
\frac{k_{1\mu}k_{2\nu}}{k_{1}^{2}k_{2}^{2}}.
\ee
In {\em both} classes of gauge, this contribution (\ref{agh}) then 
cancels against the component $G^{P}G^{P}$ of Fig.\ 6(a),
involving the contraction of the longitudinal parts $\Gamma^{P}$ of the triple gluon
vertices via the terms $g_{\rho'\rho}$, $g_{\sigma'\sigma}$ in the
propagators $iD_{\rho'\rho}(k_{1})$, $iD_{\sigma'\sigma}(k_{2})$
(cf.\ the second remark above).
\item An exact cancellation occurs between the contributions to
the PT self-energy due to i) the $b$ terms in the propagators
$iD_{\rho'\rho}(k_{1})$, $iD_{\sigma'\sigma}(k_{2})$
in the amplitude (\ref{fig6abc}) for the diagrams
Figs.\ 6(a), (b) and (c), and ii) the $b$ terms in the gluon propagators in
the amplitudes for the diagrams Figs.\ 2(c), (d) and (e).
This cancellation can be expressed as the gauge-independent statement
\be\label{b}
i\hat{\Pi}_{\mu\nu}^{{\rm (b)}mn}(q) 
=
0.
\ee
In {\em both} classes of gauge, the $b$ terms in the tree level gluon
propagators thus make no net contribution to $i\hat{\Pi}_{\mu\nu}^{mn}(q)$.

\end{itemize}

It is emphasized that these results are due purely to the Ward
identities Eqs.~(\ref{fermionwid}) and (\ref{ggffwid}), and are entirely
independent of whether the external fields are on- or off-shell.
In this way, all dependence of the PT ``effective'' two-point component
(\ref{ptPidef1}) of the off-shell
four-fermion scattering processs 
$\psi_{i}^{\scriptscriptstyle{(f)}}(p_{1})
\psi_{i'}^{\scriptscriptstyle{(f')}}(p_{1}') 
\rightarrow 
\psi_{j}^{\scriptscriptstyle{(f)}}(p_{2})
\psi_{j'}^{\scriptscriptstyle{(f')}}(p_{2}')$
on the longitudinal $a$ and $b$ terms
in the tree level gluon propagators in the diagrams of Fig.~2 cancels.
The gauge-independence of the one-loop gluon self-energy
$i\hat{\Pi}_{\mu\nu}^{mn}(q)$ obtained in 
the off-shell PT is thus proved.


\pagebreak

{\bf 4.3. Universality of $i\hat{\Pi}_{\mu\nu}^{mn}(q)$}

\noindent
In this section, it is shown how the gluon self-energy
defined in the off-shell PT is universal, i.e.\ independent of the
the particular choice of off-shell external fields in
the definition (\ref{ptPidef1}). 

For any pair of the sets of external fields appearing in
the definition (\ref{ptPidef1}), we consider first
the set of one-loop diagrams formed from the ``product'' of the 
corresponding pair of tree level connected $n$-point functions
defined in Eqs.\ (\ref{ggff})-(\ref{ggggg}). The amplitude, 
in any gauge, for the
given set of diagrams can be written  as the appropriate term from
\be\label{GG}
{\textstyle \frac{1}{2}}
\mu^{2\epsilon}\int\frac{d^{d}k}{(2\pi)^{d}}
\left\{
\begin{array}{c}
ig^{2}G_{\rho'\sigma' i'j'}^{rs(f')} \\
\\
ig^{2}G_{\rho'\sigma'\alpha'\beta'}^{rsa'b'} \\
\\
-g^{3}G_{\rho'\sigma'\alpha'\beta'\gamma'}^{rsa'b'c'} \\
\end{array}
\right\}
iD_{\rho'\rho}(k_{1})\,iD_{\sigma'\sigma}(k_{2})
\left\{
\begin{array}{c}
ig^{2}G_{\rho\sigma ij}^{rs(f)} \\
\\
ig^{2}G_{\rho\sigma\alpha\beta}^{rsab} \\
\\
-g^{3}G_{\rho\sigma\alpha\beta\gamma}^{rsabc} \\
\end{array}
\right\}
\ee
(arguments for the various $n$-point functions are again omitted for
brevity). For example, for the interaction
$A_{\alpha}^{a}A_{\alpha'}^{a'}\rightarrow A_{\beta}^{b}A_{\beta'}^{b'}$
between two pairs of off-shell gluons,
the diagrams corresponding to the ``product'' of $n$-point functions
$G_{\rho'\sigma'\alpha'\beta'}^{rsa'b'}$
and $G_{\rho\sigma\alpha\beta}^{rsab}$
in (\ref{GG}) are shown in Fig.\ 8. 
In all cases, the factor $\frac{1}{2}$ in (\ref{GG}) results in the
correct symmetry factors for the corresponding diagrams,
since each of the tree level $n$-point functions in
(\ref{GG}) is symmetric under interchange of the pair of gluons
propagating in the loops. In the most complicated case, i.e.\ that of
the interaction 
$A_{\alpha}^{a}A_{\alpha'}^{a'} \rightarrow
A_{\beta}^{b}A_{\gamma}^{c}A_{\beta'}^{b'}A_{\gamma'}^{c'}$
between six external gluons,
the corresponding amplitude from (\ref{GG}) is
represented by 25$\times$25 = 625 individual one-loop
diagrams (although many of these
are in fact identical by symmetry). It is important to
be clear that, in each case, the diagrams 
representing the amplitudes in (\ref{GG}) still
only represent a subset of the complete set of
one-loop corrections to the given tree level interaction. 

i) {\em Feynman gauge}

\noindent
To begin with, we again consider the Feynman gauge. Exactly as
in the case of the four-fermion process in the Feynman gauge,
we first decompose each of
the four- and five-point functions in (\ref{GG}) as
$G = G^{F} + G^{P}$, where for the functions on the r.h.s.\
(cf.~\eq{ggffP})
\bea \!\!
\left.
\begin{array}{r}
G_{\rho\sigma ij}^{Prs(f)}(-k_{1},k_{2};p_{1},p_{2}) \\
\\
G_{\rho\sigma\alpha\beta}^{Prsab}(-k_{1},k_{2};p_{1},p_{2}) \\
\\
G_{\rho\sigma\alpha\beta\gamma}^{Prsabc}
(-k_{1},k_{2};p_{1},p_{2},p_{3}) \\
\end{array} \!\!
\right\}
=
f^{nrs}\Gamma_{\nu'\rho\sigma}^{P}(-q;-k_{1},k_{2})D_{\nu'\nu}(q)
\left\{ \!
\begin{array}{l}
i\gamma_{\nu}T_{ji}^{n} \\
\\
\Gamma_{\nu\alpha\beta}^{nab}(q,p_{1},p_{2}) \\
\\
G_{\nu\alpha\beta\gamma}^{nabc}(q,p_{1},p_{2},p_{3}). \\
\end{array}
\right.
\eea
It should be noted that
$G_{\rho\sigma\alpha\beta\gamma}^{Prsabc}$ involves the
full four-point function
$G_{\nu\alpha\beta\gamma}^{nabc}$, and not just
the quadruple gluon vertex
$\Gamma_{\nu\alpha\beta\gamma}^{nabc}$
(cf.\ the definition Eq.\ (\ref{gggg})). 
This is in order to keep

\begin{center}
\begin{picture}(400,450)(-200,-270)

\put(-175,140){\makebox(0,0)[c]{\Large $\frac{1}{2}$}}
\Photon(-160, 120)(-160,160){2}{4}
\Photon(-160,140)(-142,140){2}{2}
\PhotonArc(-130,140)(10,34,394){2}{8}
\Photon(-118,140)(-100,140){2}{2}
\Photon(-100, 120)(-100,160){2}{4}
\put(-130,105){\makebox(0,0)[c]{\footnotesize (a)}}

\Photon(-160,50)(-160,90){2}{4}
\Photon(-160,70)(-130,70){2}{3}
\Photon(-130,70)(-100, 80){2}{3.5}
\Photon(-130,70)(-100, 60){-2}{3.5}
\Photon(-100,50)(-100,90){2}{4}
\put(-130,47){\makebox(0,0)[c]{\footnotesize + reversed}}
\put(-130,35){\makebox(0,0)[c]{\footnotesize (b)}}

\put(-175,0){\makebox(0,0)[c]{\Large $\frac{1}{2}$}}
\Photon(-160,-20)(-160,20){2}{4}
\Photon(-160,0)(-132,0){2}{3}
\PhotonArc(-120,0)(10,50,310){2}{6.5}
\Photon(-113.6, 7.7)(-100,0){-2}{2}
\Photon(-113.6,-7.7)(-100,0){2}{2}
\Photon(-100,-20)(-100,20){2}{4}
\put(-130,-23){\makebox(0,0)[c]{\footnotesize + reversed}}
\put(-130,-35){\makebox(0,0)[c]{\footnotesize (c)}}

\put(-175,-70){\makebox(0,0)[c]{\Large $\frac{1}{2}$}}
\Photon(-160,-90)(-160,-50){2}{4}
\PhotonArc(-147,-70)(10,90,270){2}{4.5}
\Photon(-147,-60)(-113,-60){-2}{4.5}
\Photon(-147,-80)(-113,-80){2}{4.5}
\PhotonArc(-113,-70)(10,-90,90){2}{4.5}
\Photon(-100,-90)(-100,-50){2}{4}
\put(-130,-105){\makebox(0,0)[c]{\footnotesize (d)}}

\Photon(-160,-160)(-160,-120){2}{4}
\Photon(-160,-140)(-100,-130){2}{6}
\Photon(-160,-140)(-100,-150){2}{6}
\Photon(-100,-160)(-100,-120){2}{4}
\put(-130,-163){\makebox(0,0)[c]{\footnotesize + reversed}}
\put(-130,-175){\makebox(0,0)[c]{\footnotesize (e)}}

\Photon(-160,-230)(-160,-190){2}{4}
\Photon(-160,-220)(-100,-220){2}{6}
\Photon(-160,-200)(-100,-200){2}{6}
\Photon(-100,-230)(-100,-190){2}{4}
\put(-130,-233){\makebox(0,0)[c]{\footnotesize + crossed}}
\put(-130,-245){\makebox(0,0)[c]{\footnotesize (f)}}

\put(-80,-35){\makebox(0,0)[c]{$ \left. \begin{array}{c} 
\\ \\ \\ \\ \\ \\ \\ \\ \\ \\ \\ \\   
\\ \\ \\ \\ \\ \\ \\ \\ \\ \\ \\ \\   \\
\end{array} \right\} $}}
\put(-55,-35){\makebox(0,0)[c]{\large =}}
\put(-40,-35){\makebox(0,0)[c]{\Large $\frac{1}{2}$}}
\put(-20,-35){\makebox(0,0)[c]{$ \left\{\begin{array}{c} 
\\ \\ \\ \\ \\ \\ \\ \\ \\ \\ \\ \\   \\ \\ \\ \\
\end{array} \right. $}}

\Photon(0, 50)(0,90){2}{4}
\Photon(0,70)(8,70){2}{1}
\PhotonArc(20,70)(10,90,-90){2}{4.5}
\put(10,35){\makebox(0,0)[c]{\footnotesize (g)}}

\Photon(0,-20)(0,20){2}{4}
\Photon(0,-10)(20,-10){2}{2}
\Photon(0, 10)(20, 10){2}{2}
\put(10,-35){\makebox(0,0)[c]{\footnotesize (h)}}

\Photon(0,-90)(0,-50){2}{4}
\Photon(0,-80)(20,-60){-2}{3}
\Photon(0,-60)(7, -67){2}{1}
\Photon(14,-73)(20,-80){2}{1}
\put(10,-105){\makebox(0,0)[c]{\footnotesize (i)}}

\Photon(0,-160)(0,-120){2}{4}
\Photon(0,-140)(20,-150){2}{2}
\Photon(0,-140)(20,-130){2}{2}
\put(10,-175){\makebox(0,0)[c]{\footnotesize (j)}}

\put( 40,-35){\makebox(0,0)[c]{$ \left. \begin{array}{c} 
\\ \\ \\ \\ \\ \\ \\ \\ \\ \\ \\ \\   \\ \\ \\ \\
\end{array} \right\} $}}
\Photon(70,-45)(90,-45){2}{2}
\Photon(70,-25)(90,-25){2}{2}
\put(120,-35){\makebox(0,0)[c]{$ \left\{ \begin{array}{c} 
\\ \\ \\ \\ \\ \\ \\ \\ \\ \\ \\ \\    \\ \\ \\ \\
\end{array} \right. $}}

\PhotonArc(140,70)(10,-90,90){2}{4.5}
\Photon(152,70)(160,70){2}{1}
\Photon(160,50)(160, 90){2}{4}
\put(150,35){\makebox(0,0)[c]{\footnotesize (k)}}

\Photon(140,-10)(160,-10){2}{2}
\Photon(140, 10)(160, 10){2}{2}
\Photon(160,-20)(160, 20){2}{4}
\put(150,-35){\makebox(0,0)[c]{\footnotesize (l)}}

\Photon(140,-80)(160,-60){-2}{3}
\Photon(140,-60)(147, -67){2}{1}
\Photon(154,-73)(160,-80){2}{1}
\Photon(160,-90)(160,-50){2}{4}
\put(150,-105){\makebox(0,0)[c]{\footnotesize (m)}}

\Photon(140,-150)(160,-140){2}{2}
\Photon(140,-130)(160,-140){2}{2}
\Photon(160,-160)(160,-120){2}{4}
\put(150,-175){\makebox(0,0)[c]{\footnotesize (n)}}

\put(0,-270){\makebox(0,0)[c]{\footnotesize
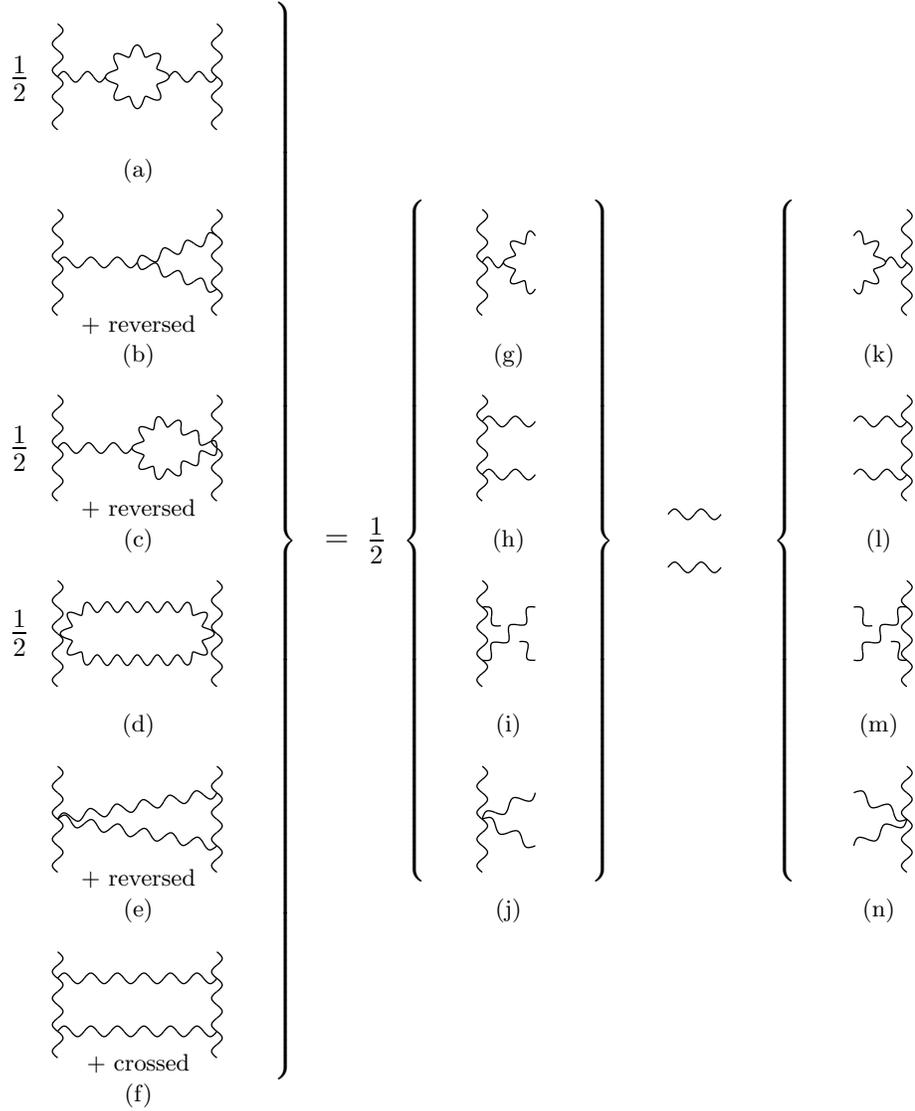
Fig.\ 8. The ten one-loop diagrams formed from the ``product'' of
four-point functions 
$G_{\rho'\sigma'\alpha'\beta'}^{rsa'b'}$, 
$G_{\rho\sigma\alpha\beta}^{rsab}$.}}

\end{picture}
\end{center}

\noindent
simple the Ward identity for
$G_{\rho\sigma\alpha\beta\gamma}^{Frsabc}$.
An exactly similar decomposition can be made for the four- and five-point
functions on the l.h.s.\ of (\ref{GG}). 
Then, in the Feynman gauge, just as in 
Eq.\ (\ref{fig6abcxi=1}), we can write the various terms in
(\ref{GG}) as products $(G^{F} + G^{P})(G^{F} + G^{P})$. 
The isolation of the PT self-energy component of
the various diagrams in (\ref{GG}) then again involves
contracting together these terms and identifying the
resulting ``effective'' 
two-point components according to the definition
(\ref{ptPidef1})-(\ref{ptPidef3}) of 
$i\hat{\Pi}_{\mu\nu}^{mn}(q)$.

For the terms $G^{P}G^{P}$, clearly in all cases we obtain an
expression identical to Eq.\ (\ref{GPGP}) except for
the appropriate
substitution of the corresponding tree-level
$n$-point function
$\Gamma_{\nu\alpha\beta}^{mab}$ or 
$igG_{\nu\alpha\beta\gamma}^{mabc}$ for
$i\gamma_{\nu}T_{ji}^{m}$ on the
r.h.s., and
$\Gamma_{\mu\alpha'\beta'}^{ma'b'}$ or 
$igG_{\mu\alpha'\beta'\gamma'}^{ma'b'c'}$ for
$i\gamma_{\mu}T_{j'i'}^{m}$ on the l.h.s.
For the cases involving the four-point functions, we then simply
retain the quadruple gluon vertex parts
$-ig\Gamma_{\nu\alpha\beta\gamma}^{mabc}$, 
$-ig\Gamma_{\mu\alpha'\beta'\gamma'}^{ma'b'c'}$
respectively. Thus, the contribution 
of the $G^{P}G^{P}$ term to the PT self-energy
is in all cases identical to that in the four-fermion case.

For the terms $G^{P}G^{F}$, we need to contract
$\Gamma_{\mu\rho\sigma}^{P}$
with $G_{\rho\sigma\alpha\beta}^{Frsab}$
and
$G_{\rho\sigma\alpha\beta\gamma}^{Frsabc}$.
Writing in each case $G^{F} = G - G^{P}$ and using
the Ward identities Eqs.\ (\ref{ggggwid}) and
(\ref{gggggwid}) respectively, we easily obtain
\bea
\lefteqn{
f^{mrs}\Gamma_{\mu\rho\sigma}^{P}(q;k_{1},-k_{2})\,
G_{\rho\sigma\alpha\beta}^{Frsab}(-k_{1},k_{2};p_{1},p_{2})
= }\nonumber\\
& &
N\Bigl( (k_{1}+k_{2})_{\mu}(k_{1} + k_{2})_{\nu'}D_{\nu'\nu}(q)
\,-\, 2\tilde{G}_{\mu}(q)\,q_{\nu}\Bigr)
\Gamma_{\nu\alpha\beta}^{mab}(q,p_{1},p_{2}) 
\nonumber \\
& &
+{\textstyle \frac{1}{2}}N\Bigl[
\Bigl(
p_{1}^{2}t_{\alpha\nu'}(p_{1})D_{\nu'\nu}(p_{1}\!-\!k_{1}) 
\,+\,\tilde{G}_{\alpha}(p_{1}\!-\!k_{1})\,(p_{1}\!-\!k_{1})_{\nu} \Bigr)
\Gamma_{\nu\beta\mu}^{abm}(p_{1}\!-\!k_{1},p_{2},k_{2})
\nonumber \\
& &
\phantom{{\textstyle \frac{1}{2}}N}
+\Bigl(
p_{1}^{2}t_{\alpha\nu'}(p_{1})D_{\nu'\nu}(p_{1}\!+\!k_{2}) 
\,+\,\tilde{G}_{\alpha}(p_{1}\!+\!k_{2})\,(p_{1}\!+\!k_{2})_{\nu} \Bigr)
\Gamma_{\nu\beta\mu}^{abm}(p_{1}\!+\!k_{2},p_{2},-k_{1})
\nonumber \\
& &
\phantom{{\textstyle \frac{1}{2}}N}
+\Bigl(
p_{2}^{2}t_{\beta\nu'}(p_{2})D_{\nu'\nu}(p_{2}\!-\!k_{1}) 
\,+\,\tilde{G}_{\beta}(p_{2}\!-\!k_{1})\,(p_{2}\!-\!k_{1})_{\nu} \Bigr)
\Gamma_{\nu\mu\alpha}^{bma}(p_{2}\!-\!k_{1},k_{2},p_{1} )
\nonumber \\
& &
\phantom{{\textstyle \frac{1}{2}}N}
+\Bigl(
p_{2}^{2}t_{\beta\nu'}(p_{2})D_{\nu'\nu}(p_{2}\!+\!k_{2}) 
\,+\,\tilde{G}_{\beta}(p_{2}\!+\!k_{2})\,(p_{2}\!+\!k_{2})_{\nu} \Bigr)
\Gamma_{\nu\mu\alpha}^{bma}(p_{2}\!+\!k_{2},-k_{1},p_{1} ) \,\Bigr]
\hspace{25pt}
\label{GammaPggggF}
\eea
and
\bea
\lefteqn{
f^{mrs}\Gamma_{\mu\rho\sigma}^{P}(q;k_{1},-k_{2})\,
G_{\rho\sigma\alpha\beta\gamma}^{Frsabc}(-k_{1},k_{2};p_{1},p_{2},p_{3})
=} \nonumber\\
& &\hspace{-20pt}
N\Bigl( (k_{1}+k_{2})_{\mu}(k_{1} + k_{2})_{\nu'}D_{\nu'\nu}(q)
\,-\, 2\tilde{G}_{\mu}(q)\,q_{\nu}\Bigr)
G_{\nu\alpha\beta\gamma}^{mabc}(q,p_{1},p_{2},p_{3}) 
\nonumber \\
& &\hspace{-20pt}
-f^{mrs}\Bigl[
f^{nra}\Bigl(
p_{1}^{2}t_{\alpha\nu'}(p_{1})D_{\nu'\nu}(p_{1}\!-\!k_{1}) 
\,+\,\tilde{G}_{\alpha}(p_{1}\!-\!k_{1})\,(p_{1}\!-\!k_{1})_{\nu} \Bigr)
G_{\nu\beta\gamma\mu}^{nbcs}(p_{1}\!-\!k_{1},p_{2},p_{3},k_{2})
\nonumber \\
& &\hspace{-20pt}
\phantom{f^{mrs}}
+f^{nsa}\Bigl(
p_{1}^{2}t_{\alpha\nu'}(p_{1})D_{\nu'\nu}(p_{1}\!+\!k_{2}) 
\,+\,\tilde{G}_{\alpha}(p_{1}\!+\!k_{2})\,(p_{1}\!+\!k_{2})_{\nu} \Bigr)
G_{\nu\beta\gamma\mu}^{nbcr}(p_{1}\!+\!k_{2},p_{2},p_{3},-k_{1})
\nonumber \\
& &\hspace{-20pt}
\phantom{f^{mrs}}
+f^{nrb}\Bigl(
p_{2}^{2}t_{\beta\nu'}(p_{2})D_{\nu'\nu}(p_{2}\!-\!k_{1}) 
\,+\,\tilde{G}_{\beta}(p_{2}\!-\!k_{1})\,(p_{2}\!-\!k_{1})_{\nu} \Bigr)
G_{\nu\gamma\mu\alpha}^{ncsa}(p_{2}\!-\!k_{1},p_{3},k_{2},p_{1})
\nonumber \\
& &\hspace{-20pt}
\phantom{f^{mrs}}
+f^{nsb}\Bigl(
p_{2}^{2}t_{\beta\nu'}(p_{2})D_{\nu'\nu}(p_{2}\!+\!k_{2}) 
\,+\,\tilde{G}_{\beta}(p_{2}\!+\!k_{2})\,(p_{2}\!+\!k_{2})_{\nu} \Bigr)
G_{\nu\gamma\mu\alpha}^{ncra}(p_{2}\!+\!k_{2},p_{3},-k_{1},p_{1})
\nonumber \\
& &\hspace{-20pt}
\phantom{f^{mrs}}
+f^{nrc}\Bigl(
p_{3}^{3}t_{\gamma\nu'}(p_{3})D_{\nu'\nu}(p_{3}\!-\!k_{1}) 
\,+\,\tilde{G}_{\gamma}(p_{3}\!-\!k_{1})\,(p_{3}\!-\!k_{1})_{\nu} \Bigr)
G_{\nu\mu\alpha\beta}^{nsab}(p_{3}\!-\!k_{1},k_{2},p_{1},p_{2})
\nonumber \\
& &\hspace{-20pt}
\phantom{f^{mrs}}
+f^{nsc}\Bigl(
p_{3}^{3}t_{\gamma\nu'}(p_{3})D_{\nu'\nu}(p_{3}\!+\!k_{2}) 
\,+\,\tilde{G}_{\gamma}(p_{3}\!+\!k_{2})\,(p_{3}\!+\!k_{2})_{\nu} \Bigr)
G_{\nu\mu\alpha\beta}^{nrab}(p_{3}\!+\!k_{2},-k_{1},p_{1},p_{2})\,\Bigr]
\nn \\
\label{GammaPgggggF}
\eea
(in Eq.\ (\ref{GammaPggggF}), we have used
$f^{alm}f^{bmn}f^{cnl} = \frac{1}{2}Nf^{abc}$). 
Comparing each of these expressions with that Eq.\ (\ref{GPGF})
for the four-fermion case, and using the definition
(\ref{ptPidef1})-(\ref{ptPidef3}), we see that in all cases only the
term proportional to $(k_{1}+k_{2})_{\mu}(k_{1}+k_{2})_{\nu'}$
contributes to the PT self-energy. Thus,
the contribution of the $G^{P}G^{F}$ terms is also in all
cases identical to that in the four-fermion case. Clearly, the
same holds for the $G^{F}G^{P}$ terms.

Lastly, there are the $G^{F}G^{F}$ terms. In the Feynman gauge, it is
not difficult to see that, despite the presence of various
longitudinal factors from the triple gluon vertices involved,
in all cases
the only contribution to the PT self-energy is from the
part $\Gamma^{F}\Gamma^{F}$ 
of the conventional gluon loop (shown in Fig.\ 8(a) for the 
case of external gluon pairs).
Thus, in all cases the contribution of the $G^{F}G^{F}$
terms is as in Eq.\ (\ref{GFGF}) in the four-fermion case.

Thus, in the Feynman gauge, the PT one-loop
``effective'' two-point component of the diagrams
corresponding to {\em any} of the
combinations of external fields in the definition
(\ref{ptPidef1})-(\ref{ptPidef3}) is identical. 
This is a direct result of the same fundamental
cancellation mechanism in each case, expressed in the Ward identities
Eqs.\ (\ref{ggffwid})-(\ref{gggggwid}).

\vspace{5pt}

ii) {\em Arbitrary gauge}

\noindent
In order to complete the demonstration of the universality of
$i\hat{\Pi}_{\mu\nu}^{mn}(q)$, it remains to be shown that, for
any of the pairs of sets of external fields in the definition
(\ref{ptPidef1}), the additional contributions to 
$i\hat{\Pi}_{\mu\nu}^{mn}(q)$ which occur when one moves away
from the Feynman gauge exactly cancel among themselves.
The analysis proceeds in exactly the same as for the four-fermion case,
and is given in App.~A. 
It is shown there that, despite the apparent dissimilarity
among the various sets of diagrams, exactly the same 
gauge-independent statements Eqs.~(\ref{agh}) and (\ref{b})
hold in all cases. Again, this is a direct result of the tree level Ward
identities given in Sec.~3.
In this way, the universality of the one-loop gauge-independent
gluon self-energy 
$i\hat{\Pi}_{\mu\nu}^{mn}(q)$ obtained in the off-shell PT is proved.

\vspace{10pt}

{\bf 4.4. Dyson Summation of $i\hat{\Pi}_{\mu\nu}^{mn}(q)$}

\noindent
The final task is to show how the PT one-loop gluon self-energy
may be summed in a Dyson series. This task involves
i) identifying the subclasses of $n$-loop diagrams, 
$n = 1,2\ldots \infty$, which 
contain the chains of $n$ PT one-loop self-energies implicitly, 
and ii) showing how, in an arbitrary gauge,
the fundamental cancellation mechanism
which operates at one loop generalizes to $n$ loops, so that
the chains of $n$ PT one-loop gauge-independent
self-energies can be isolated explicitly and unambiguously
in the corresponding Feynman integrands. 

The first step has been made by Papavassiliou and Pilaftsis
\cite{papapil1}, who gave an heuristic outline of the Dyson summation. 
This outline involved enumerating the pinch contributions needed 
in the Feynman gauge to convert
chains of conventional one-loop self-energies into chains
of PT self-energies, and then re-allocating ``by hand'' the
required contributions from conventional multi-loop self-energy,
vertex and box diagrams. Extending this approach to the 
standard electroweak model, these authors were able to show 
that the Dyson summation of the PT $W$ and $Z$ boson
self-energies does not shift the pole of the corresponding
propagators. However, no attempt was made to rearrange
directly the multi-loop diagrams at the Feynman integrand level,
as at one loop,
nor to demonstrate the required gauge cancellation mechanism.
This second step initially appears formidably difficult,
since it involves the consideration in an arbitrary gauge of
diagrams occurring at all orders in perturbation theory. 
However, it will be shown here how,
for the Dyson summation, the one-loop gauge cancellation mechanism
iterates, enabling the contributions to
the Dyson chain of $n$ PT gauge-independent one-loop self-energies to be 
isolated explicitly from the relevant subclass of diagrams for all $n$.

We shall restrict the analysis to the class of non-covariant
gauges $n\sdot A^{a} = 0$. In this way, we avoid
all diagrams involving ghosts.\footnote{
In the absence of the explicit
formulation of the PT beyond the one-loop level, 
the general procedure for dealing with multi-loop diagrams
in the PT has yet to be elucidated.
For the special case of the one-loop Dyson summation, however,
it is possible to circumvent
this difficulty for the multi-loop diagrams involving  
gluons and/or fermions (and/or scalars), as will be shown,
but {\em not} for those involving ghosts.
For this (and {\em only} this) reason, we make the above
restriction for the proof of the one-loop Dyson summation.}
To begin with, we assume the absence of fermions ($n_{\!f} = 0$),
and consider the set of purely gluonic $n$-loop diagrams
formed from chains of $n+1$ tree level gluon four-point
functions $G_{\alpha\beta\gamma\delta}^{abcd}$, each joined
to the next by a pair of tree level gluon propagators:
schematically, suppressing all indices and arguments
and all factors except that due to symmetry, this amplitude has the form
\be\label{nloops}
\Bigl({\textstyle \frac{1}{2}}\Bigr)^{n}
\int d^{d}k_{1}\ldots\int d^{d}k_{n}\,\,
G\,\underbrace{DD\,\,G\,\,DD\,\,G\,\ldots\, DD}_{
n\,\mbox{\scriptsize loops}}\,G
\ee
where $D$ is the gluon propagator and
$G$ is the gluon four-point function.  
For $n=1$, the corresponding
diagrams are just those shown in Fig.~8. For $n=2$, the corresponding
diagrams are shown in Fig.~9. 
For any $n$, the factor $(\frac{1}{2})^{n}$ results 
in all cases in the correct
symmetry factors for the $n$-loop diagrams.

We start at $n = 1$ by considering again the one-loop
diagrams shown in Figs.\ 8(a)--(f). Having isolated the PT self-energy
component of the diagrams in Fig.~8, we can go on to isolate the 
one-loop vertex-like component on the r.h.s., defined from the
remaining part of the corresponding Feynman integrands 
proportional to the product of tree level functions
$\Gamma_{\mu'\alpha'\beta'}^{ma'b'}(-q,p_{1}',p_{2}')\,iD_{\mu'\mu}(q)$
on the l.h.s. Thus, we can write
\bea
\mbox{Figs.\ 8(a)--(f)}
&=&
{\textstyle \frac{1}{2}}g^{4}
\mu^{2\epsilon}\int\frac{d^{d}k}{(2\pi)^{d}}
\,G_{\rho'\sigma'\alpha'\beta'}^{rsa'b'}
\,D_{\rho'\rho}(k_{1})\,D_{\sigma'\sigma}(k_{2})
\,G_{\rho\sigma\alpha\beta}^{rsab} \label{GDDG2} \\
&=&
g\Gamma_{\mu'\alpha'\beta'}^{ma'b'}\,iD_{\mu'\mu}(q) 
\Bigl\{ i\hat{\Pi}_{\mu\nu}(q)\,iD_{\nu\nu'}(q)
\,g\Gamma_{\nu'\alpha\beta}^{mab}
\,+\, g\hat{\Gamma}_{\mu\alpha\beta}^{mab}
\,-\, g\Delta\hat{\Gamma}_{\mu\alpha\beta}^{mab} \Bigr\} 
+ \ldots \nn \\
\label{GDDG3}
\eea
In \eq{GDDG3}, the one-loop 
vertex-like component has been written as the difference
between the full PT gauge-independent one-loop triple gluon vertex
$\hat{\Gamma}_{\mu\alpha\beta}^{mab}(q,p_{1},p_{2})$ \cite{cornwall4}
and a remaining (gauge-dependent) contribution
$\Delta\hat{\Gamma}_{\mu\alpha\beta}^{mab}(q;p_{1},p_{2})$. 
This decomposition is illustrated in Fig.~10, where the blob
marked ``R'' represents 
$-\Delta\hat{\Gamma}_{\mu\alpha\beta}^{mab}(q;p_{1},p_{2})$. 
Just as in the case of $i\hat{\Pi}_{\mu\nu}^{mn}(q)$, the isolation
of this ``vertex-like'' component involves using the Ward 

\begin{center}
\begin{picture}(400,595)(-45,0)


\put(-20,570){\makebox(0,0)[c]{\large +}}
\put(-10,570){\makebox(0,0)[c]{\large $\frac{1}{4}$}}
\Photon(  0,550)(  0,590){2}{4}
\Photon( 80,550)( 80,590){2}{4}
\Photon(  0,570)(  8,570){2}{1}
\PhotonArc(20,570)(10,34,394){2}{8}
\Photon( 32,570)( 48,570){2}{2}
\PhotonArc(60,570)(10,34,394){2}{8}
\Photon( 72,570)( 80,570){2}{1}

\put(-20,520){\makebox(0,0)[c]{\large +}}
\put(-10,520){\makebox(0,0)[c]{\large $\frac{1}{2}$}}
\Photon(  0,500)(  0,540){2}{4}
\Photon( 80,500)( 80,540){2}{4}
\Photon(  0,520)(  8,520){2}{1}
\PhotonArc(20,520)(10,34,394){2}{8}
\Photon( 32,520)( 50,520){2}{2}
\Photon( 50,520)( 80,530){2}{3.5}
\Photon( 50,520)( 80,510){-2}{3.5}
\put(40,500){\makebox(0,0)[c]{\scriptsize + reversed}}

\put(-20,470){\makebox(0,0)[c]{\large +}}
\put(-10,470){\makebox(0,0)[c]{\large $\frac{1}{4}$}}
\Photon(  0,450)(  0,490){2}{4}
\Photon( 80,450)( 80,490){2}{4}
\Photon(  0,470)(  8,470){2}{1}
\PhotonArc(20,470)(10,34,394){2}{8}
\Photon( 32,470)( 55,470){2}{2.5}
\PhotonArc( 67,470)(10,34,394){2}{8}
\put(40,450){\makebox(0,0)[c]{\scriptsize + reversed}}

\put(-20,420){\makebox(0,0)[c]{\large +}}
\put(-10,420){\makebox(0,0)[c]{\large $\frac{1}{2}$}}
\Photon(  0,400)(  0,440){2}{4}
\Photon( 80,400)( 80,440){2}{4}
\Photon(  0,420)( 20,420){2}{2}
\Photon( 20,420)( 40,430){2}{2.5}
\Photon( 20,420)( 40,410){-2}{2.5}
\Photon( 40,430)( 60,420){2}{2.5}
\Photon( 40,410)( 60,420){-2}{2.5}
\Photon( 40,410)( 40,430){-2}{2}
\Photon( 60,420)( 80,420){2}{2}

\put(-20,370){\makebox(0,0)[c]{\large +}}
\Photon(  0,350)(  0,390){2}{4}
\Photon( 80,350)( 80,390){2}{4}
\Photon(  0,370)( 20,370){2}{2}
\Photon( 20,370)( 40,380){2}{2.5}
\Photon( 20,370)( 40,360){-2}{2.5}
\Photon( 40,380)( 80,380){2}{4}
\Photon( 40,360)( 80,360){2}{4}
\Photon( 40,360)( 40,380){2}{2}
\put(40,350){\makebox(0,0)[c]{\scriptsize + reversed}}

\put(-20,320){\makebox(0,0)[c]{\large +}}
\put(-10,320){\makebox(0,0)[c]{\large $\frac{1}{2}$}}
\Photon(  0,300)(  0,340){2}{4}
\Photon( 80,300)( 80,340){2}{4}
\Photon(  0,320)( 20,320){2}{2}
\Photon( 20,320)( 40,330){2}{2.5}
\Photon( 20,320)( 40,310){-2}{2.5}
\Photon( 40,330)( 80,320){2}{4}
\Photon( 40,310)( 80,320){2}{4}
\Photon( 40,310)( 40,330){2}{2}
\put(40,300){\makebox(0,0)[c]{\scriptsize + reversed}}


\put(100,570){\makebox(0,0)[c]{\large +}}
\put(110,570){\makebox(0,0)[c]{\large $\frac{1}{4}$}}
\Photon(120,550)(120,590){2}{4}
\Photon(200,550)(200,590){2}{4}
\Photon(120,570)(136,570){2}{2}
\PhotonArc(148,570)(10,34,394){2}{8}
\PhotonArc(172,570)(10,34,394){2}{8}
\Photon(184,570)(200,570){2}{2}

\put(100,520){\makebox(0,0)[c]{\large +}}
\put(110,520){\makebox(0,0)[c]{\large $\frac{1}{2}$}}
\Photon(120,500)(120,540){2}{4}
\Photon(200,500)(200,540){2}{4}
\Photon(120,520)(138,520){2}{2}
\PhotonArc(150,520)(10,34,394){2}{8}
\Photon(162,520)(200,530){2}{4}
\Photon(162,520)(200,510){2}{4}
\put(160,500){\makebox(0,0)[c]{\scriptsize + reversed}}

\put(100,470){\makebox(0,0)[c]{\large +}}
\put(110,470){\makebox(0,0)[c]{\large $\frac{1}{4}$}}
\Photon(120,450)(120,490){2}{4}
\Photon(200,450)(200,490){2}{4}
\Photon(120,470)(151,470){2}{4}
\PhotonArc(163,470)(10,34,394){2}{8}
\PhotonArc(187,470)(10,34,394){2}{8}
\put(160,450){\makebox(0,0)[c]{\scriptsize + reversed}}

\put(100,420){\makebox(0,0)[c]{\large +}}
\Photon(120,400)(120,440){2}{4}
\Photon(200,400)(200,440){2}{4}
\Photon(120,430)(150,420){2}{3.5}
\Photon(120,410)(150,420){-2}{3.5}
\Photon(150,420)(170,420){2}{2}
\Photon(170,420)(200,430){2}{3.5}
\Photon(170,420)(200,410){-2}{3.5}

\put(100,370){\makebox(0,0)[c]{\large +}}
\put(110,370){\makebox(0,0)[c]{\large $\frac{1}{2}$}}
\Photon(120,350)(120,390){2}{4}
\Photon(200,350)(200,390){2}{4}
\Photon(120,380)(150,370){2}{3.5}
\Photon(120,360)(150,370){-2}{3.5}
\Photon(150,370)(175,370){2}{2.5}
\PhotonArc(187,370)(10,34,394){2}{8}
\put(160,350){\makebox(0,0)[c]{\scriptsize + reversed}}

\put(100,320){\makebox(0,0)[c]{\large +}}
\Photon(120,300)(120,340){2}{4}
\Photon(200,300)(200,340){2}{4}
\Photon(120,330)(200,330){2}{8}
\Photon(120,310)(200,310){2}{8}
\Photon(160,310)(160,330){2}{2}
\put(160,300){\makebox(0,0)[c]{\scriptsize + crossed}}


\put(220,570){\makebox(0,0)[c]{\large +}}
\Photon(240,550)(240,590){2}{4}
\Photon(320,550)(320,590){2}{4}
\Photon(240,580)(280,580){2}{4}
\Photon(240,560)(280,560){2}{4}
\Photon(280,560)(280,580){2}{2}
\Photon(280,580)(320,570){2}{4}
\Photon(280,560)(320,570){2}{4}
\put(280,550){\makebox(0,0)[c]{\scriptsize + reversed}}

\put(220,520){\makebox(0,0)[c]{\large +}}
\Photon(240,500)(240,540){2}{4}
\Photon(320,500)(320,540){2}{4}
\Photon(240,530)(320,510){2}{8}
\Photon(240,510)(320,530){2}{8}

\put(220,470){\makebox(0,0)[c]{\large +}}
\put(230,470){\makebox(0,0)[c]{\large $\frac{1}{2}$}}
\Photon(240,450)(240,490){2}{4}
\Photon(320,450)(320,490){2}{4}
\Photon(240,480)(280,470){2}{4}
\Photon(240,460)(280,470){-2}{4}
\Photon(280,470)(293.6,462.3){2}{2.5}
\Photon(280,470)(293.6,477.7){-2}{2.5}
\PhotonArc(300,470)(10, 50,130){2}{1.5}
\PhotonArc(300,470)(10,230,310){2}{1.5}
\Photon(306.4,477.7)(320,470){-2}{2}
\Photon(306.4,462.3)(320,470){2}{2}
\put(280,450){\makebox(0,0)[c]{\scriptsize + reversed}}

\put(220,420){\makebox(0,0)[c]{\large +}}
\put(230,420){\makebox(0,0)[c]{\large $\frac{1}{4}$}}
\Photon(240,400)(240,440){2}{4}
\Photon(320,400)(320,440){2}{4}
\PhotonArc(253,420)(10,34,394){2}{8}
\Photon(265,420)(295,420){2}{4}
\PhotonArc(307,420)(10,34,394){2}{8}

\put(220,370){\makebox(0,0)[c]{\large +}}
\put(230,370){\makebox(0,0)[c]{\large $\frac{1}{2}$}}
\Photon(240,350)(240,390){2}{4}
\Photon(320,350)(320,390){2}{4}
\Photon(240,370)(280,380){2}{4}
\Photon(240,370)(280,360){2}{4}
\Photon(280,360)(280,380){2}{2}
\Photon(280,380)(320,370){2}{4}
\Photon(280,360)(320,370){2}{4}

\put(220,320){\makebox(0,0)[c]{\large +}}
\put(230,320){\makebox(0,0)[c]{\large $\frac{1}{4}$}}
\Photon(240,300)(240,340){2}{4}
\Photon(320,300)(320,340){2}{4}
\Photon(240,320)(253.6,327.7){2}{2}
\Photon(240,320)(253.6,312.3){-2}{2}
\PhotonArc(260,320)(10, 50,130){2}{1.5}
\PhotonArc(260,320)(10,230,310){2}{1.5}
\Photon(266.4,327.7)(293.6,312.3){-2}{4}
\Photon(266.4,312.3)(293.6,327.7){2}{4}
\PhotonArc(300,320)(10, 50,130){2}{1.5}
\PhotonArc(300,320)(10,230,310){2}{1.5}
\Photon(306.4,327.7)(320,320){-2}{2}
\Photon(306.4,312.3)(320,320){2}{2}


\put(-50,145){\makebox(0,0)[c]{\LARGE =}}
\put(-30,145){\makebox(0,0)[c]{\LARGE $\frac{1}{4}$}}
\put(-10,145){\makebox(0,0)[c]{$ \left\{\begin{array}{c}
\\ \\ \\ \\ \\ \\ \\ \\ \\ \\ \\ \\ \\
\end{array} \right. $}}

\Photon( 10,200)( 10,240){2}{4}
\Photon( 10,220)( 18,220){2}{1}
\PhotonArc( 30,220)(10,90,-90){2}{4.5}

\Photon( 10,150)( 10,190){2}{4}
\Photon( 10,180)( 30,180){2}{2}
\Photon( 10,160)( 30,160){2}{2}

\Photon( 10,100)( 10,140){2}{4}
\Photon( 10,110)( 30,130){-2}{3}
\Photon( 10,130)(17,123){2}{1}
\Photon( 23,117)(30,110){2}{1}

\Photon( 10, 50)( 10, 90){2}{4}
\Photon( 10, 70)( 30, 80){2}{2}
\Photon( 10, 70)( 30, 60){2}{2}

\put( 50,145){\makebox(0,0)[c]{$ \left.\begin{array}{c}
\\ \\ \\ \\ \\ \\ \\ \\ \\ \\ \\ \\ \\
\end{array} \right\} $}}
\Photon( 80,155)(100,155){2}{2}
\Photon( 80,135)(100,135){2}{2}
\put(130,145){\makebox(0,0)[c]{$ \left\{\begin{array}{c}
\\ \\ \\ \\ \\ \\ \\ \\ \\ \\ \\ \\ \\
\end{array} \right. $}}

\PhotonArc(145,220)(10,-90,90){2}{4.5}
\Photon(157,220)(163,220){2}{1}
\PhotonArc(175,220)(10,90,-90){2}{4.5}

\Photon(145,180)(175,180){2}{3}
\Photon(145,160)(175,160){2}{3}
\Photon(160,160)(160,180){-2}{2}

\Photon(145,130)(160,130){2}{1.5}
\Photon(145,110)(160,110){2}{1.5}
\Photon(160,110)(160,130){-2}{2}
\Photon(160,110)(175,130){-2}{2}
\Photon(160,130)(165,123){ 2}{1}
\Photon(170,117)(175,110){2}{1}

\Photon(145, 60)(175, 80){2}{4}
\Photon(145, 80)(175, 60){2}{4}

\put(190,145){\makebox(0,0)[c]{$ \left.\begin{array}{c}
\\ \\ \\ \\ \\ \\ \\ \\ \\ \\ \\ \\ \\
\end{array} \right\} $}}
\Photon(220,155)(240,155){2}{2}
\Photon(220,135)(240,135){2}{2}
\put(270,145){\makebox(0,0)[c]{$ \left\{\begin{array}{c}
\\ \\ \\ \\ \\ \\ \\ \\ \\ \\ \\ \\ \\
\end{array} \right. $}}

\Photon(310,200)(310,240){2}{4}
\PhotonArc(290,220)(10,-90,90){2}{4.5}
\Photon(302,220)(310,220){2}{1}

\Photon(310,150)(310,190){2}{4}
\Photon(310,180)(290,180){2}{2}
\Photon(310,160)(290,160){2}{2}

\Photon(310,100)(310,140){2}{4}
\Photon(290,110)(310,130){-2}{3}
\Photon(290,130)(297,123){2}{1}
\Photon(303,117)(310,110){2}{1}

\Photon(310, 50)(310, 90){2}{4}
\Photon(290, 80)(310, 70){2}{2}
\Photon(290, 60)(310, 70){2}{2}

\put(330,145){\makebox(0,0)[c]{$ \left.\begin{array}{c}
\\ \\ \\ \\ \\ \\ \\ \\ \\ \\ \\ \\ \\
\end{array} \right\} $}}

\put(160,0){\makebox(0,0)[c]{\footnotesize
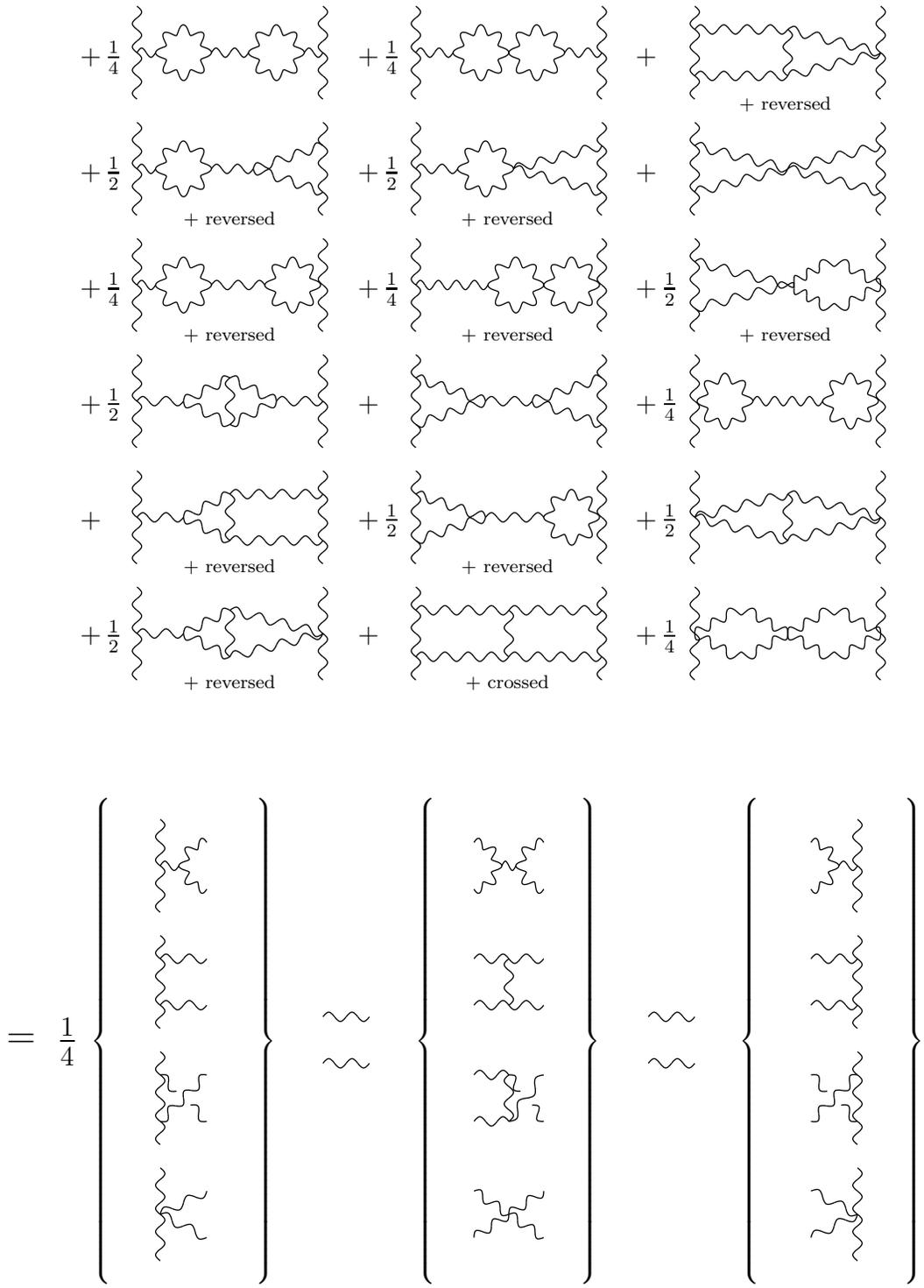
Fig.~9. The twenty-eight two-loop diagrams formed
from the ``product'' of three
gluon four-point functions.}}

\end{picture}
\end{center}

\begin{center}
\begin{picture}(400,250)(-50,10)


\put(-30,145){\makebox(0,0)[c]{\LARGE $\frac{1}{2}$}}
\put(-10,145){\makebox(0,0)[c]{$ \left\{\begin{array}{c}
\\ \\ \\ \\ \\ \\ \\ \\ \\ \\ \\ \\ \\
\end{array} \right. $}}

\Photon( 10,200)( 10,240){2}{4}
\Photon( 10,220)( 18,220){2}{1}
\PhotonArc( 30,220)(10,90,-90){2}{4.5}

\Photon( 10,150)( 10,190){2}{4}
\Photon( 10,180)( 30,180){2}{2}
\Photon( 10,160)( 30,160){2}{2}

\Photon( 10,100)( 10,140){2}{4}
\Photon( 10,110)( 30,130){-2}{3}
\Photon( 10,130)(17,123){2}{1}
\Photon( 23,117)(30,110){2}{1}

\Photon( 10, 50)( 10, 90){2}{4}
\Photon( 10, 70)( 30, 80){2}{2}
\Photon( 10, 70)( 30, 60){2}{2}

\put( 50,145){\makebox(0,0)[c]{$ \left.\begin{array}{c}
\\ \\ \\ \\ \\ \\ \\ \\ \\ \\ \\ \\ \\
\end{array} \right\} $}}
\Photon( 80,155)(100,155){2}{2}
\Photon( 80,135)(100,135){2}{2}
\put(130,145){\makebox(0,0)[c]{$ \left\{\begin{array}{c}
\\ \\ \\ \\ \\ \\ \\ \\ \\ \\ \\ \\ \\
\end{array} \right. $}}

\Photon(170,200)(170,240){2}{4}
\PhotonArc(150,220)(10,-90,90){2}{4.5}
\Photon(162,220)(170,220){2}{1}

\Photon(170,150)(170,190){2}{4}
\Photon(170,180)(150,180){2}{2}
\Photon(170,160)(150,160){2}{2}

\Photon(170,100)(170,140){2}{4}
\Photon(150,110)(170,130){-2}{3}
\Photon(150,130)(157,123){2}{1}
\Photon(163,117)(170,110){2}{1}

\Photon(170, 50)(170, 90){2}{4}
\Photon(150, 80)(170, 70){2}{2}
\Photon(150, 60)(170, 70){2}{2}

\put(190,145){\makebox(0,0)[c]{$ \left.\begin{array}{c}
\\ \\ \\ \\ \\ \\ \\ \\ \\ \\ \\ \\ \\
\end{array} \right\} $}}
\put(220,145){\makebox(0,0)[c]{\LARGE $=$}}
\put(250,145){\makebox(0,0)[c]{$ \left\{\begin{array}{c}
\\ \\ \\ \\ \\ \\ \\ \\ \\ \\ \\ \\ \\
\end{array} \right. $}}

\Photon(270,200)(270,240){2}{4}
\Photon(270,220)(290,220){2}{2}
\GCirc(300,220){10}{0.75}
\put(301, 221){\makebox(0,0)[c]{PT}}
\Photon(310,220)(330,220){2}{2}
\Photon(330,200)(330,240){2}{4}

\Photon(270,150)(270,190){2}{4}
\Photon(270,170)(320,170){2}{5}
\GCirc(330,170){10}{0.75}
\put(331, 171){\makebox(0,0)[c]{PT}}
\Photon(330,150)(330,160){2}{1}
\Photon(330,180)(330,190){2}{1}

\Photon(270,100)(270,140){2}{4}
\Photon(270,120)(320,120){2}{5}
\GCirc(330,120){10}{0.75}
\put(331, 121){\makebox(0,0)[c]{R}}
\Photon(330,100)(330,110){2}{1}
\Photon(330,130)(330,140){2}{1}

\put(300, 70){\makebox(0,0)[c]{\huge $\vdots$}}

\put(150,10){\makebox(0,0)[c]{\footnotesize 
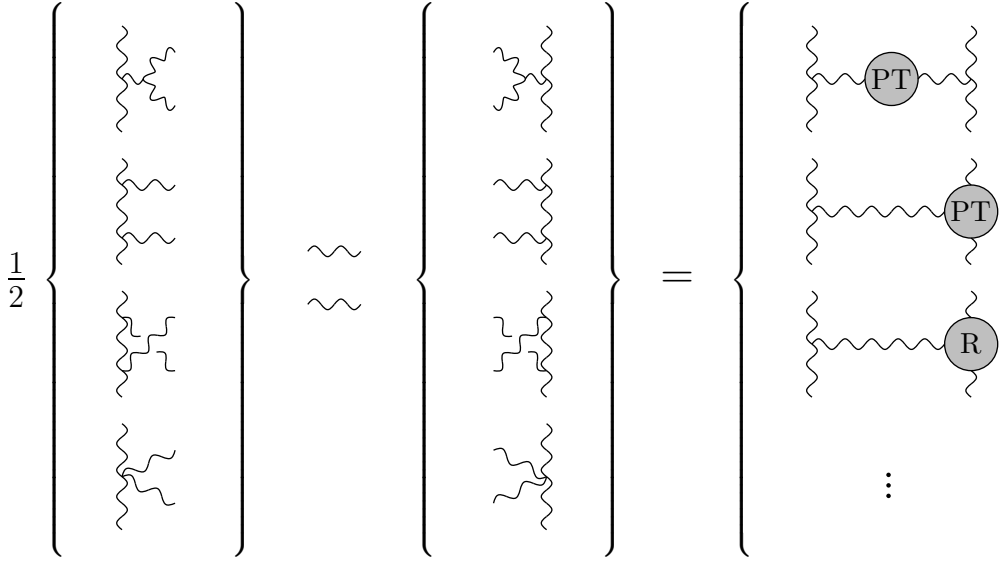
Fig.~10. The PT rearrangement \eq{GDDG3} of the conventional 
perturbation theory diagrams in Fig.~8. }}

\end{picture}
\end{center}

\noindent
identity \eq{ggggwid}
for {\em all} of the longitudinal factors which occur in order
to identify the terms in the integrand of \eq{GDDG2} proportional to
$\Gamma_{\mu'\alpha'\beta'}^{ma'b'}(-q,p_{1}',p_{2}')iD_{\mu'\mu}(q)$.
The calculation is significantly more involved than that
for $i\hat{\Pi}_{\mu\nu}^{mn}(q)$.

For the moment, we consider just the PT gauge-independent
one-loop triple gluon vertex $\hat{\Gamma}_{\mu\alpha\beta}^{mab}$
in \eq{GDDG3}. 
The crucial point is that this PT one-loop three-point function
satisfies the same Ward identity involving
the difference of two PT one-loop two-point functions as the
corresponding tree level quantities \cite{cornwall4}:
\be
q_{1\alpha}\hat{\Gamma}_{\alpha\beta\gamma}^{abc}(q_{1},q_{2},q_{3})
=
f^{abc}\Bigl( 
\hat{\Pi}_{\beta\gamma}(q_{2}) - \hat{\Pi}_{\beta\gamma}(q_{3})
\Bigr).
\label{PTwid}
\ee
We can form the one-loop improper three-point function
$\hat{G}_{\alpha\beta\gamma}^{abc}(q_{1},q_{2},q_{3})$,
consisting of the PT one-loop proper three-point function together
with the PT one-loop self-energy corrections to the three
external legs $A_{\alpha}^{a}(q_{1})$, 
$A_{\beta}^{b}(q_{2})$, $A_{\gamma}^{c}(q_{3})$:
\bea
-g\hat{G}_{\alpha\beta\gamma}^{abc}(q_{1},q_{2},q_{3})
&=&
+g\hat{\Gamma}_{\alpha\beta\gamma}^{abc}(q_{1},q_{2},q_{3}) \nn \\
& &
+g\Gamma_{\rho'\beta\gamma}^{rbc}(q_{1},q_{2},q_{3})
\,iD_{\rho'\rho}(q_{1})\,i\hat{\Pi}_{\rho\alpha}^{ra}(q_{1}) \nn \\
& &
+g\Gamma_{\alpha\rho'\gamma}^{arc}(q_{1},q_{2},q_{3})
\,iD_{\rho'\rho}(q_{2})\,i\hat{\Pi}_{\rho\beta}^{rb}(q_{2}) \nn \\
& &
+g\Gamma_{\alpha\beta\rho'}^{abr}(q_{1},q_{2},q_{3}) 
\,iD_{\rho'\rho}(q_{3})\,i\hat{\Pi}_{\rho\gamma}^{rc}(q_{3}).
\label{gggoneloop}
\eea
This improper three-point function then obeys the Ward identity
\bea
q_{1\alpha}\hat{G}_{\alpha\beta\gamma}^{abc}(q_{1},q_{2},q_{3})
&=&
-f^{rab}q_{2}^{2}t_{\beta\rho'}(q_{2})D_{\rho'\rho}(q_{3})
\,\hat{\Pi}_{\rho\gamma}^{rc}(q_{3})  \nonumber \\
& &
-f^{rac}q_{3}^{2}t_{\gamma\rho'}(q_{3})D_{\rho'\rho}(q_{2})
\,\hat{\Pi}_{\rho\beta}^{rb}(q_{2}).
\label{gggwid}
\eea

\begin{center}
\begin{picture}(400,210)(-50,50)


\Photon(-40,220)(-20,220){3}{2}
\GCirc( -7.5,220){12.5}{0.75}
\put( -7.5, 221){\makebox(0,0)[c]{PT}}
\Photon(  5,220)(25,220){3}{2}
\Photon( 25,220)(45,250){3}{3.5}
\Photon( 25,220)(45,190){3}{3.5}
\put(75,250){\makebox(0,0)[l]{$p_{2\beta}$}}
\put(70,250){\vector(-1, 0){10}}
\put(75,190){\makebox(0,0)[l]{$p_{1\alpha}$}}
\put(70,190){\vector(-1, 0){10}}

\Photon(-40,120)(-10,120){3}{3}
\Photon( 2.5,120)(45,150){3}{4.5}
\Photon( 2.5,120)(45, 90){-3}{4.5}
\GCirc(  2.5,120){12.5}{0.75}
\put(  2.5, 121){\makebox(0,0)[c]{PT}}
\put(75,150){\makebox(0,0)[l]{$p_{2\beta}$}}
\put(70,150){\vector(-1, 0){10}}
\put(75, 90){\makebox(0,0)[l]{$p_{1\alpha}$}}
\put(70, 90){\vector(-1, 0){10}}


\put(150,235){\makebox(0,0)[c]{\small pinch}}
\put(135,220){\vector(1, 0){30}}
\put(150,135){\makebox(0,0)[c]{\small pinch}}
\put(135,120){\vector(1, 0){30}}


\put(200,220){\makebox(0,0)[c]{\Large $+$}}
\Photon(215,220)(235,220){3}{2}
\GCirc(247.5,220){12.5}{0.75}
\put(247.5, 221){\makebox(0,0)[c]{PT}}
\Photon(261,220)(300,250){3}{4.5}
\Photon(261,220)(300,190){-3}{4.5}
\Vertex(260,220){2}
\put(315,220){\makebox(0,0)[l]{\Large $+ \cdots$}}

\put(200,120){\makebox(0,0)[c]{\Large $-$}}
\Photon(215,120)(235,120){3}{2}
\GCirc(247.5,120){12.5}{0.75}
\put(247.5, 121){\makebox(0,0)[c]{PT}}
\Photon(261,120)(300,150){3}{4.5}
\Photon(261,120)(300, 90){-3}{4.5}
\Vertex(260,120){2}
\put(315,120){\makebox(0,0)[l]{\Large $+ \cdots$}}

\put(150,50){\makebox(0,0)[c]{\footnotesize 
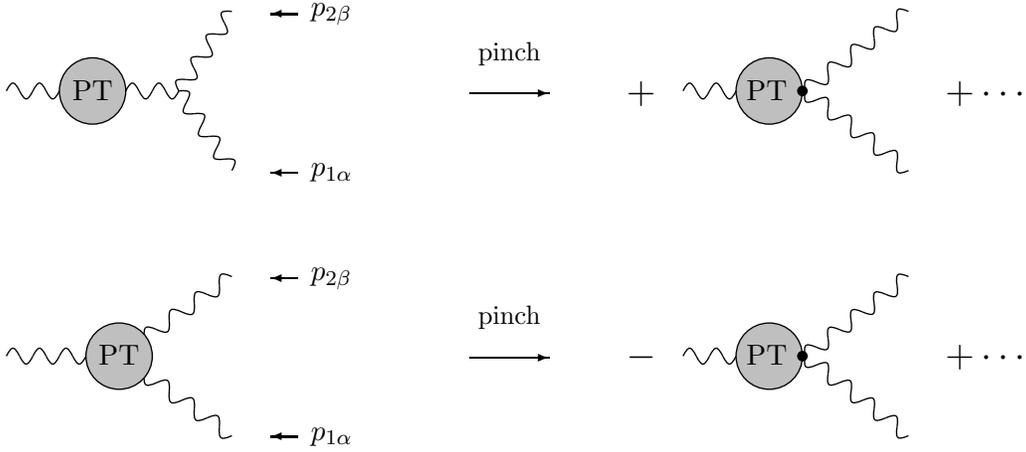
Fig.~11. The fundamental PT cancellation, expressed 
at one loop in the identities
Eqs.~(\ref{gggpartp}) and (\ref{gggpartpp}).}}

\end{picture}
\end{center}

\noindent
This Ward identity is precisely the one-loop 
analogue of those Eqs.~(\ref{ggffwid})-(\ref{gggggwid}) for
the tree level four- and five-point functions
$G_{\alpha\beta ij}^{ab(f)}$,
$G_{\alpha\beta\gamma\delta}^{abcd}$ and 
$G_{\alpha\beta\gamma\delta\epsilon}^{abcde}$.

In \eq{GDDG3}, there do not appear all four components
of this improper one-loop three-point function
Eq.~(\ref{gggoneloop}), but just one PT self-energy
correction $i\hat{\Pi}_{\mu\nu}^{mn}(q)$
together with the PT vertex
$\hat{\Gamma}_{\mu\alpha\beta}^{mab}(q,p_{1},p_{2})$. 
These two components are however {\em just sufficient} 
to give the part of the Ward identity \eq{gggwid} required for the PT
cancellation mechanism to occur when 
the two components are contracted with 
longitudinal factors $p_{1\alpha}$, $p_{2\beta}$ associated with
the external gluons $A_{\alpha}^{a}(p_{1})$, $A_{\beta}^{b}(p_{2})$: 
for the linear factor $p_{1\alpha}$,
\bea
\lefteqn{\Bigl(i\hat{\Pi}_{\mu\nu}(q)\,iD_{\nu\nu'}(q)
\,g\Gamma_{\nu'\alpha\beta}^{mab}(q,p_{1},p_{2})
\,+\, g\hat{\Gamma}_{\mu\alpha\beta}^{mab}(q,p_{1},p_{2}) \Bigr)
\,p_{1\alpha}
=}\nonumber \\
& &
\hspace{100pt}
-gf^{mab}  
\Bigl(i\hat{\Pi}_{\mu\nu}(q)\,iD_{\nu\nu'}(q)\,p_{2}^{2}t_{\nu'\beta}(p_{2})
\,-\, \hat{\Pi}_{\mu\beta}(p_{2}) \Bigr)
\label{gggpartp}
\eea
and similarly for $p_{2\beta}$, while for the quadratic
factor $p_{1\alpha}p_{2\beta}$,
\be
\Bigl(i\hat{\Pi}_{\mu\nu}(q)\,iD_{\nu\nu'}(q)
\,g\Gamma_{\nu'\alpha\beta}^{mab}(q,p_{1},p_{2})
\,+\, g\hat{\Gamma}_{\mu\alpha\beta}^{mab}(q,p_{1},p_{2}) \Bigr)
\,p_{1\alpha}p_{2\beta}
\,= \,
0.
\label{gggpartpp}
\ee
The term proportional to
$D_{\nu\nu'}(q)\,p^{2}t_{\nu'\beta}(p_{2})$
on the r.h.s.\ of the identity \eq{gggpartp} for a factor
$p_{1\alpha}$
is the exact analogue of those proportional to
$k_{2}^{2}t_{\sigma\nu'}(k_{2})D_{\nu'\nu}(q)$
on the r.h.s.\ of the Ward identities
Eqs.~(\ref{kggff})-(\ref{kggggg}) for a factor $k_{1\rho}$,
except for the PT one-loop ``effective'' two-point function
in place of the tree level three- and four-point functions.
The term proportional to
$\hat{\Pi}_{\mu\beta}(p_{2})$
on the r.h.s.\ of \eq{gggpartp}
is associated with the one-loop self-energy correction to the
external gluon $A_{\beta}^{b}(p_{2})$.
The PT cancellation encoded in Eqs.~(\ref{gggpartp}) and (\ref{gggpartpp})
is illustrated schematically in Fig.~11 (cf.\ Fig.~7).

We now contract the external legs on the r.h.s.\ of 
Eq.\ (\ref{GDDG2}) via propagators
$iD_{\alpha\alpha''}(p_{1})$, $iD_{\beta\beta''}(p_{2})$
with the tree level gluon four-point function
$ig^{2}G_{\alpha''\beta''\gamma\delta}^{abcd}(p_{1},p_{2},p_{3},p_{4})$ 
to form 

\begin{center}
\begin{picture}(400,250)(-50,0)


\put(-60,145){\makebox(0,0)[c]{\LARGE $\frac{1}{2}$}}
\put(-40,145){\makebox(0,0)[c]{$ \left\{\begin{array}{c}
\\ \\ \\ \\ \\ \\ \\ \\ \\ \\ \\ \\ \\
\end{array} \right. $}}

\Photon(-20,200)(-20,240){2}{4}
\Photon(-20,220)(-10,220){2}{1}
\GCirc(  0,220){10}{0.75}
\put(  1, 221){\makebox(0,0)[c]{PT}}
\Photon( 10,220)( 18,220){2}{1}
\PhotonArc( 30,220)(10,90,-90){2}{4.5}

\Photon(-20,150)(-20,190){2}{4}
\Photon(-20,170)(  0,170){2}{2}
\Photon( 10,170)( 30,180){2}{2.5}
\Photon( 10,170)( 30,160){-2}{2.5}
\GCirc(  10,170){10}{0.75}
\put( 11, 171){\makebox(0,0)[c]{PT}}

\put(0,120){\makebox(0,0)[c]{\huge $\vdots$}}

\put(0, 70){\makebox(0,0)[c]{\huge $\vdots$}}

\put( 50,145){\makebox(0,0)[c]{$ \left.\begin{array}{c}
\\ \\ \\ \\ \\ \\ \\ \\ \\ \\ \\ \\ \\
\end{array} \right\} $}}
\Photon( 80,155)(100,155){2}{2}
\Photon( 80,135)(100,135){2}{2}
\put(130,145){\makebox(0,0)[c]{$ \left\{\begin{array}{c}
\\ \\ \\ \\ \\ \\ \\ \\ \\ \\ \\ \\ \\
\end{array} \right. $}}

\PhotonArc(145,220)(10,-90,90){2}{4.5}
\Photon(157,220)(163,220){2}{1}
\PhotonArc(175,220)(10,90,-90){2}{4.5}

\Photon(145,180)(175,180){2}{3}
\Photon(145,160)(175,160){2}{3}
\Photon(160,160)(160,180){2}{2}

\Photon(145,130)(160,130){2}{1.5}
\Photon(145,110)(160,110){2}{1.5}
\Photon(160,110)(160,130){-2}{2}
\Photon(160,110)(175,130){-2}{2}
\Photon(160,130)(165,123){2}{1}
\Photon(170,117)(175,110){2}{1}

\Photon(145, 80)(175, 60){2}{4}
\Photon(145, 60)(175, 80){2}{4}

\put(190,145){\makebox(0,0)[c]{$ \left.\begin{array}{c}
\\ \\ \\ \\ \\ \\ \\ \\ \\ \\ \\ \\ \\
\end{array} \right\} $}}
\put(220,145){\makebox(0,0)[c]{\LARGE $=$}}
\put(250,145){\makebox(0,0)[c]{$ \left\{\begin{array}{c}
\\ \\ \\ \\ \\ \\ \\ \\ \\ \\ \\ \\ \\
\end{array} \right. $}}

\Photon(270,200)(270,240){2}{4}
\Photon(270,220)(280,220){2}{1}
\GCirc(290,220){10}{0.75}
\put(291, 221){\makebox(0,0)[c]{PT}}
\Photon(300,220)(310,220){2}{1}
\GCirc(320,220){10}{0.75}
\put(321, 221){\makebox(0,0)[c]{PT}}
\Photon(330,220)(338,220){2}{1}
\PhotonArc(350,220)(10,90,-90){2}{4.5}

\Photon(270,150)(270,190){2}{4}
\Photon(270,170)(280,170){2}{1}
\GCirc(290,170){10}{0.75}
\put(291, 171){\makebox(0,0)[c]{PT}}
\Photon(300,170)(320,170){2}{2}
\Photon(330,170)(350,180){2}{2.5}
\Photon(330,170)(350,160){-2}{2.5}
\GCirc(330,170){10}{0.75}
\put(331, 171){\makebox(0,0)[c]{PT}}

\Photon(270,100)(270,140){2}{4}
\Photon(270,120)(280,120){2}{1}
\GCirc(290,120){10}{0.75}
\put(291, 121){\makebox(0,0)[c]{PT}}
\Photon(300,120)(320,120){2}{2}
\Photon(330,120)(350,130){2}{2.5}
\Photon(330,120)(350,110){-2}{2.5}
\GCirc(330,120){10}{0.75}
\put(331,121){\makebox(0,0)[c]{R}}

\put(315, 70){\makebox(0,0)[c]{\huge $\vdots$}}

\put(150,0){\makebox(0,0)[c]{\footnotesize 
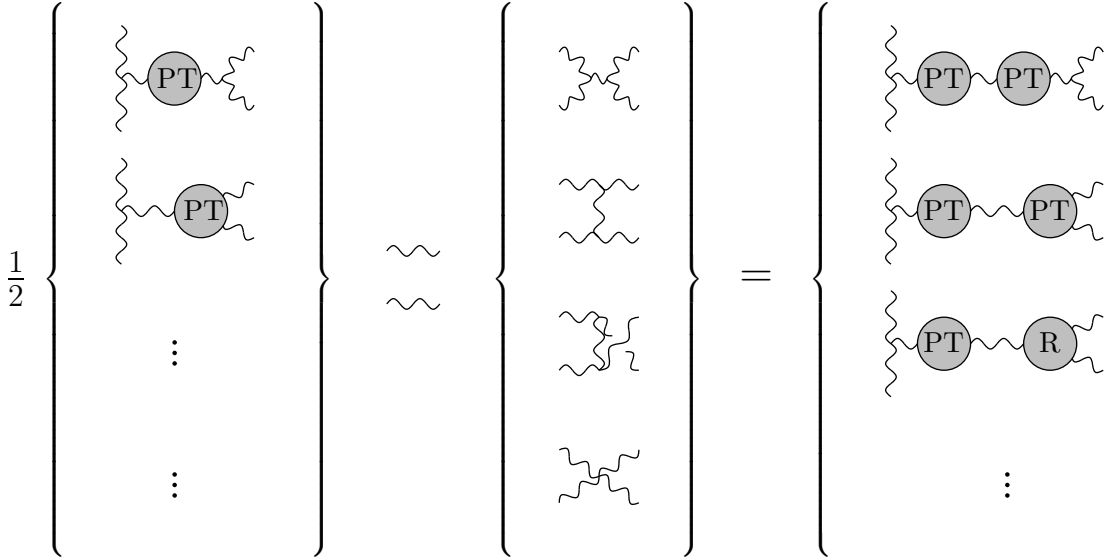
Fig.~12. The iteration of the PT one-loop cancellation mechanism.}}

\end{picture}
\end{center}

\noindent
the two-loop diagrams shown in Fig.~9.
Then, as a result of the identities 
Eqs.~({\ref{gggpartp}) and (\ref{gggpartpp}),
{\em exactly the same} PT cancellation mechanism
occurs among the diagrams of the second loop
associated with the terms 
$i\hat{\Pi}_{\mu\nu'}(q)\,iD_{\nu'\nu}(q)
\,g\Gamma_{\nu\alpha\beta}^{mab}(q,p_{1},p_{2})$ and
$g\hat{\Gamma}_{\mu\alpha\beta}^{mab}(q,p_{1},p_{2})$
in \eq{GDDG3} as occurred among the diagrams of the first loop.
We may thus follow exactly the same procedure as 
described in Secs.~4.2 and 4.3 and App.~A\footnote{
In contrast to the case at one loop, we make no attempt
in the Dyson summation to cancel explicitly the tadpole-like
terms, proportional to $\int d^{d}k\,(n\cdot k)^{-2}k^{-2}$,
generated by the $b$ terms in the gluon propagators (cf.\ \eq{b2}):
these $q$-independent factors can arise from additional
one-loop corrections to {\em any} type of $n$-loop
diagram, so that it is meaningless to attempt to
cancel them in only a subclass of diagrams.}
to isolate explicitly in the class of gauges $n\sdot A^{a} = 0$
the components of the diagrams in Fig.~9 consisting of
i) the chain of two PT gauge-independent one-loop self-energies
and ii) a single PT self-energy attached to a one-loop vertex-like
function on the r.h.s., exactly similar to that in \eq{GDDG3}. 
This procedure is illustrated in Fig.~12.

There remains the term proportional to 
$\Delta\hat{\Gamma}_{\mu\alpha\beta}^{mab}$
in \eq{GDDG3}. The explicit expression for this 
(gauge-dependent) function is given in App.~B. 
It is shown there that the internal propagator structure of 
$\Delta\hat{\Gamma}_{\mu\alpha\beta}^{mab}$ is such that 
it can never contribute to the Dyson chain of two PT one-loop
self-energies. Instead, it contributes to the two-loop 
one-particle-irreducible self-energy and vertex functions, not
considered here.

We have therefore succeeded in showing explicitly how
the fundamenal PT gauge cancellation mechanism which operates
at one loop occurs also at two loops, enabling the Dyson
chain of two PT one-loop self-energies to be isolated
explicitly in the class of non-covariant gauges $n\sdot A^{a} = 0$. 
Clearly, this process can be iterated to all orders in the amplitude
(\ref{nloops}). Furthermore, as a result of the universality
of $i\hat{\Pi}_{\mu\nu}^{mn}(q)$, together with the fact that
the fermion components of the one-loop gluon two- and three-point
functions also obey the Ward identity Eq.\ (\ref{PTwid}),
exactly the same
process occurs when fermion loops are included in
the amplitude (\ref{nloops}), i.e.\ so that $G$ represents
either the gluon four-point function \eq{gggg}
as above or the gluon pair--fermion pair four-point
function \eq{ggff}, and $D$ represents the
tree level gluon or fermion propagator as appropriate.
Also, this iterative process is clearly independent 
of the particular four-  and five-point functions
Eqs.~(\ref{ggff})-(\ref{ggggg}) which occur as
the 1st and the $(n\!+\!1)$th  $n$-point functions
at each end of (\ref{nloops}).

We conclude that, in the class of non-covariant gauges
$n\sdot A^{a} = 0$, the PT gauge-independent
one-loop self-energy $i\hat{\Pi}_{\mu\nu}^{mn}(q)$
can be summed {\em explicitly} in a Dyson series.
This infinite, gauge-independent subset of radiative
corrections to the tree level gluon propagator
may therefore be fully accounted for by the 
gauge-, scale- and scheme-independent effective
charge \eq{qcdgeff} at the tree level vertices at
each end of this Dyson series.


\setcounter{equation}{0}
\def\theequation{5.\arabic{equation}}

\vspace{10pt}

{\Large\bf 5. Summary and Conclusions}

\vspace{5pt}

\noindent
In QED, the concept of an effective charge provides an
extremely simple way of accounting for a well-defined, infinite,
gauge-independent subset of radiative corrections
to interactions mediated at tree level by a single photon.
In this paper, it has been shown how, despite appearances to the
contrary (Sec.\ 1), the QED concept of a
gauge-, scale- and scheme-independent 
effective charge may be extended directly 
at the diagrammatic level to QCD.

It was first argued (Sec.\ 2)
that the basic concept of an effective charge 
depends not on the conventionally-defined gauge boson
two-point function per se, but rather
on the existence of a universal, gauge-independent
subset of radiative corrections 
which couple to the various sets of fields from $\Lclint$
at two points $x_{1}$, $x_{2}$
in precisely the same way as the gauge boson
which mediates their interaction at tree level.
Thus, we were led to the idea of the Dyson-summed
gauge boson ``effective'' two-point function
$i\hat{\Delta}_{R\mu\nu}^{mn}(x_{1}-x_{2})$. 
In QED, this 
function is identical to the conventional two-point
function Eq.\ (\ref{qedprop}) only because the theory is abelian.
But in QCD, the non-abelian symmetry of the theory
results in contributions to 
$i\hat{\Delta}_{R\mu\nu}^{mn}$
not only from the conventionally-defined two-point function, 
but also from conventionally-defined vertex and box functions.
The existence of such contributions is the fundamental
observation upon which the
pinch technique (PT) of Cornwall and Papavassiliou is based: 
at the strictly one-loop level 
(i.e.\ without any Dyson summation) and for on-shell 
external fields from $\Lclint$, this ``effective'' 
two-point function 
is precisely the gauge-independent self-energy
$i\hat{\Pi}_{R\mu\nu}^{mn}$ 
obtained in the PT
rearrangement of one-loop contributions to S-matrix elements.

However, in QED the photon self-energy is gauge-independent
regardless of whether the fields to which the photon
couples are on- or off-shell. Furthermore,
for the effective charge to obey at high energies
the constraints of the renormalization group,
$i\hat{\Delta}_{R\mu\nu}^{mn}$ must involve the
infinite Dyson series in
$i\hat{\Pi}_{R\mu\nu}^{mn}$.
Thus, we were led (Sec.\ 4) to extend the PT
to the general case of explicitly off-shell processes,
independent of any reference to S-matrix elements or any other
a priori gauge-independent quantity.
It was shown explicitly how
the PT self-energy constitutes
a well-defined ``effective'' two-point component
of one-loop interactions which
remains gauge-independent and universal, independent
of whether the ``external'' fields are on- or off-shell.
This demonstration was carried out in both the class of linear
covariant gauges and the class of non-covariant gauges $n\cdot A^{a} = 0$.
The simultaneous gauge indendence and universality
of the PT self-energy was shown to be due 
to a fundamental cancellation among contributions 
to the ``effective'' two-point function
from the conventional gauge-dependent
self-energy  and the pinch parts of conventional
gauge-dependent vertex and box diagrams,
leaving always the gauge-independent and universal
PT self-energy
$i\hat{\Pi}_{R\mu\nu}^{mn}$.
This cancellation
is encoded in the tree level Ward identities of the theory,
and is entirely independent both of the species of the ``external''
fields in the process and of whether they are on- or off-shell. 
In the usual arrangement of perturbation theory,
this cancellation is obscured
in the complicated gauge-dependence of conventional
one-loop diagrams, and typically 
only emerges ``miraculously'' at the end of a calculation.
However, by writing subsets of one-loop diagrams in terms
of the ``products'' of tree level four- and five-point
functions and exploiting directly the Ward identities of these functions, 
we were able to make these cancellations simply
and immediately, thereby isolating,
in an arbitrary gauge and with a minimum of effort, the one-loop
gauge-independent ``effective'' two-point component 
from up to several hundred diagrams at once.

Furthermore, it was shown explicitly how the PT one-loop
``effective'' two-point function sums in a Dyson series.
In the absence of an all-orders formulation of the PT
(and only for this reason), we were restricted to the class
of non-covariant gauges in order to avoid ghosts.
Within this class of gauges, it was then shown how the
one-loop cancellation mechanism iterates
for the subclasses of $n$-loop diagrams containing implicitly
the chains of $n$ PT one-loop self-energies.
In this way, we were able
to isolate explicitly the chains of $n$ PT
self-energies required to form the Dyson series, and to
prove their gauge-independence.

We therefore conclude that the two questions posed in the introduction 
in the context of renormalon calculus are answered as follows:
\begin{enumerate}
\item The gauge-independent combination of
fields one is summing to obtain the 
QCD effective charge with the full one-loop $\beta$-function
coefficient $\beta_{1} = -\frac{11}{3}N + \frac{2}{3}n_{\!f}$
as the coefficient of the logarithm is that
included in the PT Dyson-summed ``effective''
two-point function $\hat{\Delta}_{R\mu\nu}^{mn}(q)$
Eq.\ (\ref{qcdprop}). This function
is obtained from the subclass
of diagrams corresponding to the amplitude (\ref{nloops})
after effecting
the fundamental PT cancellation
among the conventionally-defined, gauge-dependent
self-energy, vertex and box diagrams occurring
in (\ref{nloops})
at $n$ loops, $n = 1,2\ldots\infty$.
\item The value of the constant term in the 
corresponding self-energy-like
function is that given in Eq.\ (\ref{ptPi}) for the
unrenormalized PT self-energy $\hat{\Pi}(q^{2})$.
Thus, in the $\MSbar$ scheme, the value is 
$(\frac{67}{9}N - \frac{10}{9}n_{\!f})(g^{2}/16\pi^{2})$.
\end{enumerate}

{}From the point of view of renormalon analyses, 
having identified explicitly the 
infinite, gauge-independent subclass 
of radiative corrections accounted
for by the QCD effective charge, the important
question is: in what approximation or limit, if any, 
do these contributions dominate over all other classes
of diagrams? In QED, the $1/n_{\!f}$ expansion provides
a well-defined framework in which the Dyson chains
of fermion loops represent the leading (non-trivial) 
contributions to, for example, the correlation function of
two fermion currents.
In QCD however, it appears that there is no analogous
parameter or limit, so that any dominance of 
the radiative corrections involved in the Dyson
chains accounted for by the QCD effective charge
would have to be dynamical in origin.

{}From a more general point of view, the work described here
shows that the PT algorithm is in fact more elegant 
than the usual statement that it consists in rearranging
one-loop S-matrix elements into 
components with distinct kinematical properties
which are then individually gauge-independent.
Rather, it amounts to the recognition of 
a fundamental cancellation mechanism among the
underlying perturbation theory diagrams,
independent of any embedding in S-matrix elements.
Here, we have been concerned almost exclusively with the
simplest $n$-point function, viz.\ the gluon self-energy.
However, the issues dealt with in the particular
context of an
effective charge for QCD---off-shell gauge independence,
universality, multi-loop diagrams---encourage the idea
that the PT may provide the basis for a 
well-defined and complete
reorganisation of perturbation theory, to all orders, in terms
of gauge-independent $n$-point functions, with
the only gauge dependence occurring in the longitudinal
part of the tree level gluon propagator.
In order, however, to put such an approach on a firm
field-theoretic footing, it would be necessary to 
go beyond the crude diagrammatics considered here and
formulate the PT directly
at the level of the path integral.

\vspace{10pt}

{\bf Acknowledgements}

\noindent
I wish to thank Joannis Papavassiliou and
Eduardo de Rafael for many useful discussions.
This work was supported by EC HCM grant ERB4001GT933989.

\pagebreak


\setcounter{equation}{0}
\def\theequation{A.\arabic{equation}}
\appendix

{\Large\bf Appendix A. The PT Gauge Cancellation Mechanism}

\vspace{5pt}

\noindent
In this appendix, we give details of the gauge cancellation mechanism
which ensures the simultaneous 
gauge-independence and universality of the one-loop gluon
self-energy $i\hat{\Pi}_{\mu\nu}^{mn}(q)$
defined in (\ref{ptPidef1})-(\ref{ptPidef3}) in the off-shell PT.
As explained in Secs.~4.2 and 4.3, the task is to show that,
regardless of the fact that the external fields in the given process are
off-shell, the additional
contributions to $i\hat{\Pi}_{\mu\nu}^{mn}(q)$ which occur
when one moves away from the Feynman gauge exactly cancel among themselves,
leaving always the result \eq{ptPi3}.
We consider both the class of linear covariant gauges
and the class of non-covariant gauges $n\sdot A^{a} = 0$,
as described in Sec.~3.
The correspnding terms $a$ and $b$ in the tree level gluon propagator
\eq{prop} are given in Eqs.~(\ref{abcov}) and (\ref{abaxial}).

We again consider first the amplitudes (\ref{GG}) for the one-loop diagrams
formed from the ``products'' of pairs of tree level four- and five-point
functions defined in Eqs.~(\ref{ggff})-(\ref{ggggg}).
The effect on the $n$-point functions on the r.h.s.\ of (\ref{GG})
of the longitudinal factors 
$k_{1\rho}$, $k_{2\sigma}$, $k_{1\rho'}$ and $k_{2\sigma'}$
which occur in the propagators
$iD_{\rho'\rho}(k_{1})$, $iD_{\sigma'\sigma}(k_{2})$
when one moves away from the Feynman gauge
is specified by the Ward identities Eqs.~(\ref{ggffwid})--(\ref{gggggwid}).

For the linear factor $k_{1\rho}$, we obtain
\bea
\lefteqn{k_{1\rho}
G_{\rho\sigma ij}^{rs(f)}(-k_{1},k_{2},p_{1},p_{2}) 
=}  \nonumber \\
& &
+f^{nrs}\Bigl(k_{2}^{2}t_{\sigma\nu'}(k_{2})D_{\nu'\nu}(q) 
\,+\, \tilde{G}_{\sigma}(q)\,q_{\nu}\Bigr)
\,i\gamma_{\nu}T_{ji}^{n}  \nonumber \\
& &
+i\gamma_{\sigma}T_{jk}^{s}\,S(p_{1}\!-\!k_{1},m_{\!f})\,
S^{-1}(p_{1},m_{\!f})\,iT_{ki}^{r}
\nonumber\\ 
& &
-iT_{jk}^{r}\,S^{-1}(p_{2},m_{\!f})\,
S(p_{2}\!+\!k_{1},m_{\!f})\,i\gamma_{\sigma}T_{ki}^{s}, \label{kggff} \\
\nonumber \\ 
\lefteqn{k_{1\rho}
G_{\rho\sigma\alpha\beta}^{rsab}(-k_{1},k_{2},p_{1},p_{2})
=}  \nonumber \\
& &
+f^{nrs}\Bigl(k_{2}^{2}t_{\sigma\nu'}(k_{2})D_{\nu'\nu}(q) 
\,+\, \tilde{G}_{\sigma}(q)\,q_{\nu}\Bigr)\,
\Gamma_{\nu\alpha\beta}^{nab}(q,p_{1},p_{2}) 
\nonumber \\
& &
+f^{nra}\Bigl(p_{1}^{2}t_{\alpha\nu'}(p_{1})D_{\nu'\nu}(p_{1}\!-\!k_{1}) 
\,+\, \tilde{G}_{\alpha}(p_{1}\!-\!k_{1})\,(p_{1}\!-\!k_{1})_{\nu} \Bigr)\,
\Gamma_{\nu\beta\sigma}^{nbs}(p_{1}\!-\!k_{1},p_{2},k_{2})
\nonumber \\
& &
+f^{nrb}\Bigl(p_{2}^{2}t_{\beta\nu'}(p_{2})D_{\nu'\nu}(p_{2}-k_{1}) 
\,+\, \tilde{G}_{\beta}(p_{2}\!-\!k_{1})\,(p_{2}\!-\!k_{1})_{\nu} \Bigr)\,
\Gamma_{\nu\sigma\alpha}^{nsa}(p_{2}\!-\!k_{1},k_{2},p_{1} ), \label{kgggg} \\
\nonumber \\ 
\lefteqn{k_{1\rho}
G_{\rho\sigma\alpha\beta\gamma}^{rsabc}(-k_{1},k_{2},p_{1},p_{2},p_{3})
=} \nonumber \\
& &
+f^{nrs}\Bigl(k_{2}^{2}t_{\sigma\nu'}(k_{2})D_{\nu'\nu}(q) 
\,+\, \tilde{G}_{\sigma}(q)\,q_{\nu}\Bigr)\,
G_{\nu\alpha\beta\gamma}^{nabc}(q,p_{1},p_{2},p_{3}) 
\nonumber \\
& &
+f^{nra}\Bigl(p_{1}^{2}t_{\alpha\nu'}(p_{1})D_{\nu'\nu}(p_{1}\!-\!k_{1}) 
\,+\, \tilde{G}_{\alpha}(p_{1}\!-\!k_{1})\,(p_{1}\!-\!k_{1})_{\nu} \Bigr)\,
G_{\nu\beta\gamma\sigma}^{nbcs}(p_{1}\!-\!k_{1},p_{2},p_{3},k_{2})
\nonumber \\
& &
+f^{nrb}\Bigl(p_{2}^{2}t_{\beta\nu'}(p_{2})D_{\nu'\nu}(p_{2}\!-\!k_{1}) 
\,+\, \tilde{G}_{\beta}(p_{2}\!-\!k_{1})\,(p_{2}\!-\!k_{1})_{\nu} \Bigr)\,
G_{\nu\gamma\sigma\alpha}^{ncsa}(p_{2}\!-\!k_{1},p_{3},k_{2},p_{1} )
\nonumber \\
& &
+f^{nrc}\Bigl(p_{3}^{3}t_{\gamma\nu'}(p_{3})D_{\nu'\nu}(p_{3}\!-\!k_{1}) 
\,+\, \tilde{G}_{\gamma}(p_{3}\!-\!k_{1})\,(p_{3}\!-\!k_{1})_{\nu} \Bigr)\,
G_{\nu\sigma\alpha\beta}^{nsab}(p_{3}\!-\!k_{1},k_{2},p_{1},p_{2} ).
\label{kggggg}\qquad\,
\eea
From the definition (\ref{ptPidef1})-(\ref{ptPidef3}),
only the very first term, proportional to
$k_{2}^{2}t_{\sigma\nu'}(k_{2})D_{\nu'\nu}(q)$, in each of
the above three expressions can contribute to the PT self-energy:
in each case, the term proportional to
$\tilde{G}_{\sigma}(q)\,q_{\nu}$,
although contracted with the required tree level vertices,
cannot result in the required tensor structure (\ref{ptPidef2}) 
regardless of the form of the function carrying the $\mu$ index;
and clearly none of the remaining terms has the required structure.
Exactly similar expressions result for the linear factor $k_{2\sigma}$.

For the quadratic factor $k_{1\rho}k_{2\sigma}$, we obtain
\bea
\lefteqn{k_{1\rho}k_{2\sigma}
G_{\rho\sigma ij}^{rs(f)}(-k_{1},k_{2},p_{1},p_{2})
\,\,=\,\, {\textstyle \frac{1}{2}}f^{nrs}(k_{1}\!+\!k_{2})_{\nu'}
\tilde{G}_{\nu'}(q)\,q_{\nu}\,i\gamma_{\nu}T_{ji}^{n} }\nonumber \\
& &
+{\textstyle \frac{1}{2}}\{T^{r},T^{s}\}_{ji} 
\Bigl(S^{-1}(p_{1},m_{\!f})+S^{-1}(p_{2},m_{\!f})\Bigr)
\nonumber\\
& &
-(T^{s}T^{r})_{ji}S^{-1}(p_{2},m_{\!f})\,
S(p_{1}\!-\!k_{1},m_{\!f})\,S^{-1}(p_{1},m_{\!f}) 
\nonumber\\
& &
-(T^{r}T^{s})_{ji}S^{-1}(p_{2},m_{\!f})\,
S(p_{2}\!+\!k_{1},m_{\!f})\,S^{-1}(p_{1},m_{\!f}),
\label{kkggff} \\
\nn \\
\lefteqn{k_{1\rho}k_{2\sigma}
G_{\rho\sigma\alpha\beta}^{rsab}(-k_{1},k_{2},p_{1},p_{2})
\,\,=\,\,{\textstyle \frac{1}{2}}f^{nrs}(k_{1}\!+\!k_{2})_{\nu'}
\tilde{G}_{\nu'}(q)\,q_{\nu}\,\Gamma_{\nu\alpha\beta}^{nab}(q,p_{1},p_{2}) }
\nonumber \\
& &
-f^{nra}f^{nsb} \Bigl(
[ \,{\textstyle \frac{1}{2}}g_{\nu'\nu} 
+ \tilde{G}_{\nu'}(p_{1}\!-\!k_{1})\,(p_{1}\!-\!k_{1})_{\nu}\,]
[\,p_{1}^{2}t_{\alpha\nu}(p_{1})g_{\beta\nu'}
 + p_{2}^{2}t_{\beta\nu}(p_{1})g_{\alpha\nu'}\,]      \qquad\qquad
\nonumber \\
& &\hspace{60pt}
\mbox{}+p_{1}^{2}t_{\alpha\nu'}(p_{1})
\,D_{\nu'\nu}(p_{1}\!-\!k_{1})\,
p_{2}^{2}t_{\nu\beta}(p_{2})\Bigr)
\nonumber \\
& &
-f^{nrb}f^{nsa} \Bigl(
[\, {\textstyle \frac{1}{2}}g_{\nu'\nu} 
+ \tilde{G}_{\nu'}(p_{2}\!-\!k_{1})\,(p_{2}\!-\!k_{1})_{\nu}\,]
[\,p_{1}^{2}t_{\alpha\nu}(p_{1})g_{\beta\nu'}
 + p_{2}^{2}t_{\beta\nu}(p_{1})g_{\alpha\nu'}\,]
\nonumber \\
& &\hspace{60pt}
\mbox{}+p_{1}^{2}t_{\alpha\nu'}(p_{1})
\,D_{\nu'\nu}(p_{2}\!-\!k_{1})\,
p_{2}^{2}t_{\nu\beta}(p_{2})\Bigr),
\label{kkgggg} \\
\nn \\
\lefteqn{
k_{1\rho}k_{2\sigma}
\,G_{\rho\sigma\alpha\beta\gamma}^{rsabc}(-k_{1},k_{2},p_{1},p_{2},p_{3})
\,\,=\,\,{\textstyle \frac{1}{2}}f^{nrs}(k_{1}\!+\!k_{2})_{\nu'}
\tilde{G}_{\nu'}(q)
\,q_{\nu}\,G_{\nu\alpha\beta\gamma}^{nabc}(q,p_{1},p_{2},p_{3}) 
+\,\,\ldots }
\nn \\
\label{kkggggg}
\eea
where, in the last expression,
we have only troubled to record the term proportional
to $G_{\nu\alpha\beta\gamma}^{nabc}(q,p_{1},p_{2},p_{3})$.
Here, none of the terms in the above three expressions can contribute
to the PT self-energy defined in (\ref{ptPidef1})-(\ref{ptPidef3}).

A set of expressions exactly similar to Eqs.~(\ref{kggff})-(\ref{kkggggg})
is obtained for the factors
$k_{1\rho'}$, $k_{2\sigma'}$ and $k_{1\rho'}k_{2\sigma'}$ contracted
with the three $n$-point functions on the l.h.s.\ of (\ref{GG}).

We therefore see that it is only necessary to consider the 
terms in (\ref{GG})
in which no more than one longitudinal factor is contracted with the
$n$-point function on each side.  The product of propagators 
$iD_{\rho'\rho}(k_{1})\,iD_{\sigma'\sigma}(k_{2})$ in (\ref{GG})
may thus be written
\bea
\lefteqn{iD_{\rho'\rho}(k_{1})\,iD_{\sigma'\sigma}(k_{2})
\,\,=\,\,
\frac{1}{k_{1}^{2}k_{2}^{2}}\biggl\{ -g_{\rho'\rho}g_{\sigma'\sigma}
+\Bigl( a_{\rho'}(k_{1})k_{1\rho}g_{\sigma'\sigma} 
      + a_{\sigma'}(k_{2})k_{2\sigma}g_{\rho'\rho}
} \nn \\
& & \hspace{-20pt}
- a_{\rho'}(k_{1})k_{1\rho}k_{2\sigma'}a_{\sigma}(k_{2}) 
+ \{\rho'\sigma'\}\!\leftrightarrow\!\{\rho\sigma\} \Bigr) 
\mbox{}
+ b(k_{1})k_{1\rho'}k_{1\rho}g_{\sigma'\sigma}
+ b(k_{2})k_{2\sigma'}k_{2\sigma}g_{\rho'\rho}
+ \ldots \biggl\} \qquad
\label{DD}
\eea
i.e.\ with only the terms with no more than one factor $k_{1\rho}$ or $k_{2\sigma}$
and no more than one factor $k_{1\rho'}$ or $k_{2\sigma'}$
written explicitly. We consider the $a$ and $b$ terms in \eq{DD} separately.

\pagebreak

i) $a$ {\em terms}

\noindent
The $a$ terms in \eq{DD} occur only in the class of non-covariant
gauges [\,$a_{\mu}(q) = n_{\mu}/n\sdot q$\,].

For the term $a_{\rho'}(k_{1})k_{1\rho}g_{\sigma'\sigma}$ in \eq{DD},
the effect of the factor
$k_{1\rho}$ on the three $n$-point functions on the r.h.s.\ of (\ref{GG})
is given in Eqs.~(\ref{kggff})-(\ref{kggggg}).
In each case, as just described, only the term 
proportional to
$k_{2}^{2}t_{\sigma\nu'}(k_{2})D_{\nu'\nu}(q)$ can contribute to the
PT self-energy. Discarding the remaining terms in
Eqs.~(\ref{kggff})-(\ref{kggggg}) thus leaves for the term
$a_{\rho'}(k_{1})k_{1\rho}g_{\sigma'\sigma}$ from
$iD_{\rho\rho}(k_{1})iD_{\sigma'\sigma}(k_{2})$ in (\ref{GG})
\be\label{a1}
\mu^{2\epsilon}\int\frac{d^{d}k}{(2\pi)^{d}}
\left\{
\begin{array}{c}
ig^{2}G_{\rho'\sigma i'j'}^{rs(f')} \\
\\
ig^{2}G_{\rho'\sigma\alpha'\beta'}^{rsa'b'} \\
\\
-g^{3}G_{\rho'\sigma\alpha'\beta'\gamma'}^{rsa'b'c'} \\
\end{array}
\right\}
\frac{a_{\rho'}(k_{1})}{2k_{1}^{2}}
gf^{nrs}t_{\sigma\nu'}(k_{2})\,iD_{\nu'\nu}(q)
\left\{
\begin{array}{c}
ig\gamma_{\nu}T_{ij}^{n} \\
\\
g\Gamma_{\nu\alpha\beta}^{nab} \\
\\
-ig^{2}\Gamma_{\nu\alpha\beta\gamma}^{nabc} \\
\end{array}
\right\}.
\ee
The term $g_{\sigma\nu'}$ from $t_{\sigma\nu'}(k_{2})$ in (\ref{a1})
does not involve any further factors of longitudinal loop four-momentum.
It can thus only contribute
to the PT self-energy when contracted with the components
$ig\gamma_{\mu}T_{i'j'}^{m}iD_{\mu\mu'}(q)g\Gamma_{\mu'\rho'\sigma}^{mrs}$
[Fig.~6(d)],
$g\Gamma_{\mu\alpha'\beta'}^{ma'b'}
iD_{\mu\mu'}(q)g\Gamma_{\mu'\rho'\sigma}^{mrs}$
[Fig.~8(g)] or
$-ig^{2}\Gamma_{\mu\alpha'\beta'\gamma'}^{ma'b'c'}
iD_{\mu\mu'}(q)g\Gamma_{\mu'\rho'\sigma}^{mrs}$,
respectively, of the $n$-point functions on the l.h.s.\ of (\ref{a1}).
Writing $k_{1} = k$ and $k_{2} = k+q$ and using the dimensional
regularization rules
$\int d^{d}k\,k^{-2} =
 \int d^{d}k\,a_{\mu}(k)k_{\nu}\,k^{-2} = 0$,
the contribution of the $g_{\sigma\nu'}$ term in (\ref{a1})
to $i\hat{\Pi}_{\mu\nu}^{mn}(q)$ in fact vanishes. The term
$- k_{2\sigma}k_{2\nu'}/k_{2}^{2}$ from $t_{\sigma\nu'}(k_{2})$
in (\ref{a1}) involves a longitudinal factor $k_{2\sigma}$
contracted with the $n$-point functions on the l.h.s. Using the
Ward identities Eqs.~(\ref{kggff})-(\ref{kggggg}), the contribution
to the PT ``effective'' two-point component of the amplitudes (\ref{GG})
of the term
$a_{\rho'}(k_{1})k_{1\rho}g_{\sigma'\sigma}$ from the product
of propagators $iD_{\rho'\rho}(k_{1})iD_{\sigma'\sigma}(k_{2})$
is thus given by
\be\label{a2}
\mu^{2\epsilon}\int\frac{d^{d}k}{(2\pi)^{d}}
\left\{
\begin{array}{c}
ig\gamma_{\mu}T_{i'j'}^{m} \\
\\
g\Gamma_{\mu\alpha'\beta'}^{ma'b'} \\
\\
-ig^{2}\Gamma_{\mu\alpha'\beta'\gamma'}^{ma'b'c'} \\
\end{array}
\right\}
iD_{\mu\mu'}(q)\,A_{\mu'\nu'}^{(1)}(k_{1},k_{2})\,iD_{\nu'\nu}(q)
\left\{
\begin{array}{c}
ig\gamma_{\nu}T_{ij}^{m} \\
\\
g\Gamma_{\nu\alpha\beta}^{mab} \\
\\
-ig^{2}\Gamma_{\nu\alpha\beta\gamma}^{mabc} \\
\end{array}
\right\}
\ee
where $A_{\mu\nu}^{(1)}(k_{1},k_{2})
= \frac{1}{2}Ng^{2}t_{\mu\rho}(k_{1})a_{\rho}(k_{1})k_{2\nu}k_{2}^{-2}$.

For the term
$a_{\sigma'}(k_{2})k_{2\sigma}g_{\rho'\rho}$ in \eq{DD},
the contribution
$A_{\mu\nu}^{(2)}(k_{1},k_{2})$
to the PT self-energy is as in (\ref{a2}) except for the interchange
$k_{1}\leftrightarrow k_{2}$, i.e.\
$A_{\mu\nu}^{(2)}(k_{1},k_{2}) = A_{\mu\nu}^{(1)}(k_{2},k_{1})$.

For the term $-a_{\rho'}(k_{1})k_{1\rho}k_{2\sigma'}a_{\sigma}(k_{2})$
in \eq{DD}, the effect of the two longitudinal factors
$k_{1\rho}$, $k_{2\sigma'}$ on the $n$-point functions on each side
of (\ref{GG}) is again given by the Ward identities 
Eqs.~(\ref{kggff})-(\ref{kggggg}). These Ward identities
immediately give as the contribution of this term to 
the PT ``effective'' two-point
component of the amplitudes (\ref{GG})
\be\label{a3}
\mu^{2\epsilon}\int\frac{d^{d}k}{(2\pi)^{d}}
\left\{
\begin{array}{c}
ig\gamma_{\mu}T_{i'j'}^{m} \\
\\
g\Gamma_{\mu\alpha'\beta'}^{ma'b'} \\
\\
-ig^{2}\Gamma_{\mu\alpha'\beta'\gamma'}^{ma'b'c'} \\
\end{array}
\right\}
iD_{\mu\mu'}(q)\,A_{\mu'\nu'}^{(3)}(k_{1},k_{2})\,iD_{\nu'\nu}(q)
\left\{
\begin{array}{c}
ig\gamma_{\nu}T_{ij}^{m} \\
\\
g\Gamma_{\nu\alpha\beta}^{mab} \\
\\
-ig^{2}\Gamma_{\nu\alpha\beta\gamma}^{mabc} \\
\end{array}
\right\}
\ee
where $A_{\mu\nu}^{(3)}(k_{1},k_{2}) = \frac{1}{2}Ng^{2}
t_{\mu\rho}(k_{1})a_{\rho}(k_{1})a_{\sigma}(k_{2})t_{\sigma\nu}(k_{2})$.

{}From \eq{DD}, there are also three more $a$ term contributions to
$i\hat{\Pi}_{\mu\nu}^{mn}(q)$,
identical to the above three contributions except for the interchange
$\mu \leftrightarrow \nu$ in $A_{\mu\nu}^{(i)}(k_{1},k_{2})$,
$i = 1,2,3$.

Adding up these six contributions and substituting the explicit
form \eq{abaxial} for $a$, we obtain the overall contribution
$i\hat{\Pi}_{\mu\nu}^{{\rm (a)}mn}(q)|_{n\cdot A = 0}$ due to the
$a$ terms in the propagators $iD_{\rho'\rho}(k_{1})$,
$iD_{\sigma'\sigma}(k_{2})$ in 
the amplitudes
(\ref{GG}) in the class of non-covariant gauges $n\sdot A^{a} = 0$\,:
\be
\left.i\hat{\Pi}_{\mu\nu}^{{\rm (a)}mn}(q)\right|_{n\cdot A = 0}
=
-N\delta^{mn}g^{2}
\mu^{2\epsilon}\int\frac{d^{d}k}{(2\pi)^{d}}
\biggl\{ \frac{k_{1\mu}k_{2\nu}}{k_{1}^{2}k_{2}^{2}}
- \frac{n_{\mu}n_{\nu}}{(n\sdot k_{1})(n\sdot k_{2})} \biggl\}.
\label{acontrib}
\ee
The first term in (\ref{acontrib}) is precisely the contribution
to $i\hat{\Pi}_{\mu\nu}^{mn}(q)$ in the class of non-covariant gauges
$n\sdot A^{a} = 0$  which is supplied by the
standard ghost loop in the class of covariant gauges.
The second term in (\ref{acontrib}) vanishes identically;\footnote{
With $k_{2} - k_{1} = q$, write
$(n\sdot k_{1})^{-1}(n\sdot k_{2})^{-1} =
(n\sdot q)^{-1}[(n\sdot k_{1})^{-1} - (n\sdot k_{2})^{-1}]$.}
alternatively it is cancelled algebraically under the integral sign
by the standard (vanishing) ghost loop in the class of 
gauges $n\sdot A^{a} = 0$.
Given that in the class of linear covariant gauges
$i\hat{\Pi}_{\mu\nu}^{{\rm (a)}mn}(q)|_{\rm lin\,cov} = 0$,
we thus obtain the gauge-independent statement \eq{agh}.

\vspace{5pt}

ii) $b$ {\em terms}

\noindent
The $b$ terms in \eq{DD} occur both in the class of covariant
gauges [\,$b(q) = (1-\xi)/q^{2}$\,] and in the class of non-covariant
gauges $n\sdot A^{a} = 0$ [\,$b(q) = -n^{2}/(n\sdot q)^{2}$\,].

For the term $b(k_{1})k_{1\rho'}k_{1\rho}g_{\sigma'\sigma}$
in \eq{DD}, the effect of the factors $k_{1\rho}$, $k_{1\rho'}$
is given by the Ward identites Eqs.~(\ref{kggff})-(\ref{kggggg}).
We immediately obtain as the contribution of this term to
the PT ``effective'' two-point component of the amplitudes (\ref{GG})
\be\label{b1}
\mu^{2\epsilon}\int\frac{d^{d}k}{(2\pi)^{d}}
\left\{
\begin{array}{c}
ig\gamma_{\mu}T_{i'j'}^{m} \\
\\
g\Gamma_{\mu\alpha'\beta'}^{ma'b'} \\
\\
-ig^{2}\Gamma_{\mu\alpha'\beta'\gamma'}^{ma'b'c'} \\
\end{array}
\right\}
iD_{\mu\mu'}(q)\,
\frac{-Ng^{2}b(k_{1})\,k_{2}^{2}t_{\mu'\nu'}(k_{2})}{2k_{1}^{2}}\,
\,iD_{\nu'\nu}(q)
\left\{
\begin{array}{c}
ig\gamma_{\nu}T_{ij}^{m} \\
\\
g\Gamma_{\nu\alpha\beta}^{mab} \\
\\
-ig^{2}\Gamma_{\nu\alpha\beta\gamma}^{mabc} \\
\end{array}
\right\}.
\ee
Writing $k_{1} = k$ and $k_{2} = k+q$ and using the dimensional
regularization rules
$\int d^{d}k\,b(k) = 
\int d^{d}k\, b(k) \,k_{\mu}k_{\nu}\,k^{-2} = 0$
for $b(k)$ in both classes of gauge, (\ref{b1}) can be written
\be\label{b2}
\mu^{2\epsilon}\int\frac{d^{d}k}{(2\pi)^{d}}
\left\{
\begin{array}{c}
ig\gamma_{\mu}T_{i'j'}^{m} \\
\\
g\Gamma_{\mu\alpha'\beta'}^{ma'b'} \\
\\
-ig^{2}\Gamma_{\mu\alpha'\beta'\gamma'}^{ma'b'c'} \\
\end{array}
\right\}
iD_{\mu\mu'}(q)\,
\frac{iNg^{2}b(k)}{2k^{2}}\,iq^{2}t_{\mu'\nu'}(q)
\,iD_{\nu'\nu}(q)
\left\{
\begin{array}{c}
ig\gamma_{\nu}T_{ij}^{m} \\
\\
g\Gamma_{\nu\alpha\beta}^{mab} \\
\\
-ig^{2}\Gamma_{\nu\alpha\beta\gamma}^{mabc} \\
\end{array}
\right\}.
\ee
This contribution to $i\hat{\Pi}_{\mu\nu}^{mn}(q)$ is $q$-independent,
i.e.\ tadpole-like.

The term $b(k_{2})k_{2\sigma'}k_{2\sigma}g_{\rho'\rho}$
in \eq{DD} results in a contribution identical to (\ref{b2}).

\begin{center}
\begin{picture}(400,420)(20,-100)



\put( 80,250){\makebox(0,0)[r]{\footnotesize $A_{\nu}^{n}(q)$}}
\put(130,290){\makebox(0,0)[l]{\footnotesize 
$\psi_{j}^{\scriptscriptstyle{(f)}}(p_{2})$}}
\put(130,210){\makebox(0,0)[l]{\footnotesize 
$\psi_{i}^{\scriptscriptstyle{(f)}}(p_{1})$}}
\Photon( 90,250)(120,250){2.5}{3}
\Line(120,220)(120,235)
\ArrowLine(120,235)(120,250)
\ArrowLine(120,250)(120,265)
\Line(120,265)(120,280)
\PhotonArc(120,250)(15,-90,90){2}{6.5}
\put(145,245){\vector(0, 1){10}}
\put(155,250){\makebox(0,0)[c]{\small $k_{3}$}}
\put(110,190){\makebox(0,0)[c]{\small (a)}}

\put(220,265){\makebox(0,0)[c]{\small pinch}}
\put(205,250){\vector(1, 0){30}}

\Photon(300,250)(330,250){2.5}{3}
\ArrowLine(330,220)(330,250)
\ArrowLine(330,250)(330,280)
\PhotonArc(347,250)(15,7.5,367.5){-2}{12}
\put(330,190){\makebox(0,0)[c]{\small (d)}}


\put( 80,125){\makebox(0,0)[r]{\footnotesize $A_{\nu}^{n}(q)$}}
\put(130,165){\makebox(0,0)[l]{\footnotesize $A_{\beta}^{b}(p_{2})$}}
\put(130, 85){\makebox(0,0)[l]{\footnotesize $A_{\alpha}^{a}(p_{1})$}}
\Photon( 90,125)(120,125){2.5}{3}
\Photon(120, 95)(120,155){2.5}{6}
\PhotonArc(120,125)(15,-90,90){2}{7}
\put(145,120){\vector(0, 1){10}}
\put(155,125){\makebox(0,0)[c]{\small $k_{3}$}}
\put(110, 65){\makebox(0,0)[c]{\small (b)}}

\put(220,140){\makebox(0,0)[c]{\small pinch}}
\put(205,125){\vector(1, 0){30}}

\Photon(300,125)(330,125){2.5}{3}
\Photon(330, 95)(330,155){2.5}{6}
\PhotonArc(346,125)(15,7.5,367.5){-2}{12}
\put(330, 65){\makebox(0,0)[c]{\small (e)}}


\put( 80,  0){\makebox(0,0)[r]{\footnotesize $A_{\nu}^{n}(q)$}}
\put(105, 40){\makebox(0,0)[r]{\footnotesize $A_{\gamma}^{c}(p_{3})$}}
\put(130, 40){\makebox(0,0)[l]{\footnotesize $A_{\beta}^{b}(p_{2})$}}
\put(130,-40){\makebox(0,0)[l]{\footnotesize $A_{\alpha}^{a}(p_{1})$}}
\Photon( 90,  0)(120,  0){2.5}{3}
\Photon(120,-30)(120, 30){2.5}{6}
\Photon(120,  0)(105, 26){2.5}{3}
\PhotonArc(120,0)(15,-90,90){2}{7}
\put(145, -5){\vector(0, 1){10}}
\put(155,  0){\makebox(0,0)[c]{\small $k_{3}$}}
\put(80, -30){\makebox(0,0)[c]{\footnotesize 
3 perms of }}
\put(80, -40){\makebox(0,0)[c]{\footnotesize 
$A_{\alpha}^{a}$, $A_{\beta}^{b}$, $A_{\gamma}^{c}$.}}
\put(110,-60){\makebox(0,0)[c]{\small (c)}}

\put(220, 15){\makebox(0,0)[c]{\small pinch}}
\put(205,  0){\vector(1, 0){30}}

\Photon(300,  0)(330,  0){2.5}{3}
\Photon(330,-30)(330, 30){2.5}{6}
\Photon(330,  0)(315, 26){2.5}{3}
\PhotonArc(346,0)(15,7.5,367.5){-2}{12}
\put(330,-60){\makebox(0,0)[c]{\small (f)}}

\put(210,-100){\makebox(0,0)[c]{\footnotesize 
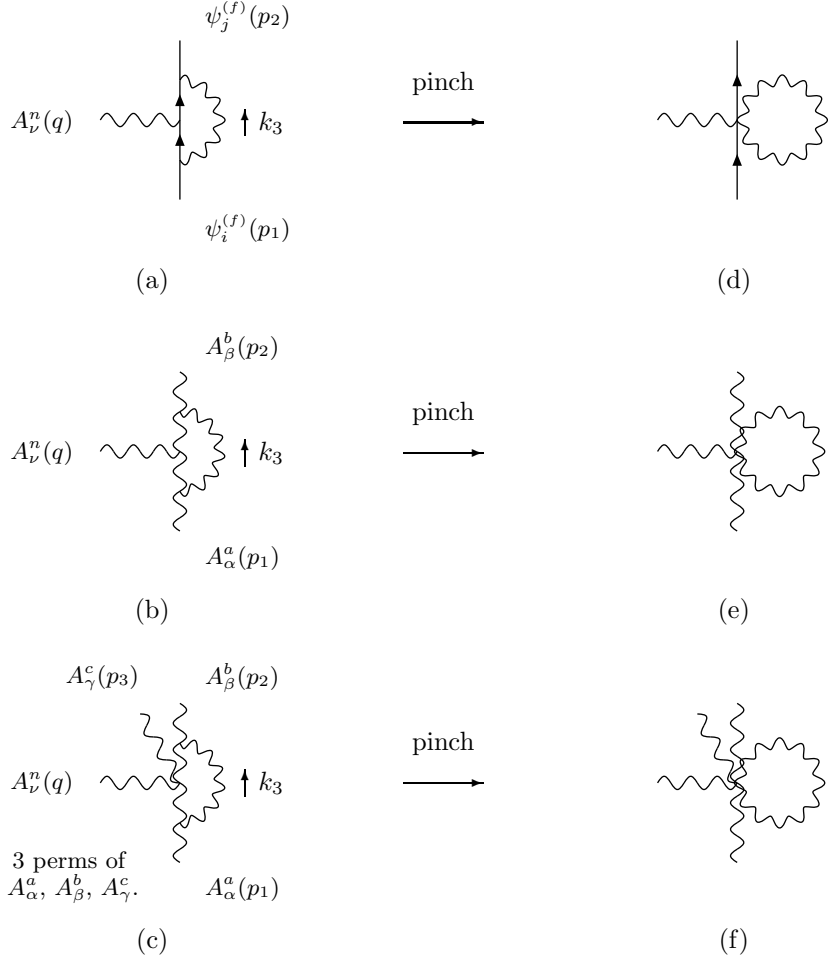
Fig.\ 13. The tadpole-like pinch parts of the one-loop vertex diagrams.}} 

\end{picture}
\end{center}

\pagebreak

This accounts for the contribution to
$i\hat{\Pi}_{\mu\nu}^{mn}(q)$
due to the $b$ terms in the product of propagators
$iD_{\rho'\rho}(k_{1})\,iD_{\sigma'\sigma}(k_{2})$
in the amplitudes (\ref{GG}). 
However, there remain the tadpole-like
pinch parts of the diagrams shown in Fig.~13. 
The amplitudes for these diagrams, in an arbitrary gauge, are given by
\bea
\mbox{Fig.~13(a)}
&=&
-ig^{3}\mu^{2\epsilon}\int\frac{d^{d}k}{(2\pi)^{d}}
\,i\gamma_{\tau'}T_{jk}^{t}\,S(k_{2},m_{\!f})
\,i\gamma_{\nu}T_{kl}^{n}
\,S(k_{1},m_{\!f})\,i\gamma_{\tau}T_{li}^{t}\,D_{\tau'\tau}(k_{3}) 
\qquad\,\,\,\,\label{fig13a} \\
\nn \\
\mbox{Fig.~13(b)}
&=&
-ig^{3}\mu^{2\epsilon}\int\frac{d^{d}k}{(2\pi)^{d}}
\,\Gamma_{\nu\rho'\sigma'}^{nrs}(q,k_{1},-k_{2}) 
\,D_{\rho'\rho}(k_{1})\,D_{\sigma'\sigma}(k_{2})\times \nn \\ 
& &
\Gamma_{\rho\tau'\alpha}^{rta}(-k_{1},-k_{3},p_{1})
\,\Gamma_{\sigma\tau\beta}^{stb}(k_{2},k_{3},p_{2})\,D_{\tau'\tau}(k_{3})
\label{fig13b} \\
\nn \\
\mbox{Fig.~13(c)}
&=&
-g^{4}\mu^{2\epsilon}\int\frac{d^{d}k}{(2\pi)^{d}}
\,\Gamma_{\nu\rho'\sigma'\gamma}^{nrsc}(q,k_{1},-k_{2},p_{3}) 
\,D_{\rho'\rho}(k_{1})\,D_{\sigma'\sigma}(k_{2})\times \nn \\ 
& &
\Gamma_{\rho\tau'\alpha}^{rta}(-k_{1},-k_{3},p_{1})
\,\Gamma_{\sigma\tau\beta}^{stb}(k_{2},k_{3},p_{2})\,D_{\tau'\tau}(k_{3})
\,\,+\,\, \mbox{c.p.}
\label{fig13c}
\eea
where in \eq{fig13c} ``c.p.'' indicates that there are two further terms
obtained from the cyclic permutation of
$\{p_{1},a,\alpha\}$, $\{p_{2},b,\beta \}$, $\{p_{3},c,\gamma\}$.
In each of Eqs.~(\ref{fig13a})-(\ref{fig13c}),
the $b$ term in the propagator $D_{\tau'\tau}(k_{3})$
contributes\footnote{
The diagram Fig.~13(b) occurs in the amplitudes (\ref{GG}) involving
$G_{\rho\sigma\alpha\beta}^{rsab}$ on the r.h.s.\ (cf.\ Fig.~8(b)).
However, the $b$ term from $D_{\tau'\tau}(k_{3})$ in \eq{fig13b}
vanishes when $D_{\tau'\tau}(k_{3})$ is pinched by factors $k_{1\rho}$,
$k_{2\sigma}$.}
a quadratic factor $k_{3\tau'}k_{3\tau}$.
Using the Ward identities Eqs.~(\ref{fermionwid}) and (\ref{tgvwid})
together with
$q^{2}t_{\mu\rho}(q)\,D_{\rho\nu}(q) = -g_{\mu\nu} - G_{\mu}(q)\,q_{\nu}$
for the gluonic cases, and choosing $k_{3} = k$,
the contribution of the $b$ term in each of the propagators
$D_{\tau'\tau}(k_{3})$ 
in Eqs.~(\ref{fig13a})-(\ref{fig13c}) is given by
\bea
\mbox{Fig.~13(a)}|_{b(k_{3})}
&=&
ig^{2}\mu^{2\epsilon}\int\frac{d^{d}k}{(2\pi)^{d}}\,\frac{b(k)}{k^{2}} 
\,\Bigl\{ 1\,-\,S^{-1}(p_{2},m_{\!f})\,S(p_{2}\!-\!k,m_{\!f}) \Bigl\}
\times  \nn \\
& &
\,ig\gamma_{\nu}(T^{t}T^{n}T^{t})_{ji}
\,\Bigl\{1\,-\,S(p_{1}\!-\!k,m_{\!f})\,S^{-1}(p_{1},m_{\!f})\Bigr\}.
\label{13ab} \\
\nn \\
\mbox{Fig.~13(b)}|_{b(k_{3})}
&=&
ig^{2}\mu^{2\epsilon}\int\frac{d^{d}k}{(2\pi)^{d}}\,\frac{b(k)}{k^{2}} 
f^{rta}f^{stb} 
g\Gamma_{\nu\rho'\sigma'}^{nrs}(q,p_{1}\!-\!k,p_{2}\!+\!k) \times  \nn \\ 
& & 
\Bigl\{ g_{\rho'\alpha}
\,+\,(p_{1}\!-\!k)_{\rho'}\tilde{G}_{\alpha}(p_{1}\!-\!k) 
\,+\,D_{\rho'\rho}(p_{1}\!-\!k)\,p_{1}^{2}t_{\rho\alpha}(p_{1}) \Bigr\} \times
\nn \\
& &
\Bigl\{ g_{\sigma'\beta}
\,+\,(p_{2}\!+\!k)_{\sigma'}\tilde{G}_{\beta}(p_{2}\!+\!k)
\,+\,D_{\sigma'\sigma}(p_{2}\!+\!k)\,p_{2}^{2}t_{\sigma\beta}(p_{2}) \Bigr\}
\label{13bb} \\
\nn \\
\mbox{Fig.~13(c)}|_{b(k_{3})}
&=&
ig^{2}\mu^{2\epsilon}\int\frac{d^{d}k}{(2\pi)^{d}}\,\frac{b(k)}{k^{2}} 
f^{rta}f^{stb} (-ig^{2})
\Gamma_{\nu\rho'\sigma'\gamma}^{nrsc}(q,p_{1}\!-\!k,p_{2}\!+\!k,p_{3})
\times \nn \\ 
& & 
\Bigl\{ g_{\rho'\alpha}
\,+\,(p_{1}\!-\!k)_{\rho'}\tilde{G}_{\alpha}(p_{1}\!-\!k) 
\,+\,D_{\rho'\rho}(p_{1}\!-\!k)\,p_{1}^{2}t_{\rho\alpha}(p_{1}) \Bigr\} \times
\nn \\
& &
\Bigl\{ g_{\sigma'\beta}
\,+\,(p_{2}\!+\!k)_{\sigma'}\tilde{G}_{\beta}(p_{2}\!+\!k)
\,+\,D_{\sigma'\sigma}(p_{2}\!+\!k)\,p_{2}^{2}t_{\sigma\beta}(p_{2}) \Bigr\}
\,+\, \mbox{c.p.} \qquad\,\,\,\, \label{13cb} 
\eea
Then in each of Eqs.~(\ref{13ab})-(\ref{13cb}), 
retaining only the very first term in each set of curly parentheses,
i.e.\ that which is independent of the external momenta,
and using the identites
\bea
T^{t}T^{n}T^{t} 
&=& 
(- {\textstyle \frac{1}{2}}C_{A} + C_{F})T^{n} \qquad \\
f^{rta}f^{stb}f^{nrs}
&=&
{\textstyle \frac{1}{2}}C_{A}f^{nab} \\
f^{rta}f^{stb}f^{unr}f^{usc} + 
f^{rtb}f^{stc}f^{una}f^{urs} +
f^{rtc}f^{sta}f^{uns}f^{ubr}
&=&
C_{A}f^{rna}f^{rbc}\qquad \,\,
\eea
where $C_{F}$ ($C_{A}$) is the quadratic Casimir coefficient for the
fundamental (adjoint) representation, we obtain\footnote{
In the case of \eq{13bb}, we have also used
$\Gamma_{\nu\alpha\beta}^{nrs}(q,p_{1}\!-\!k,p_{2}\!+\!k)
= \Gamma_{\nu\alpha\beta}^{nrs}(q,p_{1},p_{2}) +
\Gamma_{\nu\alpha\beta}^{nrs}(0,-k,k)$. 
The latter term is odd in $k$ and so vanishes in the integral
in \eq{fig13abcpinch}.}
the tadpole-like pinch parts of the diagrams in Figs.~13\,:
\be\label{fig13abcpinch}
\left.\begin{array}{c}
\mbox{Fig.~13(d)} \\
\\
\mbox{Fig.~13(e)} \\
\\
\mbox{Fig.~13(f)} \\
\end{array}\right\}
=
ig^{2}\mu^{2\epsilon}\int\frac{d^{d}k}{(2\pi)^{d}}
\,\frac{b(k)}{2k^{2}}
\left\{ \begin{array}{l}
(- C_{A} + 2C_{F})\,ig\gamma_{\nu}T_{ij}^{n} \\
\\
(- C_{A} + 2C_{A})\,g\Gamma_{\nu\alpha\beta}^{nab} \\
\\
(- C_{A} + 3C_{A})(-ig^{2})\Gamma_{\nu\alpha\beta\gamma}^{nabc}. \\
\end{array}\right.
\ee
In the above expression for Fig.~13(d), the term proportional to $2C_{F}$
exactly cancels against the tadpole-like pinch parts of the self-energy
corrections Figs.~2(c) and (e) associated with the pair of external
fermion legs. Similarly, in the expressions for Fig.~13(e) and Fig.~13(f),
the terms proportional to $2C_{A}$ and $3C_{A}$
exactly cancel against tadpole-like pinch parts of the conventional
self-energy corrections associated with the pair and triple, respectively,
of external gluon legs. In each case, putting $C_{A} = N$,
the remaining term proportional to $-C_{A}$ then cancels
against the contribution to $i\hat{\Pi}_{\mu\nu}^{mn}(q)$ in \eq{b2}.
The overall contribution $i\hat{\Pi}_{\mu\nu}^{{\rm (b)}mn}(q)$
to the PT self-energy due the $b$ terms in the 
gluon propagators therefore vanishes. We thus
obtain the gauge-independent statement \eq{b}.

\vspace{10pt}


\setcounter{equation}{0}
\def\theequation{B.\arabic{equation}}
\appendix

{\Large\bf Appendix B. The Function 
$\Delta\hat{\Gamma}_{\mu\alpha\beta}^{mab}$ }

\vspace{5pt}

\noindent
In this appendix, we give the explicit form of the function
$\Delta\hat{\Gamma}_{\mu\alpha\beta}^{mab}$ appearing in \eq{GDDG3}:
\bea
\lefteqn{-\Delta\hat{\Gamma}_{\mu\alpha\beta}^{mab}(q;p_{1},p_{2})
\,\,=\,\,
{\textstyle \frac{1}{2}}ig^{2}Nf^{mab}
\mu^{2\epsilon}\int\frac{d^{d}k}{(2\pi)^{d}}
\frac{1}{k_{1}^{2}k_{2}^{2}k_{3}^{2}} \times }\nonumber\\
& &\biggl\{
  A_{\mu\rho\beta}( k_{3};p_{1},p_{2})\,p_{1}^{2}t_{\rho\alpha}(p_{1})
- A_{\mu\rho\alpha}(-k_{3};p_{2},p_{1})\,p_{2}^{2}t_{\rho\beta}(p_{2})
\nonumber \\
& &
\mbox{}- 
\frac{n^{2}}{(n\sdot k_{3})^{2}}\Gamma_{\mu\rho\sigma}^{F}(q;k_{1},-k_{2})
\,p_{1}^{2}t_{\rho\alpha}(p_{1})\,p_{2}^{2}t_{\sigma\beta}(p_{2})
\nonumber \\
& &
\mbox{}+ 
k_{1}^{2}B_{\mu\alpha\beta}(k_{3};p_{1},p_{2})
-k_{2}^{2}B_{\mu\beta\alpha}(-k_{3};p_{2},p_{1})
-2k_{1}^{2}k_{2}^{2}\frac{n^{2}}{(n\sdot k_{3})^{2}}
\Gamma_{\mu\alpha\beta}(q,p_{1},p_{2}) \biggl\} \qquad
\label{DeltaPTtgv} 
\eea
where $q + p_{1} + p_{2} = 0$
with, e.g., $k_{1} = k$, $k_{2} = k+q$ and $k_{3} = k-p_{1}$, and
\bea
\lefteqn{A_{\mu\rho\beta}(k_{3};p_{1},p_{2})
\,\,=\,\,
(k_{1}\!+\!k_{2})_{\mu}g_{\rho\beta}
+ \frac{1}{n\sdot k_{3}}\biggl\{
(k_{1}\!+\!k_{2})_{\mu}n_{\rho}(2k_{2}\!+\! k_{3})_{\beta}
+ 2(n\sdot q) g_{\mu\rho}(k_{2} \!+\! k_{3})_{\beta}}
\nonumber \\
& &\hspace{-20pt}
+ 2(n\sdot p_{2})\Gamma_{\mu\rho\beta}^{F}(q;k_{1},-k_{2})\biggr\}
+\frac{n^{2}}{(n\sdot k_{3})^{2}}\biggl\{
k_{2}^{2}\Gamma_{\mu\rho\beta}^{F}(q;k_{1},-k_{2})
-k_{2}^{2}g_{\mu\rho} k_{2\beta}
-(k_{1}\! +\!k_{2})_{\mu}k_{1\rho}k_{2\beta}\biggr\}
\nonumber \\
& &\hspace{-20pt}
+\frac{1}{n\sdot k_{1}}\biggl\{
(k_{1}\!+\!k_{2})_{\mu}n_{\sigma}
-2 (n\sdot q) g_{\mu\sigma}
+ \frac{n^{2}}{n\sdot k_{1}} k_{2}^{2}g_{\mu\sigma} \biggr\}
s_{\rho\tau}(k_{3})\Gamma_{\tau\beta\sigma}(-k_{3},p_{2},k_{2}) \\
\nn \\
\lefteqn{B_{\mu\alpha\beta}(k_{3};p_{1},p_{2})
\,\,=\,\,
6(p_{2\mu}g_{\alpha\beta} - p_{2\alpha}g_{\mu\beta})
+2\biggl\{\frac{1}{n\sdot k_{2}} + \frac{1}{n\sdot k_{3}}\biggr\}
g_{\mu\alpha}
\biggl\{ (n\sdot p_{1}) k_{2\beta} 
- (n\sdot q) k_{3\beta} \biggr\} }
\nonumber \\
& &\hspace{-20pt}
+2(n\sdot p_{2})
\biggl\{ \frac{1}{n\sdot k_{2}}\Gamma_{\mu\alpha\beta}(-k_{1},p_{1},k_{3})
- \frac{1}{n\sdot k_{3}}\Gamma_{\mu\alpha\beta}(q,k_{1},-k_{2})\biggr\}
\nonumber \\
& &\hspace{-20pt}
+2(n\sdot p_{1})
\biggl\{\frac{1}{n\sdot k_{2}} - \frac{1}{n\sdot k_{3}}\biggr\}
g_{\mu\alpha} p_{2\beta}
+\frac{n^{2}}{n\sdot k_{3}}
\biggl\{ k_{1}^{2}t_{\mu\alpha}(k_{1}) 
- p_{1}^{2}t_{\mu\alpha}(p_{1})      \biggr\}
\biggl\{ \frac{1}{n\sdot k_{3}}p_{2\beta} 
- \frac{2}{n\sdot k_{2}}k_{3\beta}   \biggr\}.\nn \\
\eea
We have dropped all terms proportional to
$n_{\alpha}$, $n_{\beta}$ since
in the class of non-covariant gauges $n\sdot A^{a} = 0$ they give vanishing
contribution for $A_{\alpha}^{a}(p_{1})$, $A_{\beta}^{b}(p_{2})$
on- or off-shell (the latter since $n_{\mu}D_{\mu\nu}(q) = 0$).
Also, we have used the dimensional regularization
rules \cite{liebbrandt} $\int d^{d}k\,(n\sdot k)^{-2} =
\int d^{d}k\,k_{\mu}k_{\nu}\,k^{-2}(n\sdot k)^{-2} = 0$.

The crucial feature of 
$\Delta\hat{\Gamma}_{\mu\alpha\beta}^{mab}$ 
is that, when contracted via propagators
$iD_{\alpha\alpha''}(p_{1})$,
$iD_{\beta\beta''}(p_{2})$
with the tree level gluon four-point function
$ig^{2}G_{\alpha''\beta''\gamma\delta}^{abcd}(p_{1},p_{2},p_{3},p_{4})$,
{\em none} of the terms in \eq{DeltaPTtgv} has the propagator
structure $(k_{1}^{2}k_{2}^{2})^{-1}(p_{1}^{2}p_{2}^{2})^{-1}$
in the integrand
required to give a contribution to the product of
two PT one-loop self-energies in \eq{GDDG3}: the first three terms 
in \eq{DeltaPTtgv} proportional to
$p_{1}^{2}t_{\alpha\rho}(p_{1})$, 
$p_{2}^{2}t_{\beta\rho}(p_{1})$ and
$p_{1}^{2}t_{\alpha\rho}(p_{1})\,p_{2}^{2}t_{\beta\rho}(p_{1})$,
respectively, pinch one or both of the propagators
$D_{\alpha\alpha''}(p_{1})$, $D_{\beta\beta''}(p_{2})$,
\be
p_{1}^{2}t_{\rho\alpha}(p_{1})\, D_{\alpha\alpha''}(p_{1})
=
-g_{\rho\alpha''} - \tilde{G}_{\rho}(p_{1})p_{1\alpha''},
\qquad
p_{2}^{2}t_{\rho\beta}(p_{2})\, D_{\beta\beta''}(p_{2})
=
-g_{\rho\beta''} - \tilde{G}_{\rho}(p_{2})p_{2\beta''};
\ee
the two terms in \eq{DeltaPTtgv} 
proportional to $k_{1}^{2}$ and $k_{2}^{2}$ pinch the
corresponding terms $k_{1}^{-2}$ and $k_{2}^{-2}$ 
associated with the propagators $D_{\rho'\rho}(k_{1})$ 
and $D_{\sigma'\sigma}(k_{2})$, respectively, 
from \eq{GDDG3}, so that these contributions to 
$\Delta\hat{\Gamma}_{\mu\alpha\beta}^{mab}$ are 
one-loop corrections associated purely with the
external gluon legs $A_{\beta}^{b}(p_{2})$
and $A_{\alpha}(p_{1})$, respectively; and
the last term in \eq{DeltaPTtgv},
proportional to $k_{1}^{2}k_{2}^{2}$, is the tadpole-like contribution, 
exactly cancelled (cf.\ App.~A) by
terms from the conventional self-energy corrections to
the external legs $A_{\alpha}(p_{1})$ and
$A_{\beta}^{b}(p_{2})$, not included in the amplitude \eq{GDDG2}.

Thus, the component
$\Delta\hat{\Gamma}_{\mu\alpha\beta}^{mab}$
in \eq{GDDG3} can make {\em no contribution}
to the Dyson chain of two one-loop self-energies
when the diagrams of Fig.~8 are contracted with a third
gluon tree level four-point function to form the diagrams
of Fig.~9.

\pagebreak


\setcounter{equation}{0}
\def\theequation{C.\arabic{equation}}
\appendix

{\Large\bf Appendix C. The Heavy Quark Limit}

\vspace{5pt}

\noindent
In this paper, it has been shown explicitly
how the Dyson summed ``effective'' two-point function 
obtained in the off-shell PT
constitutes the gauge-independent
and universal subset of radiative corrections
required for the effective charge defined in Sec.~2.
This has involved considerations purely at the level
of the Feynman integrands for $n$-loop diagrams.
However, in QED the vacuum polarization of the photon has a direct
physical interpretation as the correction to the Coulomb interaction
between static heavy charges. This has led, by analogy, to a popular 
definition (see e.g.\ \cite{michael}) 
of an effective charge for SU($N$) QCD in terms
of (the Fourier transform of) the
potential between a static heavy quark-antiquark pair:
\be\label{V}
V_{q\bar{q}}(\bfq^{2})
=
-\frac{g_{q\bar{q}}^{2}(-\bfq^{2})\,C_{F}}{\bfq^{2}}
\ee
where $\bfq$ is the three-momentum associated 
in the Fourier transform with the interquark separation 
$\mbox{\bf r}$, $C_{F} = (N^{2}-1)/2N$ and
$g_{q\bar{q}}$ is the static heavy quark effective charge.
The question therefore arises: what is the relation of this
heavy quark effective charge to that defined in Sec.~2?

The potential $V_{q\bar{q}}(\bfq^{2})$
may be calculated directly in perturbation theory.
At the one-loop level, and before renormalization, 
one obtains for $n_{\!f}$ flavours of massless fermion
\cite{fischbill}
\bea
V_{q\bar{q}}(\bfq^{2})
&=&
-\frac{g^{2}C_{F}}{\bfq^{2}}
\Biggl\{ 1 + 
\frac{g^{2}}{16\pi^{2}}
\Biggl( \beta_{1}
\biggl[-\CUV + 
\ln\biggl(\frac{\bfq^{2}}{\mu^{2}}\biggr)\biggr] 
+ \frac{31}{9}N - \frac{10}{9}n_{\!f}
\Biggl)
\Biggl\} \\
&\equiv &
-\frac{g^{2}C_{F}}{\bfq^{2}}
\Biggl\{ 1 + \Pi_{q\bar{q}}(-\bfq^{2}) \Biggl\}.
\eea
The corresponding one-loop self-energy-like
function $\Pi_{q\bar{q}}$ thus
{\em differs} from that 
Eq.\ (\ref{ptPi}) of the PT by a constant:
\be\label{qqbarPi}
\Pi_{q\bar{q}}(-\bfq^{2}) 
= 
\hat{\Pi}(-\bfq^{2}) - 4N(g^{2}/16\pi^{2}).
\ee
In the one-loop potential $V_{q\bar{q}}(\bfq^{2})$, 
this constant is of course irrelevant. 
But if, after renormalization, this function 
$\Pi_{q\bar{q}}$ is used to define an
effective charge 
\be\label{qqbargeff}
g_{q\bar{q}}^{2}(-\bfq^{2}) 
=
\frac{g_{R}^{2}}{1 - \Pi_{Rq\bar{q}}(-\bfq^{2})}
\ee
then, away from the asymptotic regime governed by the $\beta$-function,
this heavy quark effective charge Eq.\ (\ref{qqbargeff})
{\em differs} from that Eq.\ (\ref{qcdgeff})
obtained via the PT. 

In order to understand this difference, 
it is necessary to consider the PT in the
limit of very heavy external quark fields.  
In particular, we consider the S-matrix element
at the one-loop level for the
four-fermion scattering process 
$\psi_{i}^{\scriptscriptstyle (f)}(p_{1})
\psi_{i'}^{\scriptscriptstyle (f')}(p_{1}')
\rightarrow
\psi_{j}^{\scriptscriptstyle (f)}(p_{2})
\psi_{j'}^{\scriptscriptstyle (f')}(p_{2}')$,
with $m_{i}\,\, (=m_{j}) = m_{i'} \,\,(=m_{j'}) = M$
and $M^{2} \gg -q^{2}$ where $q^{2} < 0$ (elastic scattering). 
The diagrams for this process are just those shown in Fig.~2.
Evaluating the one-loop integrals for this on-shell process 
in dimensional regularization and then taking the heavy quark
limit, so that
\be\label{limit}
1/\epsilon \sim 
\ln(\Lambda^{2}/\mu^{2}) \gg 
\ln(M^{2}/\mu^{2}) \gg 
\ln(-q^{2}/\mu^{2})
\ee
where $\Lambda$ is the ultraviolet cutoff, one obtains
for the {\em full} (unrenormalized)
one-loop amplitude \cite{papaphilsch} 
\be\label{HQamp}
\Bigl(\overline{u}_{j'}ig\gamma_{\mu}T_{j'i'}^{m}u_{i'}\Bigr)
\,\frac{-i}{q^{2}}\Pi_{q\bar{q}}(q^{2}) \,
\Bigl(\overline{u}_{j}ig\gamma_{\mu}T_{ji}^{m}u_{i}\Bigr)
\ee
i.e.\ a two-point-like interaction with
precisely the self-energy-like function
$\Pi_{q\bar{q}}(q^{2})$ Eq.\ (\ref{qqbarPi}). 
Thus, in the heavy quark limit Eq.\ (\ref{limit}),
one not only isolates the PT ``effective'' 
two-point contribution $\hat{\Pi}(q^{2})$, 
but also additional two-point-like contributions
from the vertex, external leg and box diagrams Figs.\ 2(b)-(f)
which combine to give the term $-4N(g^{2}/16\pi^{2})$ 
in Eq.\ (\ref{qqbarPi})
(In QED, these additional contributions vanish identically
due to the abelian group structure, exactly like the pinch 
contributions, leaving just the
conventional photon self-energy.)

There are three essential observations to make concerning this result. 
First, as emphasized earlier, the PT self-energy component 
of $\Pi_{q\bar{q}}$ Eq.\ (\ref{qqbarPi}) in (\ref{HQamp}) is
obtained before carrying out the loop integration for the
one-loop interaction. By constrast, 
the additional constant component $-4N(g^{2}/16\pi^{2})$ 
is obtained only {\em after} carrying out the loop integration
for the vertex, external leg and box diagrams Figs.\ 2(b)-(e)
and then taking the limit Eq.\ (\ref{limit}).
Thus, after extracting the pinch part contributions 
Figs.\ 2(g)-(i) of these diagrams
and then carrying out the loop integration,
there is a component of the remaining vertex, external leg
and box corrections which has the Lorentz and colour 
structure of a pair of tree level vertices 
$ig\gamma_{\mu}T_{j'i'}^{m}$, $ig\gamma_{\nu}T_{ji}^{n}$,
multiplied by a complicated function
of the momenta and masses of the external fermions. 
In the limit (\ref{limit}), this
function reduces to a constant, giving the additional 
component $-4N(g^{2}/16\pi^{2})$ in (\ref{HQamp}).
Second, while the PT component of $\Pi_{q\bar{q}}$
is universal, the component $-4N(g^{2}/16\pi^{2})$ 
is obtained from a particular process in a particular kinematic
limit. Clearly, for the interaction between external gluons,
there is no analogue of the limit (\ref{limit}).
Third, the amplitude (\ref{HQamp}) is an S-matrix element,
and so is known a priori to be gauge-independent. 
For the case of off-shell external fermions, there is no reason
for the contribution $-4N(g^{2}/16\pi^{2})$ to $\Pi_{q\bar{q}}$
from the vertex, external leg and box diagrams
to remain gauge-independent. This is in contrast to
the PT contribution, which remains always gauge-independent
as a result of the Ward identities of the theory.

Thus, the effective charge Eq.\ (\ref{qqbargeff})
defined from the static heavy quark potential Eq.\ (\ref{V})
does not have a direct interpretation in terms of
radiative corrections included in a renormalized
propagator like Eq.\ (\ref{qedprop}), is not universal, and
is not necessarily gauge-independent for off-shell processes.
We therefore argue that the heavy quark definition, though
appealing from the simple analogy with the QED Coulomb Law,
is inferior to the more purely field-theoretic definition 
of the QCD effective charge given here in Sec.\ 2.

\end{document}